\documentclass[preprint1]{aastex}

\shorttitle{MIR Properties of N66}
\shortauthors{Whelan et al.}

\begin{document}

\title{An In-Depth View of the Mid-Infrared Properties of Point Sources and the Diffuse ISM in the SMC Giant H{\sc ii} Region, N66}

\author{David G. Whelan}
\affil{Department of Astronomy, University of Virginia, P.O. Box
  400325, Charlottesville, VA 22904}
\email{dww7v@astro.virginia.edu}

\author{Vianney Lebouteiller}
\affil{Laboratoire AIM, CEA, Universit\'{e} Paris Diderot, IRFU/Service 
  d'Astrophysique, B\^{a}t. 709, 91191 Gif-sur-Yvette, France}
\email{vianney.lebouteiller@cea.fr}

\author{Fr\'{e}d\'{e}ric Galliano}
\affil{Service d'Astrophysique - Laboratoire AIM, CEA/Saclay, L'Orme 
  des Merisiers, 91191 Gif-sur-Yvette, France}
\email{frederic.galliano@cea.fr}

\author{Els Peeters\altaffilmark{1}}
\affil{Department of Physics \& Astronomy, University of Western
  Ontario, 1151 Richmond Street, London, ON N6A 3K7, Canada}
\email{epeeters@uwo.ca}

\author{Jeronimo Bernard-Salas\altaffilmark{2}}
\affil{IAS, Universit\'{e} Paris-Sud 11, Bat. 121, 91405 Orsay,
  France}
\email{jeronimo.bernard-salas@ias.u-psud.fr}

\author{Kelsey E. Johnson\altaffilmark{3}}
\affil{Department of Astronomy, University of Virginia, P.O. Box
  400325, Charlottesville, VA 22904}
\email{kej7a@virginia.edu}

\author{R\'{e}my Indebetouw\altaffilmark{4}}
\affil{Department of Astronomy, University of Virginia, P.O. Box
  400325, Charlottesville, VA 22904}
\email{ri3e@virginia.edu}

\and

\author{Bernhard R. Brandl}
\affil{Leiden Observatory, Leiden University, P.O. Box 9513, 2300 RA
  Leiden, The Netherlands}
\email{brandl@strw.leidenuniv.nl}

\altaffiltext{1}{SETI Institute, 189 Bernardo Avenue, Suite 100,
  Mountain View, CA 94043, USA}
\altaffiltext{2}{Department of Physical Sciences, Open University, Milton
  Keynes, MK7 6AA, UK}
\altaffiltext{3}{Adjunct at National Radio Astronomy Observatory, 520
   Edgemont Road, Charlottesville, VA 22904, USA}
\altaffiltext{4}{Assistant Staff Scientist, National Radio Astronomy
   Observatory, Charlottesville, VA 22904, USA}

\begin{abstract}

The focus of this work is to study mid-infrared point sources and the
diffuse interstellar medium (ISM) in the low-metallicity
($\sim$~0.2~Z$_{\sun}$) giant H{\sc ii} region N66 in order to
determine properties that may shed light on star formation in these
conditions. Using the {\it Spitzer Space Telescope}'s Infrared
Spectrograph, we study polycyclic aromatic hydrocarbon (PAH), dust
continuum, silicate, and ionic line emission from 14 targeted infrared
point sources as well as spectra of the diffuse ISM that is
representative of both the photodissociation regions (PDRs) and the
H{\sc ii} regions. Among the point source spectra, we
spectroscopically confirm that the brightest mid-infrared point source
is a massive embedded young stellar object, we detect silicates in
emission associated with two young stellar clusters, and we observe
spectral features of a known B[e] supergiant that are more commonly
associated with Herbig Be stars. In the diffuse ISM, we provide
additional evidence that the very small grain population is being
photodestroyed in the hard radiation field. The 11.3~\micron\ PAH
complex emission exhibits an unexplained centroid shift in both the
point source and ISM spectra that should be investigated at higher
signal-to-noise and resolution. Unlike studies of other regions, the
6.2~\micron\ and 7.7~\micron\ band fluxes are decoupled; the data
points cover a large range of I$_{7.7}$/I$_{11.3}$ PAH ratio values
within a narrow band of I$_{6.2}$/I$_{11.3}$ ratio
values. Furthermore, there is a spread in PAH ionization, being more
neutral in the dense PDR where the radiation field is relatively soft,
but ionized in the diffuse ISM/PDR. By contrast, the PAH size
distribution appears to be independent of local ionization
state. Important to unresolved studies of extragalactic
low-metallicity star-forming regions, we find that emission from the
infrared-bright point sources accounts for only 20-35\% of the PAH
emission from the entire region. These results make a comparative
dataset to other star-forming regions with similarly hard and strong
radiation fields.

\end{abstract}


\section{INTRODUCTION \label{introduction}}

The Small Magellanic Cloud (SMC) is an excellent test-bed for studying
star formation in a low-metallicity environment. Its low metallicity
\citep[$\sim$~0.2~Z$_{\sun}$ determined from numerous elemental
  abundances;][]{Russell1992} and strong interstellar radiation field
\citep[ISRF; 4-10 G$_0$,][]{Cox2007} make it an important contrasting
environment to star forming environments in the Milky Way or the Large
Magellanic Cloud (LMC). The SMC is also a good comparative theater to
studies of `passive' star formation in blue compact dwarf galaxies
\citep[BCDs; see][for the distinction between ``active'' and
  ``passive'']{Thuan2008}, because their star-forming regions have
similar densities ($\sim$~100~cm$^{-3}$), star formation rates
($\sim$~0.1~M$_{\sun}$~yr$^{-1}$), radiation field hardnesses, and the
SMC is the lowest-metallicity nearby star-forming region
\citep{Wilke2004, Madden2006}.

N66 \citep{Henize1956} is the largest H{\sc ii} region in the SMC,
covering an area on the sky of approximately 180\arcsec $\times$
300\arcsec, and therefore offers the best view of large-scale star
formation in the SMC. It surrounds a large stellar association known
as NGC~346. N66 contains 33 O stars distributed across the H{\sc ii}
region, which is about half the number for the entire SMC, and 11 of
them are earlier than type O7 \citep{Massey1989}. The most massive
star is of O3{\sc iii}(f*) ($\sim$ 100 M$_{\sun}$) or O3{\sc v}{f*}
($\sim$ 90 M$_{\sun}$) type \citep{Walborn1986,Massey2005}. The
O~stars illuminate the surrounding ISM and are responsible for an
H$\alpha$\ luminosity of about 60 times that of the Orion nebula
\citep{Kennicutt1984}. UV and optical spectra have been used to derive
an age of about 3~Myr for the O~stars in N66 and a metallicity of
0.2~Z$_{\sun}$ \citep[the metallicity has been determined
  independently for individual O stars, forbidden line emission
  originating in the gas, and spectral
  models;][]{Haser1998,Lebouteiller2008,Bouret2003}.

N66 is experiencing ongoing star formation. \citet{Simon2007}
identified about 100 embedded YSOs with {\it Spitzer} IRAC and MIPS
photometry, and \citet{Gouliermis2010} found a further 263 candidate
young stellar sources including intermediate mass pre-main sequence
and Herbig AeBe stars, as well as massive YSO candidates. The first
mid-IR study of N66, with ISOCAM, showed strong nebular [S{\sc iv}]
10.51~\micron\ and [Ne{\sc iii}] 15.56~\micron\ emission across the
region that is indicative of young and massive (O- and B-type) stars,
the presence of faint polycyclic aromatic hydrocarbon (PAH) emission
bands, a mid-infrared continuum from very small grain (VSGs) and large
thermal dust grain emission, and an ISRF at 1600~\AA~$\geq 10^{5}$
times that of solar \citep{Contursi2000}. A companion paper to
\citeauthor{Contursi2000}, \citet{Rubio2000}, included [O{\sc
    iii}]~$\lambda$5007, H$_{2}$ v(1-0) S(1) 2.12~\micron, and CO
observations to show that the peaks in H$_2$, CO, and PAH emission are
all spatially correlated across the photodissociation regions (PDRs)
in general, and further suggested that the CO has been largely
photodissociated across the H{\sc ii} region by the O~star population,
and exists only in small clumps. \citet{Sandstrom2012} included N66 as
part of a study of PAHs in H{\sc ii} regions across the SMC, and
determined that the PAH population is both smaller and less ionized
than in higher-metallicity galaxies. In two comparison studies, the
atomic/ionic gas content and the effects of metallicity on PAH
emission were studied for N66, 30 Doradus in the LMC, and NGC 3603 in
the Milky Way \citep{Lebouteiller2008, Lebouteiller2011}. The
elemental abundances were determined for each region using the ionic
forbidden lines from mid-infrared spectra; for N66, the results
confirmed that the metallicity is about 0.2~Z$_{\sun}$. It was
discovered that the PAHs are photodestroyed in radiation fields where
nebular [Ne{\sc iii}]/[Ne{\sc ii}]~$\gtrsim$~3, and that this
photodestruction law is independent of metallicity. What is still
unclear about N66 is where the PAH emission originates, and what
conditions are traced by the PAH emission. In particular, PAH
ionization state, which is a function of the ultraviolet (UV)
radiation field, is also equally sensitive to electron density
\citep[charge state Z~$\propto$~G$_0$T$^{1/2}$/n$_e$;][]{Tielensbook},
and there are cases evident in the literature that suggest that
neutral PAHs have the ability to exist inside H{\sc ii} regions
\citep[e.g. in the vicinity of the Horsehead
  Nebula:][]{Compiegne2007}.

The reason that star formation is often traced by emission from PAHs
\citep{Peeters2004} is that, while PAHs are sensitive to excitation
from a broad range of wavelengths (UV-IR), they are particularly
susceptible to excitation by UV photons.  PAH emission is commonly
observed in the PDRs around young massive clusters
\citep{LegerPuget1984,Tielens1999}. These spectral features are
predominantly present from 3-17~\micron. The molecules responsible for
this emission are typically dozens to thousands of carbon atoms
large. Following photoexcitation, they emit by fluorescence from
stretching and bending modes either from the carbon-to-hydrogen (C-H)
or carbon-to-carbon (C-C) bonds.

Due to the stochastic excitation and emission mechanism as well as the
ionization balance of PAHs, the local physical conditions have a large
impact on the observed PAH band ratios via radiation field hardness,
column density, dust temperature, and dust composition
\citep{Hony2001, Peeters2002}. Due to the relatively low ionization
potentials of PAHs \citep[about 6-8~eV for small PAHs;][Table
  $6.1$]{Tielensbook}, PDRs are expected to be dominated by ionized
PAHs whereas regions with weaker radiation fields, such as the diffuse
ISM in the Milky Way or inside molecular clouds, should have largely
neutral or negatively-charged PAHs \citep{BakesTielens1994}.

In order to study the mid-infrared properties of N66 in greater detail
with particular emphasis on the PAH emission as an independent tracer
of the physical conditions across the region, we present {\it Spitzer
  Space Telescope}/IRS spectra of a number of infrared point sources
and use the spectral information along the entire IRS long slits to
study the dust and gas emission throughout the H{\sc ii} region and
photodissociation region (PDR). We present the observations in
Section~\ref{observations}, describe our data reduction in
Section~\ref{datareduction}, present our analysis in
Section~\ref{analysis}, and summarize the results in
Section~\ref{Conclusion}.

\section{OBSERVATIONS \label{observations}}

\begin{figure}
\includegraphics[scale=0.4]{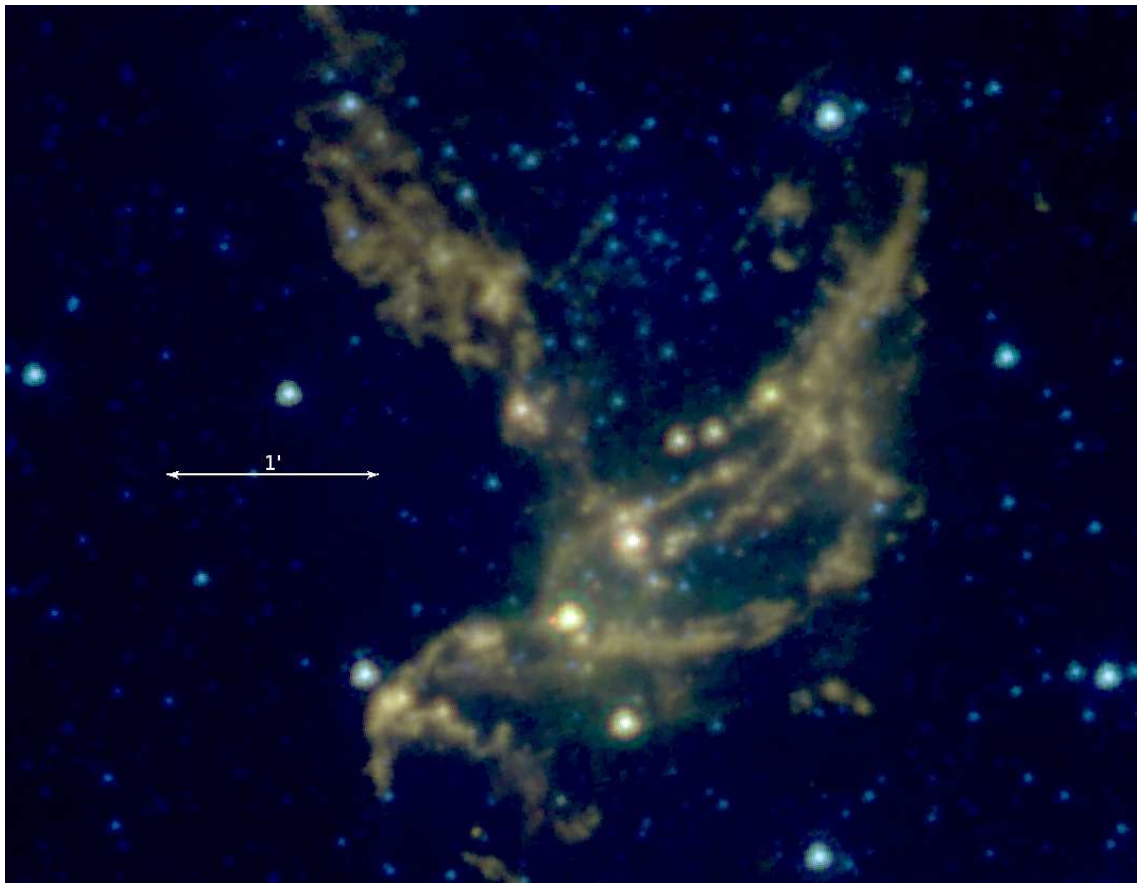}
\caption{A {\it Spitzer}/IRAC 3-color image of N66.  Blue is
  3.6~\micron, green is 5.8~\micron, and red is 8.0~\micron. The
  5.8~\micron\ and 8.0~\micron\ images trace the PAH emission in the
  region, highlighting the PDRs, while the 3.6~\micron\ image traces
  the stellar population; in particular, the old star cluster, BS 90,
  is visible to the north of NGC~346. The angular scale of
  1~\arcmin\ corresponds to a physical distance of 17.6~pc at the
  distance of NGC~346.\label{threecolor}}
\end{figure}

N66 (Figure~\ref{threecolor}) was observed as part of the IRS GTO
program to study massive star formation in giant H{\sc ii} regions in
the Local Group (PID 63). Imaging using {\it Spitzer}/IRAC
\citep{Fazio2004} of the whole region revealed a number of bright
point sources, the brightest eight of which were chosen for
spectroscopic follow-up with the low- and high-resolution modules of
the {\it Spitzer} Infrared Spectrograph \citep[IRS;][]{Houck2004}.
The present study concentrates on the short-wavelength low-resolution
spectra only ($\lambda$~$<$~14.7~\micron); the high-resolution {\it
  Spitzer} spectra have been analysed in \citet{Lebouteiller2008,
  Lebouteiller2011} to determine element abundances and to constrain
the strength of the interstellar radiation field (ISRF) using
forbidden emission lines, and the long-wavelngth low-resolution data
(15~\micron~$<$~$\lambda$~$<$~37 \micron) have not been included due
to the spatial resolution and slit orientation mismatch. This dataset
is therefore in contrast to that presented in \citet{Sandstrom2012},
in which long- and short-wavelength spectral maps are used to analyze
N66. While they analyzed full spectral maps of the entire region, the
data we present here is deeper and allows us to study the faint
extended emission around the point sources at a level of detail that
the full maps did not allow.

\begin{figure}
\includegraphics[scale=0.4]{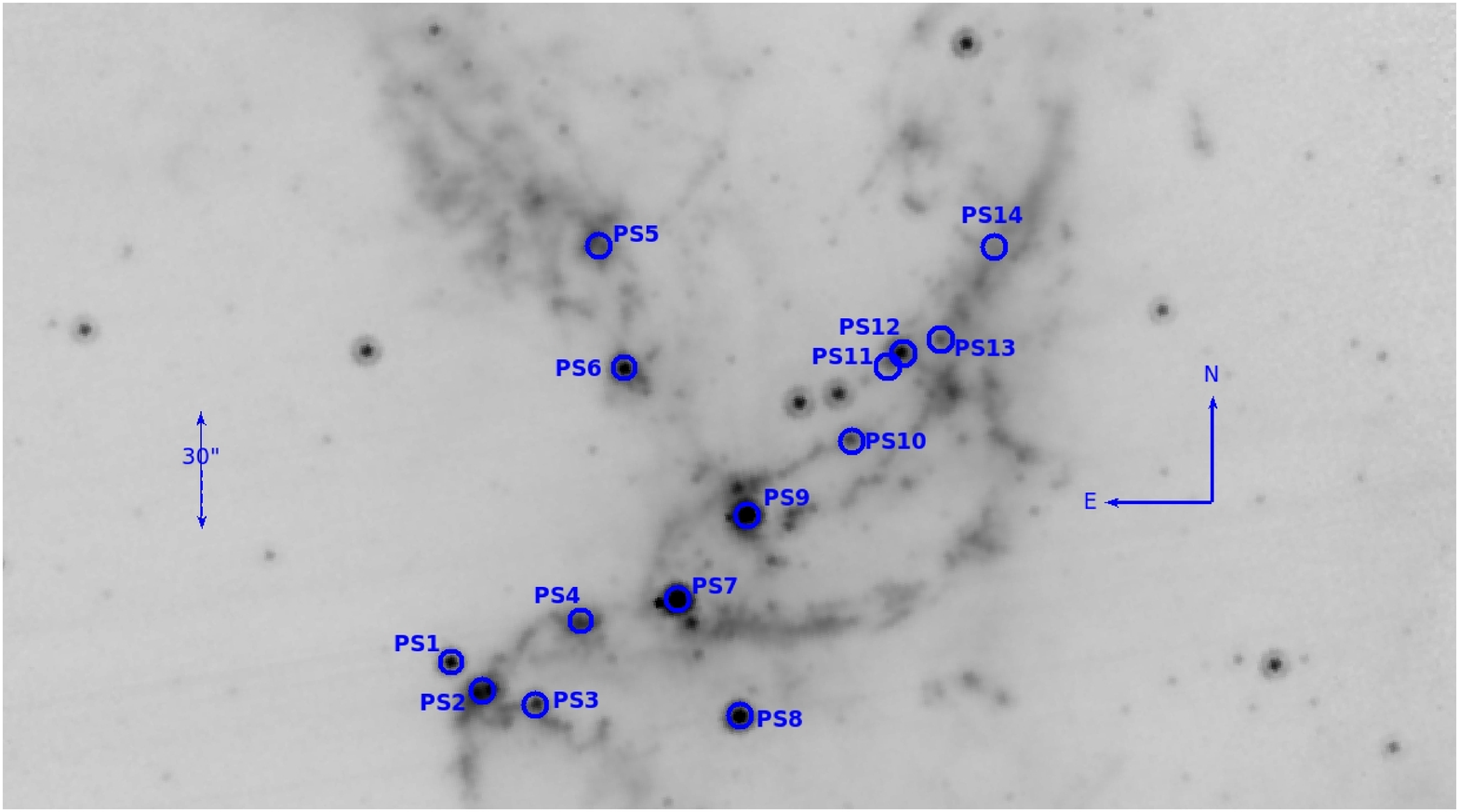}
\includegraphics[scale=0.4]{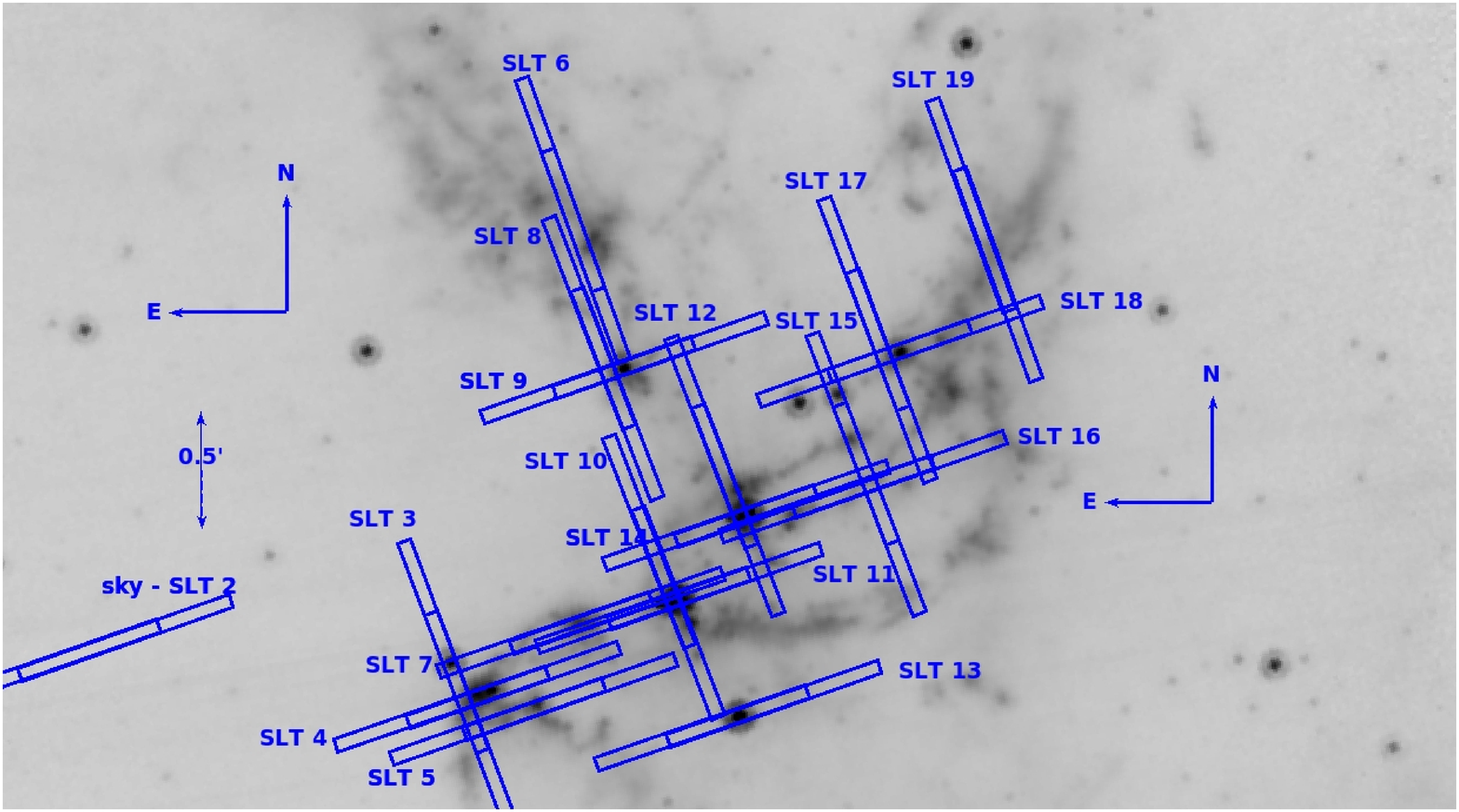}
\caption{The {\em IRAC} 8.0~\micron\ image with the observed point
  sources circled and labeled in blue, and the labeled slit positions
  across the region. See Tables~\ref{targets} and \ref{slitcentroids}
  for coordinate information.\label{psposns}}
\end{figure}

The pointings for the original eight positions (Astronomy Observing
Request 4385024) were systematically offset by about 5\arcsec\ with
respect to the actual source positions. As a result, most of the
intended point sources were partially or wholly outside of the
3.6\arcsec -wide slit. However, there is at least one point source in
each of these slits regardless, though they are not always the
intended ones. In order to target the intended point sources, N66 was
re-observed with correct astrometry (AOR 16207872). The point sources
detected in the original and re-observed samples are shown in
Figure~\ref{psposns} along with the slit orientations. See
Table~\ref{targets} for a full list of the point source (PS)
positions, labeled in order of decreasing right ascension, and
Table~\ref{slitcentroids} for a list of slit centroids.

Three exposures were taken at each nod position in staring mode (see
the {\it Spitzer/IRS} manual), where the source is positioned at 1/3
(nod1) and 2/3 (nod2) of the slit length. The original observations
had 60-second exposure times, whereas the re-observed sample have
15-second exposure times.

The low-resolution spectra ($\lambda/\Delta \lambda$ $\sim$ 65-130)
cover the spectral range 5.2-14.7~\micron\ using the short-low (SL)
module. The SL-1 slit covers the spectral range 7.4-14.7~\micron\ and
the SL-2 slit covers the spectral range 5.2-7.7~\micron. The sizes of
the two slits (3.7\arcsec~$\times$~57\arcsec\ and
3.6\arcsec~$\times$~57\arcsec\ respectively) correspond to a physical
size of 1.1~$\times$~16.7~pc at the distance of N66 \citep[we adopt a
  value of 60.6~kpc from][]{Hilditch2005}. Therefore the point
sources, with a FWHM at 6~\micron\ of 4 pixels out of 34, are resolved
down to about 2~pc along the slit.

\section{DATA REDUCTION \label{datareduction}}

\subsection{Spitzer/IRS Image Reduction \label{imageredux}}

The {\it Spitzer/IRS} images were processed with the {\it Spitzer
  Science Center (SSC)} pipeline, version S18.7. The basic calibrated
data (BCD) were used. Images were cleaned by interpolating values over
flagged bad pixels using the {\it SSC}-provided IDL package
IRSCLEAN\footnote{All {\it SSC}-provided packages may be found at
  $http://irsa.ipac.caltech.edu/data/SPITZER/docs/dataanalysistools/$}. After
cleaning, the observed sky positions were subtracted from the other
positions in their respective AORs to remove any intervening flux from
zodiacal light, the Milky Way, and foreground SMC. Since the sky
positions are relatively far from N66, we may assume that emission in
the sky spectra is primarily from the diffuse SMC and not from
N66. The spectra of these sky positions exhibit [S{\sc iv}] emission
but no dust features. The strength of the sky position's [S{\sc iv}]
emission is 7.9~$\times$~10$^{-21}$~W~cm$^{-2}$, which is almost 50\%
the value for the lowest [S{\sc iv}] flux found in the extended
emission spectra (see \S~\ref{srcext}), and less than 20\% of the flux
for the rest of the spectra.

\subsection{Spectral Extractions \label{srcext}}

\begin{figure}
\includegraphics[scale=0.7]{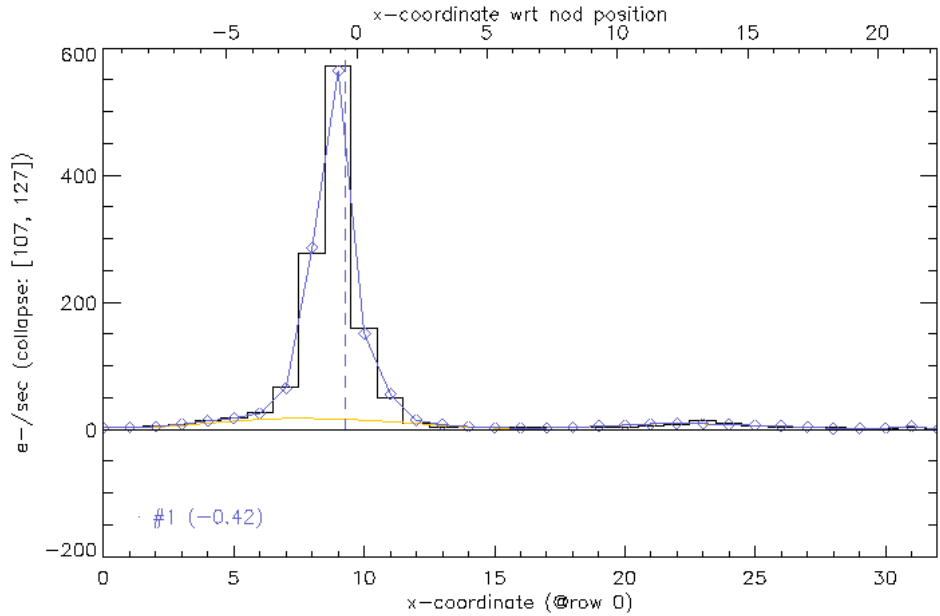}
\caption{The spatial extent of an extracted point source and the
  underlying extended emission for the 7-9~\micron\ continuum as
  output by SMART-AdOpt. The yellow line is a polynomial fit to the
  background emission in the slit and a stellar template is used to
  fit the point source emission. The IRS data are shown in black
  histogram. The point source is labeled, as would other point sources
  found, at the bottom of the plot, along with its centroid in pixels
  with respect to the nod position (on top of the plot) and with
  respect to the edge of the slit (bottom of the
  plot). \label{optext_fit}}
\end{figure}

Using the optimal extraction routines available in the {\it
  Spectroscopy Modeling Analysis and Reduction Tool}
\citep[SMART-AdOpt;][]{Lebouteiller2010, Higdon2004}, point source and
extended emission can be simultaneously extracted. SMART's optimal
extraction routines fit a polynomial function to the extended emission
as well as template supersampled point spread functions (PSFs) to
point sources in the slit, as shown in Figure~\ref{optext_fit}. The
backgrounds are relatively smooth for these spectra, and polynomials
of four orders and less were used. Details, including how to fit
partially extended point sources and measuring the residuals, are
presented in \citet{Lebouteiller2010}. However, the fundamental
operation demands that for each wavelength element, the relative
weights to the total flux assigned to the point source(s) and
background are measured simultaneously. Therefore, the flux is
distributed to each component in a way that agrees with the
combination of point source(s) and polynomially-fitted background.

\begin{figure}
\includegraphics[scale=0.9]{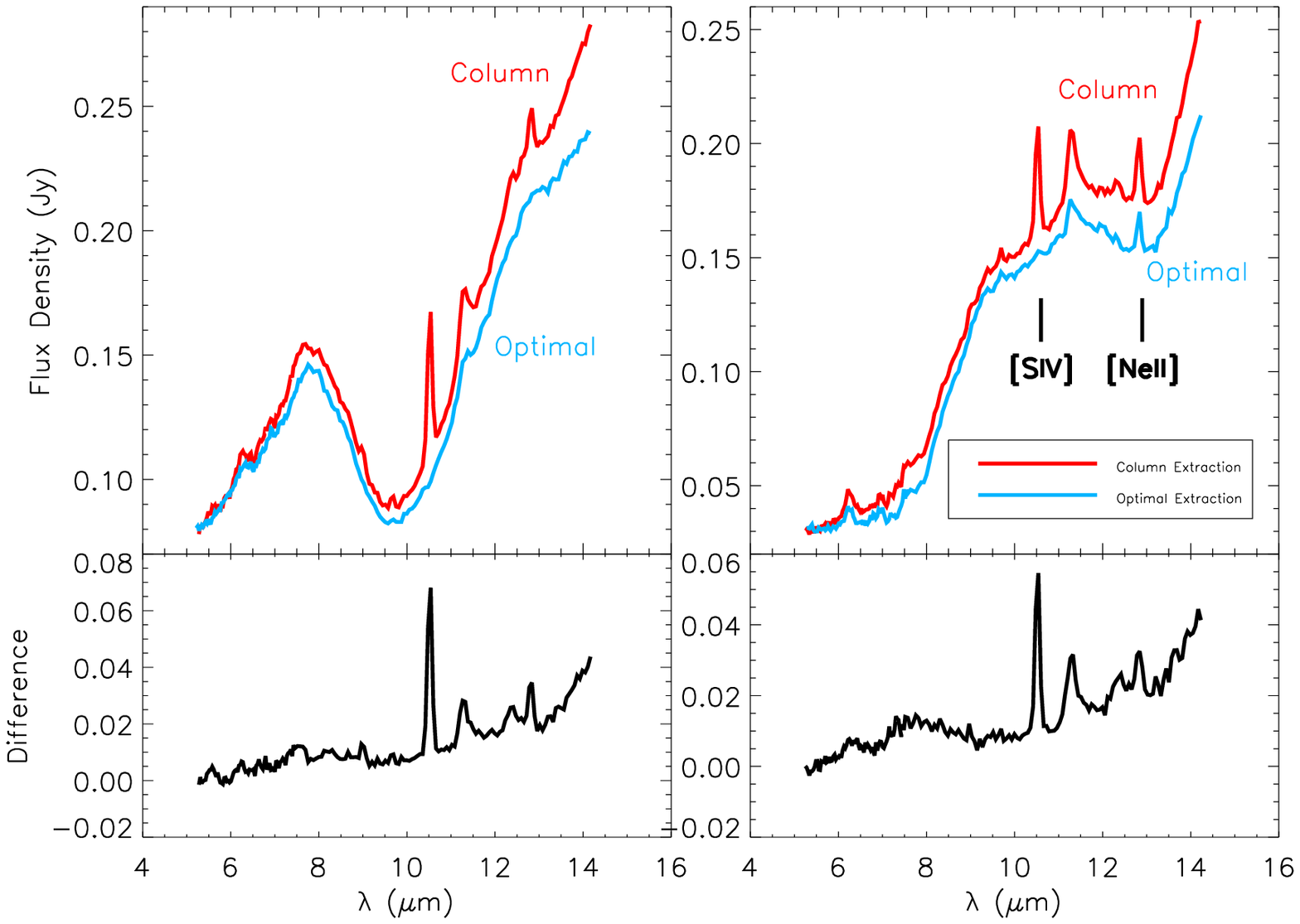}
\caption{Variable column extractions (in red) are shown overplotted
  with optimal point source extractions (in blue) for the two
  brightest IR point sources in N66: the massive YSO in N66A (left,
  PS7) and NGC~346 (right, PS9). The column extractions include a
  significant amount of emission along the line-of-sight that is not
  associated with the point source: the [S{\sc
      iv}]~10.51~\micron\ emission, a portion of the [Ne{\sc
      ii}]~12.81~\micron\ emission, and some dust continuum emission
  is not associated with these point source spectra, but is nebular in
  nature.  This nebular emission is shown in the ``Difference''
  spectra in the bottom panels.\label{optext_comp}}
\end{figure}

The advantage of simultaneous point source and extended emission
extraction compared to more traditional, variable column extractions
is illustrated in Figure~\ref{optext_comp}. Two point sources have
been extracted using two different methods as an example. The source
in the left-hand plot is a massive embedded young stellar object
(YSO), PS7, and the right-hand plot shows the spectrum of the
unresolved dust emission associated with the young star cluster NGC
346, PS9. The `column extraction' extracts all of the emission inside
a column that varies in width with the PSF. The `optimal extraction'
uses the SMART-AdOpt routines. There are significant differences
revealed by the optimal extraction method. For example, the [S{\sc
    iv}]~10.51~\micron\ emission line, which is often prominent in
H{\sc ii} regions, is not associated with the point source spectra;
the reason it appears in the column extraction is because it is
nebular emission along the line-of-sight to the point source. In fact,
we would get the wrong diagnostic with a variable column extraction,
since the nebular [S{\sc iv}] would be erroneously assigned to the
point source. The [Ne{\sc ii}]~12.81~\micron\ emission line is also
nebular in nature for the massive embedded YSO (left-hand plot). In
addition to the nebular forbidden atomic lines, a certain amount of
the dust continuum is also missing from the optimally extracted
spectra, meaning that it is also nebular in nature. By separating out
the point source from the line-of-sight nebular emission, we have
separate views of emission that is associated with the unresolved
point sources in N66 and emission that originates in the diffuse H{\sc
  ii} region and PDR.

Optimal extraction of all of the pointings was performed, producing
point source and extended emission spectra from the full slit length
for each pointing. Two of the re-observed pointings have extended
emission spectra that were dominated by noise and are not used in the
analysis.

Observations where the point source lies partially outside of the slit
in the dispersion direction were extracted with a special tool
developed for and included in SMART-AdOpt. Applying the centered PSF
template to an observation with incorrect pointing makes the spectral
shape incorrect, but accounting for the pointing error allows a
properly-shaped PSF to be fit to the point source so that regular
optimal extraction and accurate flux calibration can be done. For
details on the source extraction for sources offset in the dispersion
direction, see \citet{Lebouteiller2010}. The reduced spectra are shown
in Appendix 1 at the end of this paper.

\subsection{Spectral Decomposition \label{specfit}}

Typical mid-infrared spectra of H{\sc ii} regions and their
surrounding PDRs have the following features: a dust continuum,
several prominent PAH features, molecular hydrogen emission lines,
atomic and ionic emission lines, and absorption or emission features
due to small silicaceous grains centered at 9.8~\micron\ and
17~\micron. In order to study the dust and gas diagnostics in N66, all
of the spectra were decomposed into these components.

\begin{figure}
\includegraphics[scale=0.5]{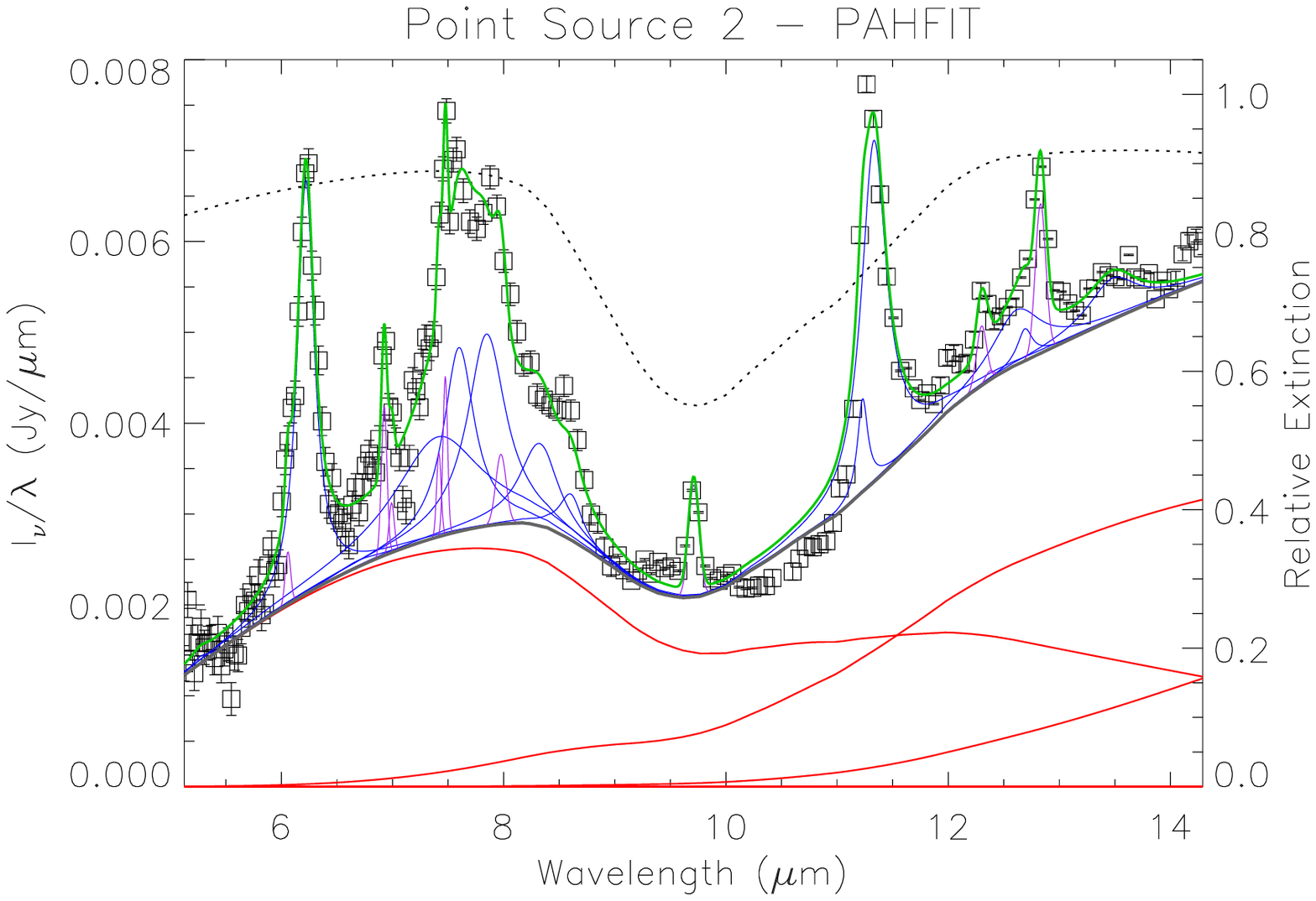}
\includegraphics[scale=0.5]{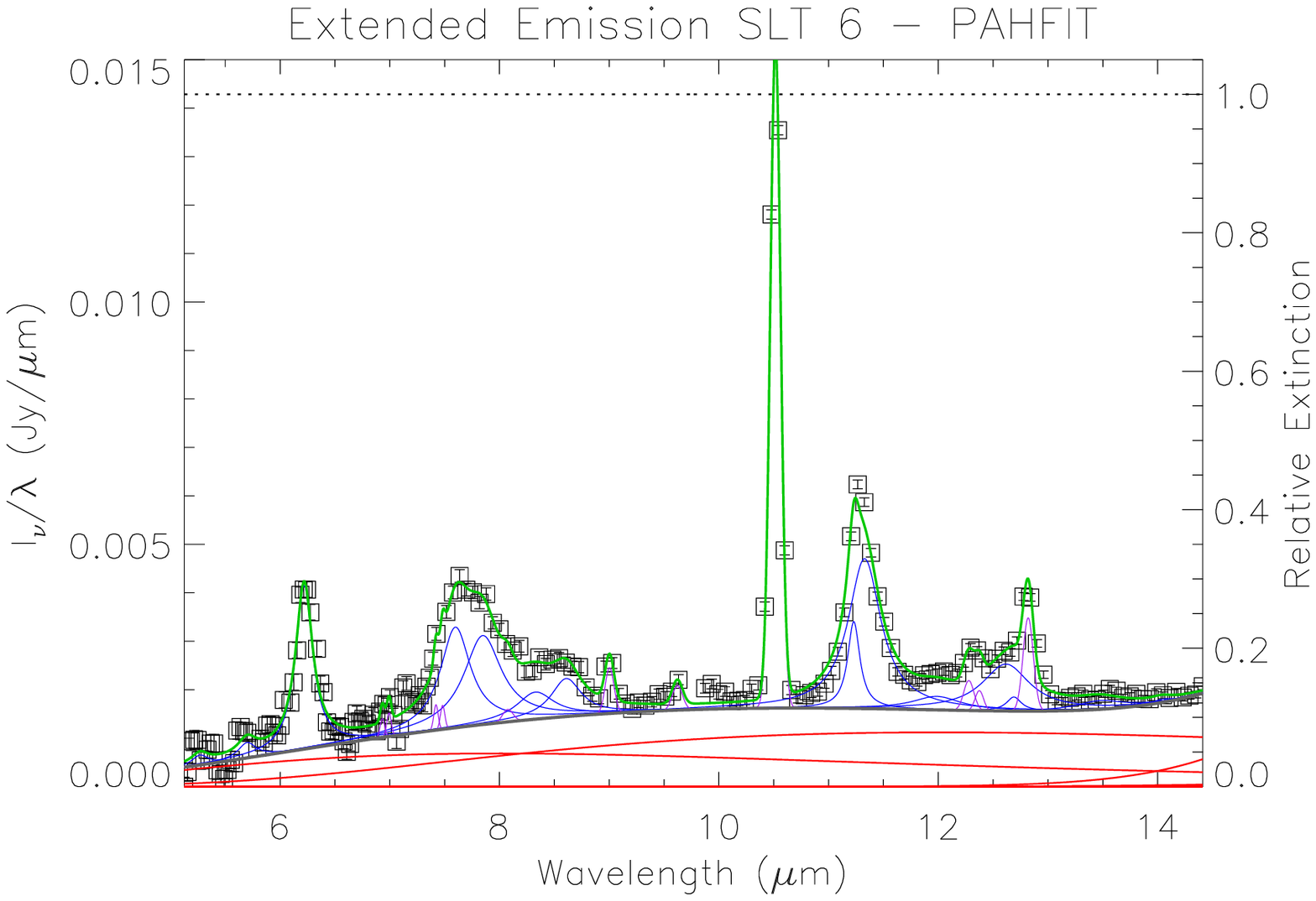}
\caption{Example spectral fits with PAHFIT are shown for two sources,
  point source PS2 and the extended emission spectrum SLT6. The red
  lines are blackbody fits to the dust continuum, the blue features
  are PAHs, the narrow features in purple are atomic and molecular
  emission lines, the dotted line is the relative extinction and the
  green line is the composite fit to the data
  (squares).\label{DecompComp}}
\end{figure}

We used the specral decomposition tool PAHFIT \citep{Smith2007}.
PAHFIT models the dust continuum that underlies the PAH features with
as many as eight blackbody continua at fixed temperatures less than
300~K, fits the silicate absorption features at 9.8~\micron\ and
17~\micron, and models the PAH features with asymmetric Drude
profiles. In order to fit the spectra for N66 with PAHFIT, a couple of
small changes were made. In some cases, the input full-width at half
maxima (FWHM) of the 6.2~\micron\ and 11.3~\micron\ features needed to
be changed in order to fit the features more precisely (see
\S~\ref{pahprofiles} below for a discussion about the
11.3~\micron\ PAH feature profiles). Second, for sources that exhibit
a silicate emission feature, two additional Drude profiles were added
to model the 9.8~\micron\ silicate emission feature: these Drude
profiles, centered at 9.2~\micron\ and 10.0~\micron, have FWHM of
0.3~\micron\ and 0.2~\micron\ respectively. For spectra with silicate
emission features included, PAHFIT did {\it not} also fit silicate
absorption features. Due to the presence of Hydrogen recombination
lines in some of the spectra, we include the H{\sc i}~6-5 7.46
\micron\ line to PAHFIT's line list. Example point source and extended
emission PAHFIT fits are shown in the top panels of
Figure~\ref{DecompComp}, where the plotting scheme is explained in the
caption. One major difference in the fits between the point source and
extended emission spectra is that the silicate absorption feature is
only fit to those point source spectra that do not explicitly require
a silicate emission feature, and none of the extended emission
features exhibit silicate absorption in the final fits.

There is some concern that PAH fitting routines like PAHFIT may be
introducing errors inadvertently into the PAH fluxes. When applying
PAHFIT, the underlying PAH ``plateaux'' are incorporated into the
wings of the Drude profiles used to fit the PAH bands, in particular
for the 7.7~\micron\ band. Peeters et al. (2013, in preparation) has
found that the spatial morphology of the plateau is quite distinct
from that of the 6.2, 7.7, and 8.6~\micron\ PAH bands in data of the
reflection nebula NGC~2023. To see if the PAH plateaux in the N66
spectra is also decoupled from the PAH features, we have fit a spline
to the underlying continuum and measured the PAH features above the
continuum, as in \citet{Galliano2008b}. We have determined that,
although the exact values for the PAH band fluxes differ depending on
the decomposition method used, the trends discussed in detail in
\S~\ref{pahratios} hold regardless. The spectral decomposition method
illustrated in Figure~\ref{DecompComp} is a reliable method for
comparing PAH band fluxes.

Due to concerns about error propagation from extraction to
decomposition, the errors of the PAH feature strengths were determined
with a Monte Carlo method. The root mean square error in the continuum
near 6~\micron\ was measured for each spectrum and then used to
randomly perturb the spectrum. 6~\micron\ was chosen because the noise
is systematically higher at the shortest wavelengths in the SL
spectrum, due to the lower signal in these sources at this
wavelength. The noise is, on average, two times greater at
6~\micron\ than it is at 14~\micron, measured as the standard
deviation of the root mean square of the difference of the unperturbed
spectrum and the best fit PAHFIT model. The spectrum, including the
random noise, was then fit with PAHFIT and the resultant strengths of
the PAH features were recorded. This was done 300 times, after which
the standard deviation of the PAH feature strength measurements was
taken as the statistical error in the 300 fitted feature strengths.

While PAHFIT also fits any number of user-requested atomic, ionic, and
molecular features in addition to the dust continuum and PAH features,
we fit the narrow lines manually with the IDEA tool in SMART. The
reason for this additional step is that PAHFIT functions first and
foremost as a fitting tool, and will therefore include all of the
parameters it is given into the final solution. This means that in the
point source spectra, for instance, PAHFIT will generally give the
[S{\sc iv}]~10.51~\micron\ line a flux even when it is clear from a
visual inspection that this line is not detected. Therefore, in order
to systematically determine narrow line fluxes and upper limits, we
chose to use a manual tool for line fits. We used a first-order local
continuum subtraction to estimate continuum flux beneath the narrow
features. The [Ne{\sc ii}]~12.81~\micron\ feature is blended with a
PAH feature and the line flux may therefore contain some PAH emission;
due to the presence of PAHs in all spectra, we believe that any offset
due to this contamination is largely systematic but may contribute
substantially to the point source spectra PS12, PS13, and PS14, which
exhibit much lower [Ne{\sc ii}] fluxes than the other spectra. The
continuum at 13.7-14.2~\micron\ contains two prominent PAH features
that, when compared to the PAHFIT dust continuum and PAH feature fits,
systematically contributed roughly twenty per-cent to the continuum at
these wavelengths; the continuum was therefore measured as the average
in the PAHFIT dust continuum beneath the PAH features. The final
atomic, ionic, PAH, and continuum measurements along with their errors
are given in Tables~\ref{lines} \& \ref{PAHs}.

\section{ANALYSIS\label{analysis}}

\subsection{Point Source Identification}

We carried out a literature search in order to identify known objects
associated with the infrared point source centroids.
Table~\ref{targets} lists the targeted point sources from this study,
corresponding labels from \citet{Lebouteiller2008} and
\citet{Contursi2000}, and additional information from the
literature. All point sources are identified as sites of recent or
ongoing star formation, and are typically young protostars or young
stellar clusters. A number of the point sources exhibit interesting
spectral features that are worth a closer look. We describe these
point sources in detail below.

\subsubsection{Young Star Clusters Exhibiting Silicates in Emission}

\begin{figure}
\includegraphics[scale=0.6,angle=90]{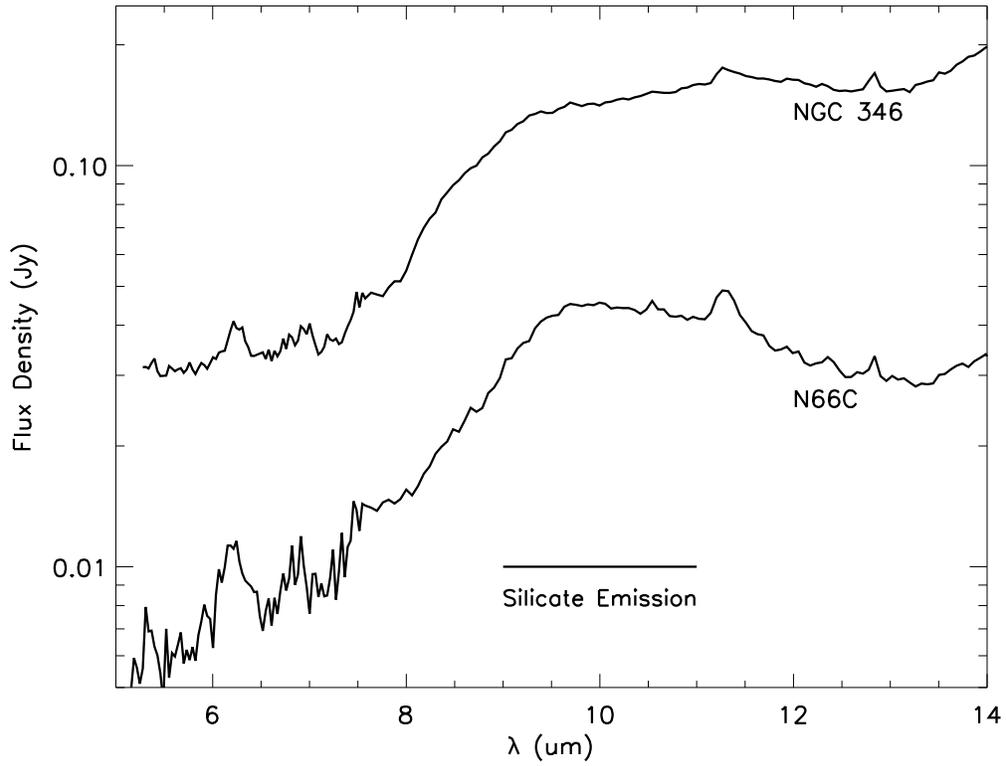}
\caption{The spectra of two stellar clusters in the N66 giant H{\sc
    ii} region: NGC~346 (PS9), the central source and brightest
  optically visible cluster, and N66B (PS6), a large stellar cluster
  to the north of NGC~346 (see Figure~\ref{psposns}). Both show
  pronounced silicate emission features.
  \label{starclusters}}
\end{figure}

Unresolved emission from two bright star clusters in N66 exhibit
silicate emission features in their spectra: NGC~346 (PS9) and N66B
(PS6); see Figure~\ref{starclusters}. Both of these point sources are
bright H$\alpha$\ sources \citep{Henize1956}, contain `blue' stars
\citep{Massey1989, Gouliermis2006}, and have been modeled as
$\sim$~3~Myr old with {\it Hubble} color-magnitude diagrams
\citep{Sabbi2007}. This age is consistent with the presence of remnant
dust from the natal cloud surrounding the cluster.

The origin of the silicate emission is most likely an optically thin
layer of intracluster and/or enveloping dust that has been heated to
$\sim$~200~K by the stars in the central clusters. The presence of
silicate emission suggests a relatively low optical depth; this is
supported by optical data of N66 \citep[see, for example, the
  high-resolution {\it Hubble} data presented in ][]{Sabbi2007}, where
the stars in the centers of the clusters are clearly visible through
the intervening dust, and supported by the measured extinction by dust
\citep{Caplan1996}. This geometry is similar to clumpy model
geometries for young star clusters presented in \citet{Whelan2011},
where silicate emission/absorption was found to be dependent on the
line-of-sight dust geometry. Neither cluster is resolved in the
spectral slit, and we therefore did not employ the IRS LL module (15
$<$ $\lambda$ $<$ 37\micron) to measure the 17~\micron\ silicate
emission as in \citet{Hao2005} and \citet{Sturm2005} because the flux
mis-match between the SL and LL {\it Spitzer}/IRS modules is very
substantial and would bias the silicate dust temperature
measurement. The presence of an O5.5V star in N66B and an O9V star in
NGC~346 means that the point source extractions show a little [S{\sc
    iv}] emission in N66B, none in NGC~346, and [Ne{\sc ii}] emission
in both. The [S{\sc iv}] emission in N66B is likely due to the O5.5V
star. See \citet{Sabbi2008}, Figure 7 for a map of the positions of
the known O stars across N66, many of which appear off of the stellar
cluster positions.

Silicates in emission associated with young star clusters have not
been observed regularly before. In NGC~3603, pointings on and near the
central star cluster show silicate emission
\citep[][]{Lebouteiller2007}. There is one pointing in 30~Doradus in
the LMC \citep[source B;][ and also found in Lebouteiller et al$.$
  2008]{Indebetouw2009} near R136 that exhibits silicate emission, but
this source has been spectroscopically identified as an M-type
supergiant star \citep{Parker1993}. While there are numerous
detections of silicates in emission among protoplanetary systems
\citep[$e.g.$][]{Furlan2011,SiciliaAguilar2007} and evolved stars such
as asymptotic giant branch stars \citep[AGBs, both galactically and
  extragalactically; see][]{Sloan1995, Lebouteiller2012}, there are
relatively few young stellar clusters that show silicate
emission. \citet{Robberto2005} show that a diffuse silicate population
is likely across the Orion nebula. Compared with those observations,
the silicate emission observed in NGC~3603, NGC~346, and N66B is
distinct in that it is clearly associated with the stellar clusters
and not visibly dispersed across the region. It seems likely that the
strong silicate emission associated with the star clusters in N66 and
NGC~3603 is tied to a relatively short period of time in the early
evolution of star clusters and will only last a short period of time;
that these clusters are definitively young
\citep[e.g.][]{Gouliermis2006} and therefore contain no AGB stars
excludes the possibility of silicate-rich winds from post-main
sequence stars contributing to the observed silicate emission.

\subsubsection{Silicate and PAH Emission Associated with a B[e] star}

There is a third spectrum exhibiting a silicate emission feature:
PS8. In this instance, the silicate feature is not as pronounced as
for the star clusters discussed above, though it has pronounced PAH
features as well (see Figure~\ref{Spec1} in the Appendix). Searching
by position, we found a Be star at those coordinates, Cl* NGC~346
KWBBE~200. \citet{Wisniewski2007} fit a UV-to-8~\micron\ spectral
energy distribution (SED) with a B-star template and a T~$\sim$~800~K
blackbody, observed P~Cygni profiles on a number of optical spectral
lines, roughly determined a luminosity of 10$^{4.4}$~L$_{\sun}$, and
concluded that this source is a B[e] supergiant. Evidence against it
being a Herbig Be system is that no inverse P~Cygni profiles
associated with infall were observed, and that the derived luminosity
is on the high end for Herbig Be stars. We note, however, that
silicate emission and strong PAH bands are often detected in Herbig
AeBe star systems \citep[e.g.][]{Keller2008}, where cooler dust and
PAHs are expected due to its young age. Additionally, PS8 exhibits a
Class~A PAH spectrum as shown in Figure~\ref{PAH7786} and discussed in
\S~\ref{pahprofiles}, which is typical for non-isolated Herbig~AeBe
stars \citep{Peeters2002}, and is a 24~\micron\ source as observed by
{\it Spitzer}/MIPS \citep{Rieke2004}, suggesting that there is cold
dust that \citeauthor{Wisniewski2007}'s SED fit did not account
for. While the absence of inverse P~Cygni profiles offers a conundrum,
the evidence from the mid-IR suggests that KWBBE~200 is, in fact, a
Herbig AeBe star, not a B[e] supergiant.

\subsubsection{An Embedded Massive Young Stellar Object at the Edge of H{\sc ii} Region N66A\label{yso}}

\begin{figure}
\includegraphics[scale=0.6,angle=90]{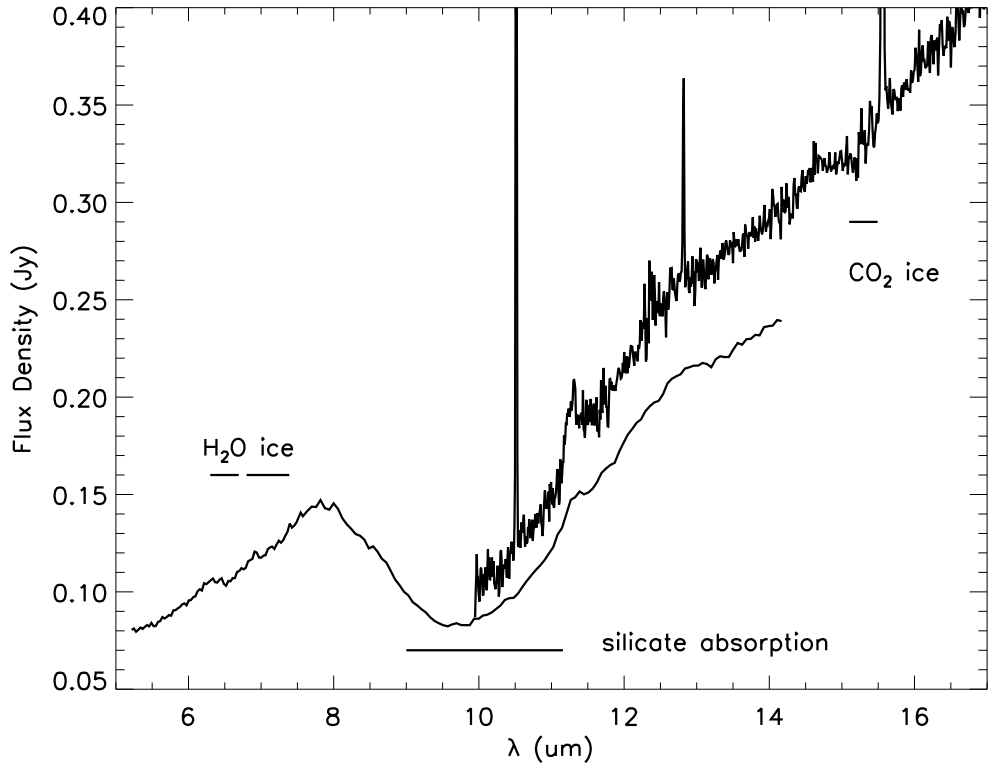}
\caption{The {\em Spitzer/IRS} SL/SH spectrum of N66A exhibits
  features commonly associated with a massive young stellar object:
  water ice features between 6-7.5~\micron, deep silicate absorption
  indicative of high optical depth, CO$_2$ ice at 15~\micron, and a
  strong mid-infrared continuum. The nebular lines seen in the
  high-resolution spectrum come from intervening diffuse
  material.  \label{ysospec}}
\end{figure}

The single brightest mid-IR point source in N66, PS7, lying at the
location of the H{\sc ii} region N66A \citep{Henize1956}, was found to
be a 16.6~M$_{\sun}$ class I young stellar object (YSO) by
\citet{Simon2007} using {\it Spitzer/IRAC} and {\it MIPS} photometry
matched to models of class I protostars presented in
\citet{Robitaille2006}. Shown in Figure~\ref{ysospec}, the spectrum
has a deep silicate feature at 9.8~\micron\ which corresponds to high
optical depth and therefore supports the class I
designation. Furthermore, the presence of H$_2$O ice in the
6-7.5~\micron\ range and a CO$_2$ ice feature at 15~\micron\ in the
high-resolution {\it Spitzer} spectrum of N66A suggests cold, dense
conditions similar to other massive class~I YSO environments
\citep[$e.g.$][]{vanLoon2005}. Lastly, there is no [S{\sc iv}] or
      [Ne{\sc ii}] detected, suggesting that the central heating
      source does not have a strong UV continuum. These spectroscopic
      signatures all confirm \citeauthor{Simon2007}'s original
      designation.

To confirm the YSO mass, we fit {\it Spitzer}/IRS data (wavelength
coverage from 5 to 35~\micron) with the \citeauthor{Robitaille2006}
models\footnote{We used the SED fitting routine described in
  \citet{Robitaille2007} and available to the public via:
  http://caravan.astro.wisc.edu/protostars/}. For these fits of an
embedded protostar, stellar temperature (and mass), disk/envelope mass
and inner/outer radii are all fit, though for embedded sources disk
masses are known to be ill-constrained. The best fit to our data was a
17.8~M$_{\sun}$ YSO embedded in an envelope of about
10$^{3}$~M$_{\sun}$. The circumstellar extinction is calculated to be
A$_V$ = 25.7, and the interstellar extinction is A$_V$ = 0.1. For
comparison, $\tau_{9.8\micron}$=1.88 from the PAHFIT parameter fit to
this source, which is consistent with the circumstellar extinction
\citep{Roche1985}. The total luminosity of this model is
3.47~$\times$~10$^4$~L$_\sun$. This result differs from the fit
presented in \citet{Simon2007} by a $\sim$~7\% increase in mass and
$\sim$~15\% increase in luminosity.

\citet{Heydari2010} studied the {\it Hubble} data for the H{\sc ii}
region N66A in great detail and determined that it is supported by an
O8 star. While the massive YSO discussed in this section dominates the
infrared emission, the fine structure emission in the extended
emission spectrum is due to the young O- and B-star population in
N66A, while the diffuse dust emission probably comes from the PDR at
the interface to the molecular cloud in which the massive YSO is
buried. There are in fact two YSOs at PS9's centroid in the
\citet{Simon2007} atlas; the more massive YSO studied here dominates
the infrared luminosity substantially: there is a factor of about 19
ratio between the luminosities of the two YSOs. At the sensitivity of
these data, the massive YSO is the only point source detected.

\subsection{Ionic Lines \label{siv}}

Ionic emission lines in an H{\sc ii} region can help quantify the
strength and hardness of the radiation \citep[see][]{Lebouteiller2011}
but may also be used to shed light on the physical characteristics of
the point sources and diffuse emission across the region. In general,
the point source spectra are all associated with sites of active star
formation. By contrast, the extended emission spectra trace the H{\sc
  ii} and PDR emission that is photoexcited by the massive stellar
population dispersed across the region. For this dataset, we find two
distinguishing features in the ionic line emission that are specially
worth noting.

(1) [S{\sc iv}] 10.51~\micron\ emission is largely undetected among
the point source spectra but is detected in all of the extended
emission spectra.

(2) [Ne{\sc ii}] 12.81~\micron\ is detected in every point source and
extended emission spectrum with only two exceptions among the point
source spectra.

The strong [S{\sc iv}] line, due to ionization of the interstellar gas
by the O star population, is seen in the extended emission spectra
across the region (see Figures~\ref{Spec3} and~\ref{Spec4}). [S{\sc
    iv}] and [Ne{\sc ii}] are often found in sites of active star
formation, e.g. giant H{\sc ii} regions \citep{Lebouteiller2008}, blue
compact dwarf galaxies \citep[BCDs;][]{Wu2006}, and starburst galaxies
\citep{Brandl2006,BernardSalas2009}. Our analysis differs from
previous works by separating the diffuse emission from the infrared
point source emission. We show that the dense regions embedded in the
PDR do not generally show [S{\sc iv}] emission. The absence of [S{\sc
    iv}] in most of the point source sample is likely because the
heating sources at the point source locations are not hard enough to
triply-ionize sulfur. The gas density at these positions can be
estimated from the [S{\sc iii}] 18.71~\micron\ and
33.48~\micron\ lines fluxes from high-resolution {\it Spitzer} data at
the point source positions published in \citet{Lebouteiller2008}. The
ratio of these two lines ranges from 0.43-2.7, suggesting densities of
n$_e$T$_4^{1/2}$~$<$~3~$\times$~10$^3$~cm$^{-3}$ for the lowest value
ratio and below about 10$^2$~cm$^{-3}$ for most of the positions, well
less than the critical densities of $n_{crit}$([S{\sc iv}])$ = 5.4
\times$~10$^{4}$~cm$^{-3}$ $n_{crit}$([Ne{\sc ii}])$ = 6.5
\times$~10$^5$~cm$^{-3}$. It should be noted that these estimates for
the [S{\sc iii}]-derived gas density are subject to the diffuse as
well as the dense material at these positions because the point source
emission cannot be treated separately from the diffuse emission in the
high-resolution modules as it can for the low-resolution modules.

\subsection{PAH Feature Profiles\label{pahprofiles}}

\begin{figure}
\includegraphics[scale=0.7]{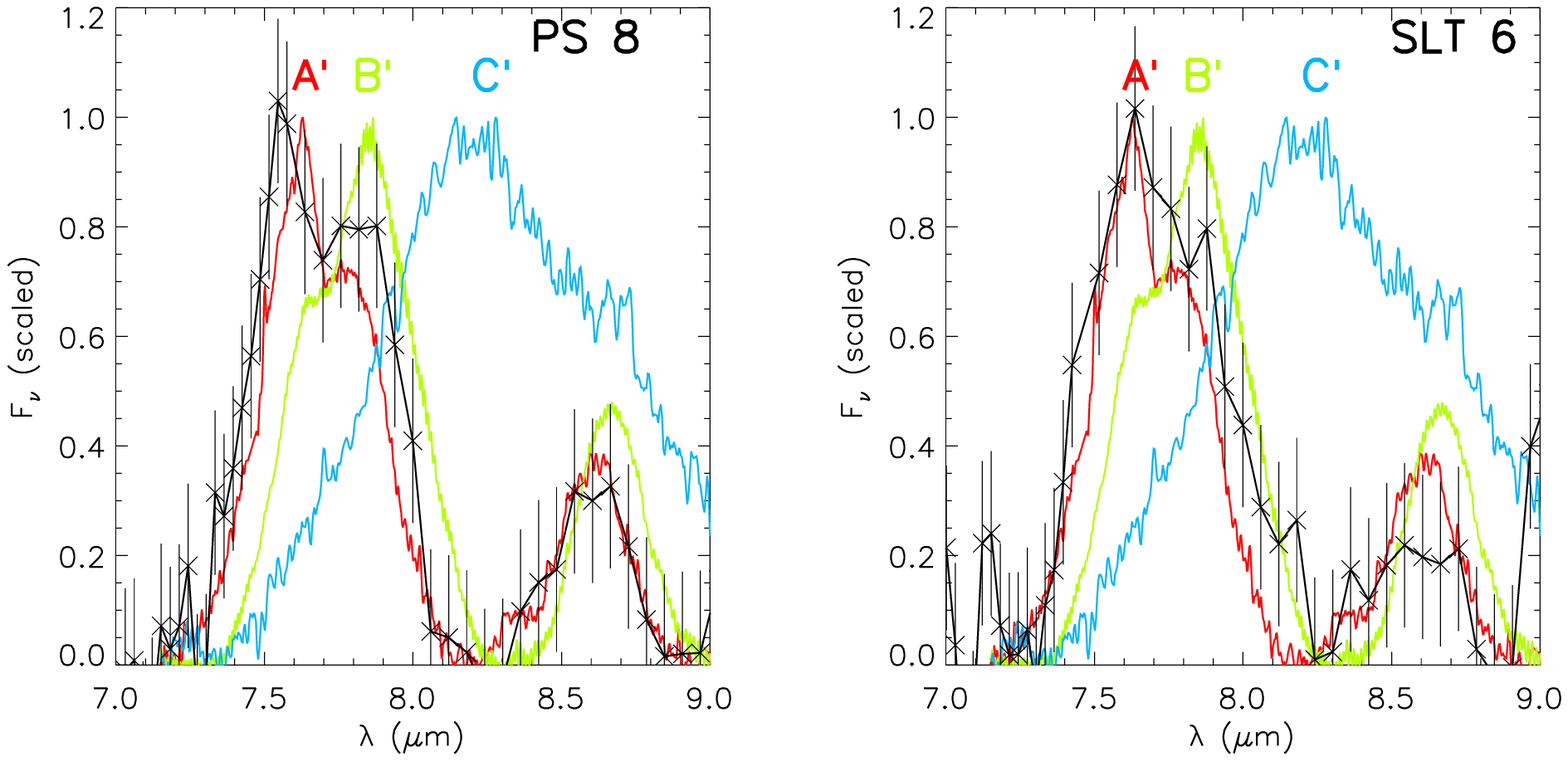}
\caption{Example 7.7~\micron\ and 8.6~\micron\ PAH features are
  plotted versus the \citet{Peeters2002} templates (templates are from
  ISO SWS, with R~$\sim$~450 for classes A and C and $\sim$~1500 for
  class B), showing both the width of the 7.7~\micron\ feature and the
  suppressed 8.6~\micron\ emission in the right panel. For the spectra
  of the extended emission, the strong lines at 7.46~\micron\ and
  8.99~\micron\ are H{\sc i}~6-5 and [Ar{\sc iii}] respectively. The
  N66 spectra by and large resemble the class A template, as expected
  for a giant H{\sc ii} region. Spectra were scaled so that the peak
  of the 7.7~\micron\ feature matched that of the
  templates. \label{PAH7786}}
\end{figure}

In order to classify the PAH spectra observed in N66, we plot example
high signal-to-noise (S/N) N66 PAH spectra versus templates in the
7-9~\micron\ range and 11-12~\micron\ range from \citet{Peeters2002}
and \citet{vanDiedenhoven2004} respectively in Figures~\ref{PAH7786}
and \ref{PAH11hi}. \citeauthor{Peeters2002} was able to classify the
6-9~\micron\ PAH features based on their peak centroids, and
discovered that there is a relationship with environment: Class A
spectra are typical for H{\sc ii} regions and non-isolated Herbig AeBe
stars, Class B spectra are more typical for isolated Herbig AeBe
stars, and Class C spectra are more common in evolved stellar
systems. \citeauthor{vanDiedenhoven2004} studied the 11.3~\micron\ PAH
complex and developed templates for Classes A and B only.

\begin{figure}
\includegraphics[scale=0.7]{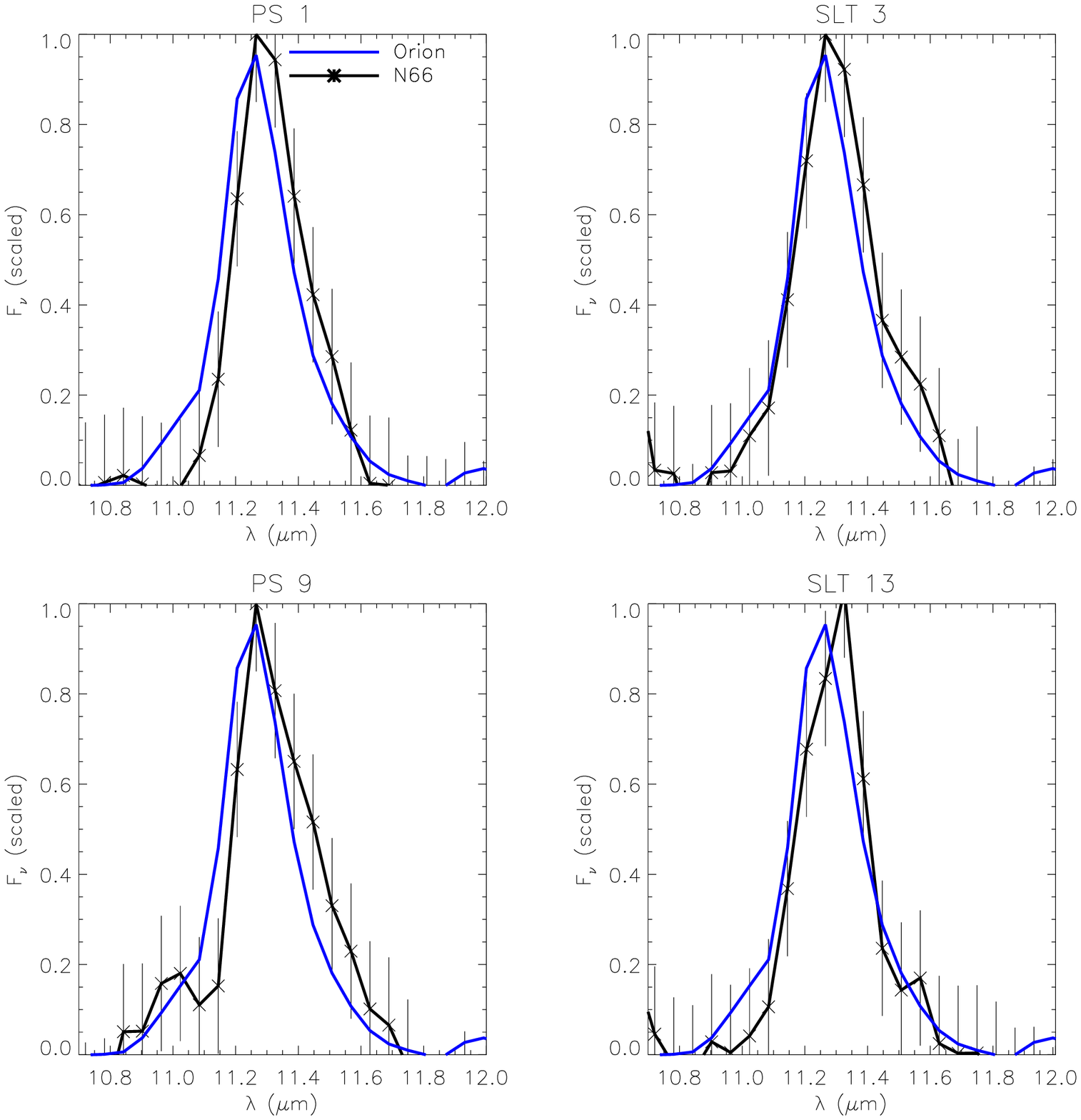}
\caption{Example 11.3~\micron\ PAH spectra in black are plotted
  against a SL spectrum of the Orion nebula. The N66 spectra appear to
  have peaks that are redward of the Orion spectra.\label{PAH11lo}}
\end{figure}

\begin{figure}
\includegraphics[scale=0.8]{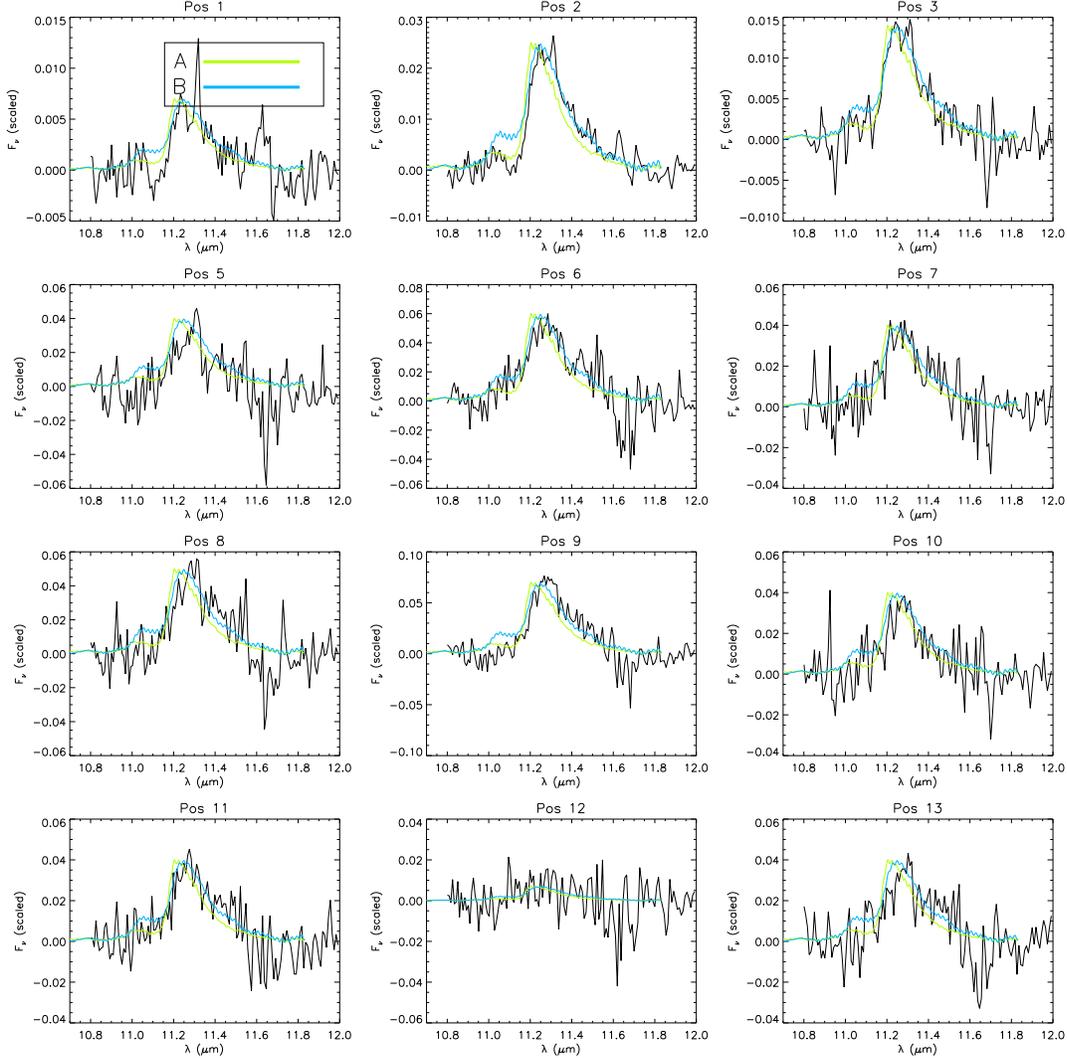}
\caption{The 11.3~\micron\ PAH for all of the {\it IRS} SH pointings
  are plotted versus class A and class B PAH template spectra from
  \citet{vanDiedenhoven2004}. The N66 spectra are scaled to best fit
  the templates, excluding the possible H{\sc i}~9-7. For several
  sources, there are prominent shifts that are independent of the HI
  line (most notably position 8, 9, 10, and 13). \label{PAH11hi}}
\end{figure}

The N66 spectra for the 7.7~\micron\ and 8.6~\micron\ features
(Figure~\ref{PAH7786}) are generally most similar to the Class A
template, as is expected for an H{\sc ii} region. However, in some of
the spectra, the 7.7~\micron\ features appear to be wider than the
templates. For the 11.3~\micron\ PAHs, the observed line centers
appear to be slightly redshifted with respect to the Orion Nebula
(Figure~\ref{PAH11lo}); i.e. less than one resolution
element. However, this is the only feature in the N66 spectra, ionic,
molecular, or PAH, which shows a centroid shift, and while such a
systematic offset could very well be due to continuum subtraction it
is deserving of closer attention.

Therefore, the available high-resolution spectra (R $\sim$~600) of N66
were reduced for comparison with the PAH templates from
\citet{vanDiedenhoven2004} (Figure~\ref{PAH11hi}). Labelling for these
spectra, first used in \citet{Lebouteiller2008}, follows that
work. Unfortunately, the SH data have a low S/N. However, in the
spectra for positions 1, 2, 3, and 5, a narrow line, 2-3 wavelength
elements wide, that corresponds to the position of the H{\sc i}~9-7
11.31~\micron\ line, is visible. Due to the fact that the raw images do
not reveal single pixels with high values at this wavelength, it is
not likely that this line is due to improperly calibrated pixels on
the array, but is an actual astronomical feature. In order to
determine whether the narrow line at 11.3~\micron\ is the H{\sc i}~9-7
line, we compared the line strength to that of the detected H{\sc
  i}~7-6 12.37~\micron\ line to see if their ratio is consistent with
case B recombination theory \citep{Hummer1987}. We used a temperature
of 12,500 K and a number density of 10$^2$~cm$^{-3}$ to compare to the
data; N66 has measured values of 12,269 K average and 50-500~cm$^{-3}$
\citep{Oliveira2008, Tsamis2003, Peimbert2000, Dufour1977}. For case
B,

\begin{equation}
\frac{HI~9-7}{HI~7-6} = 0.223
\end{equation}

For sources 1, 2, and 3, the ratio was about 0.2, and for source 5, it
was $<$ 0.1. Considering the low S/N of these data, but also
considering the other detected H{\sc i} lines (H{\sc i}~7-6 line in
hi-res and the H{\sc i}~6-5 line at lo-res), it is likely that the
detected line is the H{\sc i}~9-7 line and that this line is
contributing to the apparent redshift of the 11.3~\micron\ feature.

Leaving aside the narrow line, there are still several cases in which
the peak of 11.3~\micron\ PAHs in N66 are slightly redward ($\Delta
\lambda \sim$~0.1~\micron) of the two templates (positions {\it 8, 9,
  10} and {\it 13} show it most clearly in spite of the low
S/N). According to the {\it Spitzer} IRS instrument
handbook\footnote{The IRS instrument handbook may be found online at:
  http://irsa.ipac.caltech.edu/data/SPITZER/docs/irs/irsinstrumenthandbook/}
the wavelength calibration is good to 1/5 of a resolution element, or,
for SH, $\sim$~0.001~\micron. Pointing offsets are also a minor
concern, but could produce a 0.5-pixel shift, or
$\sim$~0.04~\micron. Both of these potential errors are small in
comparison to the $\sim$~0.1~\micron\ shift in the PAH feature
centroid, would additionally affect all features in this spectral
order, and therefore cannot account for the shift seen. We must stress
once again that continuum subtraction may play a role in the centroid
mismatch. If born out, this is the only ISM-type environment observed
so far that exhibits a profile redshifted compared to class A. We also
note the absence of the 11.0~\micron\ feature seen in the templates,
though its absence is likely due to the low S/N of these spectra.

While the observed wavelength shifts across N66 appear in both point
source and ISM spectra, a worthwhile comparison is Herbig AeBe stars,
which often exhibit 11.3~\micron PAH peaks that are shifted redward by
a similar amount. \citep{Sloan2005,Keller2008} show peak shifts in
both the 7.7~\micron\ and 11.3~\micron\ PAH complexes. It is
interesting that we do not see an appreciable shift in the
7.7~\micron\ feature, which is often shifted by as much as
0.3~\micron\ or more in Herbig AeBe environments and would therefore
be detectable with our spectra.

Because the redward shift of the 11.3~\micron\ PAH complex is small
(0.1~\micron) and the detected H{\sc i}~9-7 line must be resolved out
of the feature, we recommend that high-resolution and high S/N
follow-up observations be made. JWST's MIRI instrument, with its
spectral resolution of $\sim$~3000, is a perfect candidate instrument
for such follow-up. Such observations would be capable of not only
confirming the wavelength shift of the PAH and more accurately
determining the line flux and width of the narrow line, but, an
expanded high-resolution study could be used to determine whether
these features are peculiar to N66 (or, perhaps, the SMC), or are
systematic at low metallicity.

\clearpage

\subsection{14 \micron\ Continuum Emission}

The continuum between 13.5 and 14.2~\micron\ is due to emission by a
combination of very small carbonaceous dust grains \citep[VSGs:
  a~$>$~50~\AA;][]{LiDraine2001} and warm large dust grains. However,
we expect the VSGs to dominate the emission at these wavelengths;
\citet{DraineLi2007}, \S~5, notes that the VSG contribution to the
continuum at $\lambda$~$<$~20~\micron\ is roughly independent of the
radiation field strength because it results from single-photon
heating.

\begin{figure}
\includegraphics[scale=1.0]{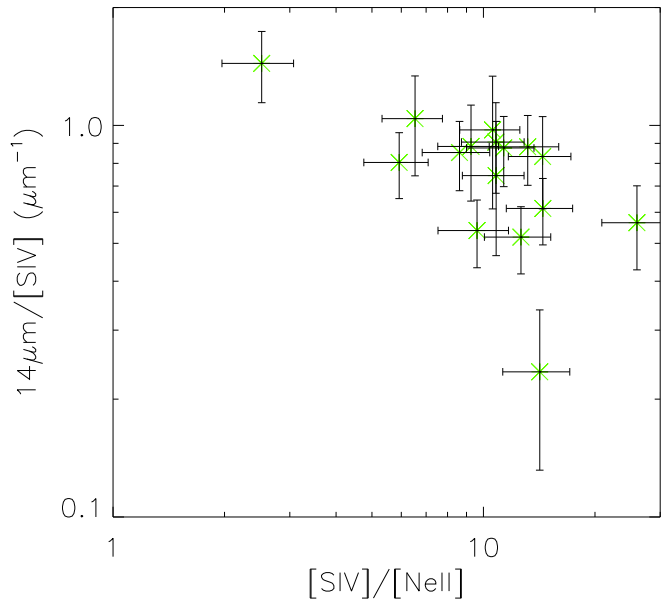}
\caption{The 14~\micron\ continuum divided by [S{\sc iv}] flux in the
  extended emission versus [S{\sc iv}]/[Ne{\sc ii}]. These data show
  an anti-correlation that suggests that the 14~\micron\ continuum is
  weakest in the hardest and strongest radiation fields.\label{VSGsb}}
\end{figure}

The 14~\micron\ continuum is measured as the continuum beneath the PAH
bands as fit with PAHFIT. Figure~\ref{VSGsb} plots the
14~\micron\ continuum over [S{\sc iv}] versus the radiation field
hardness as traced by [S{\sc iv}]/[Ne{\sc ii}] for the extended
emission spectra. By normalizing the 14~\micron\ continuum emission by
the [S{\sc iv}] line flux, we are able to compare the dust continuum
emission to the radiation field strength, and then plot that value
versus the radiation field hardness. There is an anti-correlation that
suggests that the 14~\micron\ continuum is weakest where the radiation
field is hardest and strongest. Weaker continua in harder and stronger
radiation fields supports the assertion first presented in
\citet{Contursi2000} that the VSGs are likely being photodestroyed by
the U = 10$^5$ ISRF radiation field in N66.

Because the correlation seen in Figure~\ref{VSGsb} is roughly linear
excluding the outlier, this supports the assertion that the
14~\micron\ emission is mostly due to VSGs, not large grains; if there
was a substantial contribution from large grains to the
14~\micron\ continuum, then the expected variations in large grain
dust temperature across the region would make this trend non-linear.

\subsection{PAH Ratios as Diagnostics \label{pahratios}}

There are four major PAH features in the 5-15~\micron\ wavelength
range. These features are identified by specific resonant modes inside
the PAHs \citep{Allamandola1989, DraineLi2007}. (1) The
6.2~\micron\ feature is created by an aromatic C-C stretch mode. (2)
The 7.7~\micron\ feature is also due to C-C aromatic stretch
modes. (3) The 8.6~\micron\ feature is emitted by a C-H in-plane
bending mode, while (4) the 11.3~\micron\ feature is due to a solo C-H
out-of-plane bending mode and is sensitive to edge structure
\citep{Hony2001, Allamandola1989, Puget1989}. Furthermore, the 6.2,
7.7, and 8.6~\micron\ PAH features are attributed to ionized PAHs,
whereas the 11.3~\micron\ PAH feature is attributed to neutral PAHs
\citep{Hudgins1999}, and there is believed to be a wavelength-size
dependence for PAHs, wherein larger PAHs emit most efficiently in the
longer-wavelength PAH features \citep{DraineLi2007, Schutte1993}.

\begin{figure}
\includegraphics[scale=1.5]{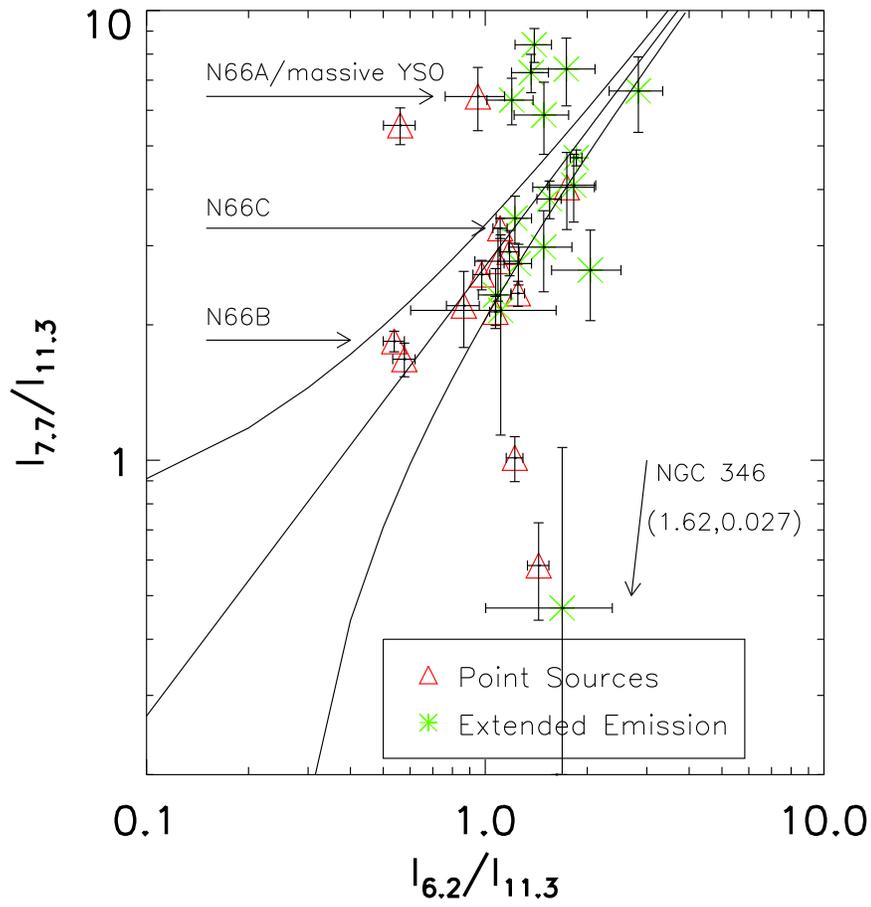}
\caption{The average I$_{7.7}$/I$_{11.3}$ versus I$_{6.2}$/I$_{11.3}$
  from \citet{Galliano2008b} Figure 3 is plotted as the solid line
  with its 1-$\sigma$ spread shown. The N66 data points are
  overplotted.\label{Galliano1}}
\end{figure}

PAH ratios offer an independent assessment of the ISM's physical
characteristics. \citet{Galliano2008b} found an average correlation
between the I$_{7.7}$/I$_{11.3}$ and I$_{6.2}$/I$_{11.3}$ for a sample
of starbursts, BCDs, H{\sc ii} regions and PDRs, reproduced in
Figure~\ref{Galliano1}, where Galliano's average is shown as a solid
line with the one-$\sigma$ standard deviation plotted on either side
of the average. The data points for N66 are also plotted, with the
point source data as red triangles and the extended emission data as
green stars. The point source data show a range of
I$_{7.7}$/I$_{11.3}$ from $\sim$~0.03-6.3, over two orders of
magnitude, in a relatively narrow range of I$_{6.2}$/I$_{11.3}$
$\sim$~0.5-1.8. The extended emission spectra have respective ranges
of 0.35-8.4 and $\sim$~1-3. Averaging together all values for the
I$_{6.2}$/I$_{11.3}$ and I$_{7.7}$/I$_{11.3}$ ratios weighted by
feature strength, the point sources have values of 1.1 and 2.7 and the
extended emission have values of 1.4 and 3.7 respectively. These
averages are relatively close together and lie within the galaxy
sample's average correlation.

However, the wide spread around those averages suggests an intriguing
variety of physical attributes. For instance, the massive embedded YSO
(PS7, discussed in detail in Section~\ref{yso}) has a very high
I$_{7.7}$/I$_{11.3}$ ratio value; see Figure~\ref{Galliano1}. This
might suggest that the PAH emission from the embedding envelope is
either more heavily ionized than other point sources, or else the
average PAH size is smaller. Having a population of ionized PAHs in
the YSO environment seems unlikely because the temperature of a Class
I YSO is probably very low \citep[$\sim$250 K; ][]{Myers1998}. The
H{\sc ii} regions N66B and N66C (PS5 \& PS6) both lie near the average
value for I$_{7.7}$/I$_{11.3}$, but NGC~346 (PS9) lies well off of
this average, with an outlying value of $\sim$~0.03. The low value for
the 7.7~\micron\ feature strength in NGC~346 is confirmed with a
visual inspection of the spectrum - see the Appendix,
Figure~\ref{Spec2}. As an opposite example to the YSO, NGC~346 may
either contain a largely neutral PAH population or else very large
PAHs; the latter solution seems most sensible given the numerous blue
stars in the cluster.

The lack of correlation seen in Figure~\ref{Galliano1} is intriguing
because the 6.2~\micron\ and 7.7~\micron\ PAH features are generally
very tightly correlated for spectra of H{\sc ii} regions and entire
galaxies \citep{Vermeij2002, Galliano2008b}, for planetary nebulae
\citep{BernardSalas2009b}, and in the reflection nebula NGC~2023
(Peeters et al. 2013 in preparation). We have already determined that
this is not due to the decomposition method (see
\S~\ref{specfit}). The same lack of trend can be seen with the
column-extracted PAH fluxes from \citet{Lebouteiller2011}, although a
weak correlation may be seen in the \citet{Sandstrom2012} results for
the high S/N data points. We are not aware of any other mention in the
literature of a decoupling of the 6.2~\micron\ and 7.7~\micron\ PAH
features in this manner.

\begin{figure}
\includegraphics[scale=0.8]{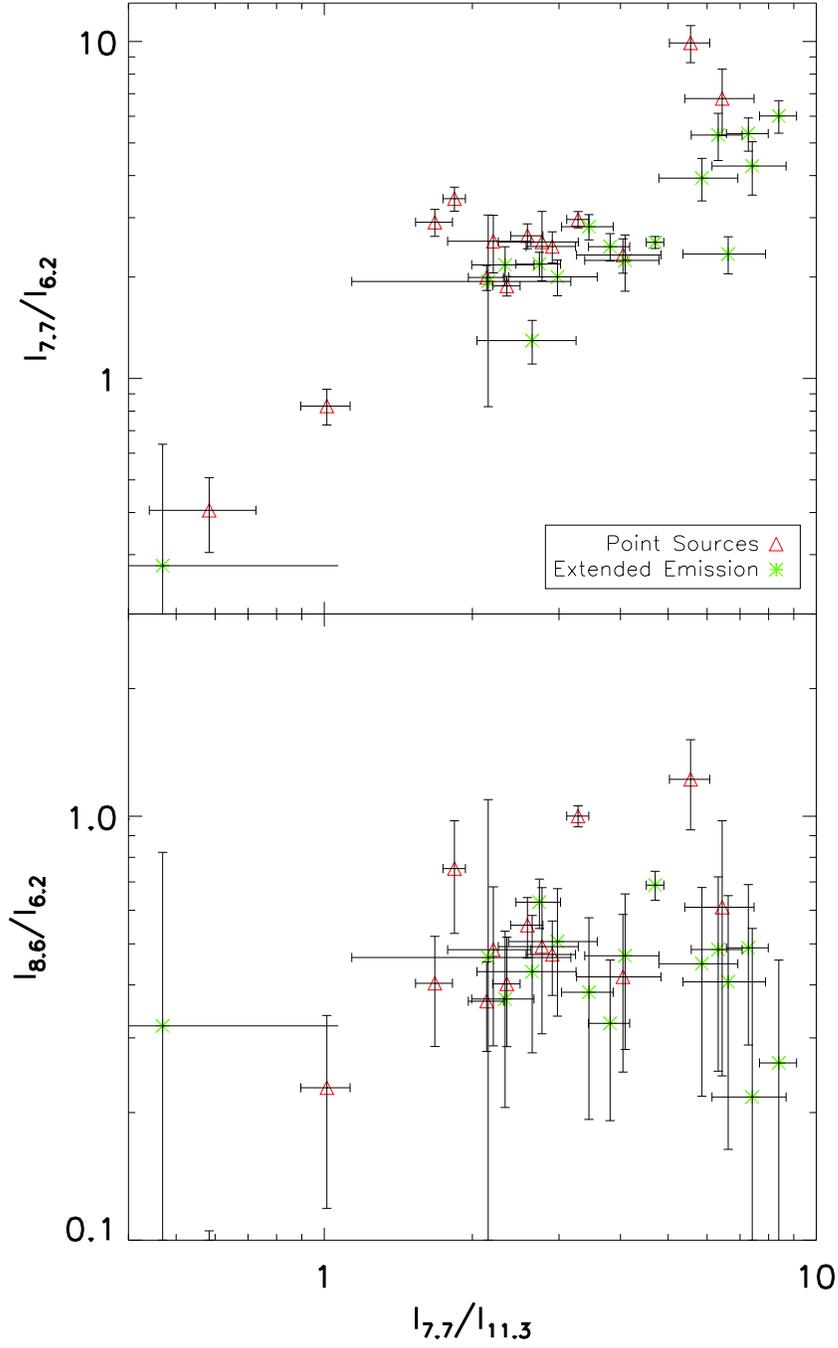}
\caption{The I$_{7.7}$/I$_{6.2}$ and I$_{8.6}$/I$_{6.2}$ versus
  I$_{7.7}$/I$_{11.3}$ PAH ratios are plotted for
  N66. \label{Galliano2}}
\end{figure}

To study general trends, the I$_{7.7}$/I$_{6.2}$ and
I$_{8.6}$/I$_{6.2}$ are plotted against I$_{7.7}$/I$_{11.3}$ in
Figure~\ref{Galliano2}. A trend is seen in the I$_{7.7}$/I$_{6.2}$
versus I$_{7.7}$/I$_{11.3}$ ratios plot that was not recorded in the
\citeauthor{Galliano2008b} dataset. The values of the
I$_{7.7}$/I$_{11.3}$ ratio seem to suggest that the areas of extended
emission (on the right) are largely photoionized whereas the point
sources (on the left) are more neutral, with some exceptions. The
variation of the I$_{7.7}$/I$_{6.2}$ ratio might suggest some
photoprocessing effect: the 6.2~\micron\ PAH flux becomes weaker in
harder radiation fields due to the destruction of the smaller PAH
molecules.

There is no trend between the I$_{8.6}$/I$_{6.2}$ and
I$_{7.7}$/I$_{11.3}$ PAH ratios. Because the I$_{8.6}$/I$_{6.2}$ ratio
should be independent of ionization state and therefore should trace
the size of the PAHs, while the I$_{7.7}$/I$_{11.3}$ ratio does trace
ionization state, it appears that the size distribution is {\it
  independent of ionization state}. This includes both point sources
and extended emission spectra, and it should be noted that a spread of
about one order of magnitude still exists in the I$_{8.6}$/I$_{6.2}$
ratio values. We therefore conclude that N66 shows no evidence of
local variation in the N$_{C}^{min}$ value due to the ionization state
for the PAH population, but it is unclear how this result should
compare to galaxies with significantly different metallicity or star
formation activity.

In \citet{Sandstrom2012}, the weak 7.7~\micron\ feature across the
entire SMC, including N66, and the relatively strong
11.3~\micron\ feature suggested to the authors that PAHs across the
SMC are both small and mostly neutral. However, the authors used a
stringent signal-to-noise cut-off, thereby excluding extended emission
from their analysis and biasing their results towards the conditions
found for our point source spectra. When our data are compared against
the grey data points in the PAH ratios plots in \citet{Sandstrom2012},
it is interesting to note that the results from the map agree well
with our targeted spectroscopy results. What our analysis is able to
show is that the point sources lie in a similar PAH ratio space to the
extended emission, despite the different physical conditions as
discussed. Furthermore, it is clear that the PAH population exists on
a spectrum, from more neutral in the dense PDR where the point sources
lie, to more ionized in the diffuse PDR and H{\sc ii} region.

The PAH population size distribution in low-metallicity environments
is a contentious topic. In a study of low-metallicity BCDs,
\citet{Hunt2010} found relatively strong I$_{8.6}$ and I$_{11.3}$
features and interpreted this finding to mean a large N$_{C}^{min}$
value for PAHs. The analysis of the I$_{17}$/I$_{11.3}$ PAH ratio by
\citet{Smith2007} suggests the opposite, that low-metallicity
environments (12+log(O/H) $\lesssim$ 8.1) host PAH populations with
small sizes on average. We do not have the long-wavelength data to
sufficient depth or spatial resolution for a I$_{17}$/I$_{11.3}$
analysis; however, that data is available in the \citet{Sandstrom2012}
maps, and they determine that the N$_C^{min}$ value is smaller in the
SMC than in the Milky Way based on the weak 17.0 \micron\ PAH
feature. Our own results, including a wide range of
I$_{8.6}$/I$_{6.2}$ ratio values, suggest that these average trends do
not hold on small size scales, where local effects likely come in to
play.

\subsection{N66 PAH Emission in Context}

\begin{figure}
\includegraphics[scale=1.5]{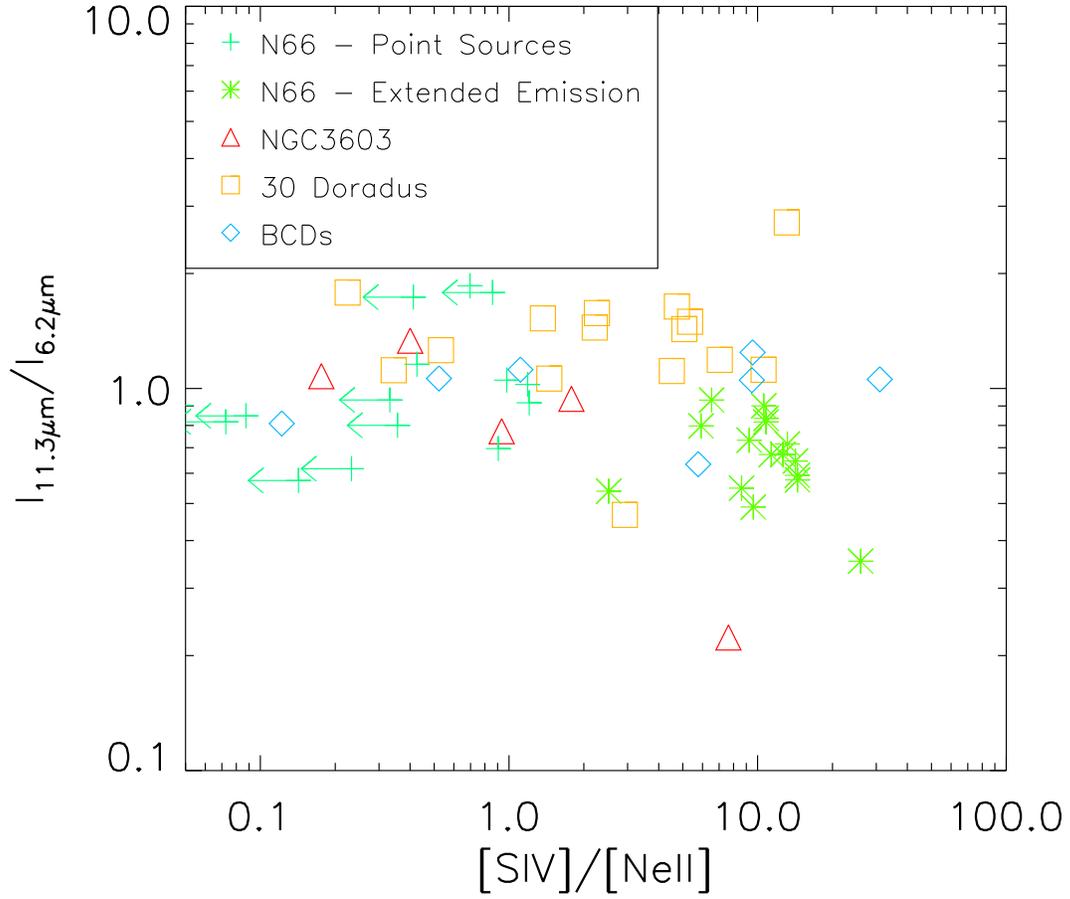}
\caption{The PAH ratio I$_{11.3}$/I$_{6.2}$ versus [S{\sc iv}/[Ne{\sc
        ii}] for N66 point sources (plus signs) and extended emission
    spectra (stars) plotted along with results for NGC 3603
    (triangles), 30 Doradus (squares), and some well-studied BCDs
    (diamonds). References for these data are given in the 
    text. \label{PAHvsRadn}}
\end{figure}

In this section, we compare the PAH and forbidden line emission for
N66 with data from other sources with similar star formation
environments to N66, including giant H{\sc ii} regions and blue
compact dwarf galaxies (BCDs). For the PAH emission, we take the ratio
of the 11.3~\micron\ and 6.2~\micron\ PAH features which will trace
the ionization state of the PAHs and also the PAH size
distribution. For the gas, we take the [S{\sc iv}]/[Ne{\sc ii}] ratio
as a tracer of the radiation field hardness. The other giant H{\sc ii}
regions studied are 30~Doradus and NGC~3603 \citep{Lebouteiller2008,
  Lebouteiller2011}, and the population of well-studied BCDs comes
from \citet{Wu2006}. The results are plotted in
Figure~\ref{PAHvsRadn}.

The N66 point sources have PAH ratio values similar to sources with
small [S{\sc iv}]/[Ne{\sc ii}] ratio values. However, the extended
emission spectra, whose [S{\sc iv}]/[Ne{\sc ii}] ratio is on average
higher than most other sources, also exhibits a lower
I$_{11.3\micron}$/I$_{6.2\micron}$ value, as was discussed in
\S~\ref{pahratios}, suggesting that the PAH population in the diffuse
ISM of N66 is more ionized than in the other H{\sc ii} regions and in
integrated spectra of BCDs, despite sometimes equally hard ISRFs as
traced by the [S{\sc iv}]/[Ne{\sc ii}] ratio.

There is a source of bias in this comparison worth noting. The spectra
of 30~Doradus, NGC~3603, and the BCD sample were not extracted using
SMART-AdOpt as described in \S~\ref{srcext}. Instead, they are all
composite spectra. For the giant H{\sc ii} regions NGC~3603 and
30~Doradus, each spectrum likely has contributions from both
unresolved point sources in the PDR and diffuse H{\sc ii}
emission. For the BCDs, each spectrum likely contains emission from
both star-forming and quiescent regions. It is therefore not
surprising that the results for N66 are both separated and
qualitatively different from those of the other regions.

\subsection{Fraction of PAH Emission Coming From the Point Sources}

The SMC (including N66) is the closest low-metallicity star forming
galaxy to the Milky Way and can therefore be used to study possible
trends to other low-metallicity star-forming environments. It is
therefore important to determine what the dominant source of the PAH
emission in N66 is, in order to properly interpret unresolved H{\sc
  ii} region studies. Does the PAH emission primarily come from the
high S/N point sources, or from the diffuse ISM in the H{\sc ii}
region and PDR?

Using the spectral maps of N66 presented in
\citet{Sandstrom2010,Sandstrom2012} in order to study a spatially
well-sampled dataset, we made maps of the 6.2~\micron\ and
11.3~\micron\ PAH features by subtracting away the continuum measured
locally. The 7.7~\micron\ feature straddles the SL1 and SL2 orders and
the 8.6~\micron\ feature is very hard to pick out from the continuum,
so these two features were excluded from the analysis. The average
brightness in the PDR was measured and then point sources were defined
as anything that was 3-$\sigma$ above the average; the annulus
selected around each point source was 6 pixels across, or
10.8~$\arcsec$. Using this metric, we summed up the emission for point
sources as well as the diffuse emission from the entire map in the
6.2~\micron\ and 11.3~\micron\ PAH features. There are a total of 13
selected point sources in the 6.2~\micron\ map, and 14 in the
11.3~\micron\ map, and the map is 300$\arcsec$~$\times$~430$\arcsec$.
We find that the fraction of the PAH emission coming from the point
sources is 28~$\pm$~8\% for the 6.2~\micron\ PAH feature and
21~$\pm$~6\% for the 11.3~\micron\ PAH feature. Due to the relatively
high level of variance between these two PAH features, we conclude
that, as a rough proxy for studies of unresolved star formation at
low-metallicity, the point sources (resolved down to $\sim$~2 pc)
contribute roughly $\sim$~20-35\% of the PAH emission across the
region. This means that it is the extended emission which dominates
the observed PAH brightness in N66, and possibly low-metallicity giant
H{\sc ii} regions generally.

\section{CONCLUSION\label{Conclusion}}

N66 is the largest starburst in the SMC and contains half of the
entire galaxy's O~stars. Its hard radiation field and low metallicity
make it an excellent corollary to studies of BCDs and high-redshift
starbursts. Our analysis of {\it Spitzer/IRS} spectra from
5-14~\micron\ makes use of powerful optimal extraction routines that
separate emission from the dense PDR's unresolved point sources and
the surrounding ISM's more diffuse emission. The point sources are all
associated with sites of active star formation in the region. Of
special note are the spectroscopic confirmation of a massive embedded
class~I YSO, the detection of silicates associated with young
($\sim$~3~Myr) stellar clusters, and the detection of PAHs and
silicates associated with a B[e] star that indicates it may in fact be
a Herbig AeBe star. In the diffuse ISM, we find evidence that the very
small grains are being photodestroyed, consistent with previous
studies of N66. The PAH emission shows several interesting
features. The 11.3~\micron\ PAH complex has an unexplained centroid
shift for both point sources and the diffuse ISM. In general, the PAHs
are best classified as type A. However, a slight widening of the
7.7~\micron\ feature in some of the spectra is either due to the
7.46~\micron H{\sc i} 6-5 line in the diffuse emission spectra, or
else the removal of ISM contamination that reveals type B-like
circumstellar features. The 6.2~\micron\ and 7.7~\micron\ PAH band
fluxes in N66 do not correlate with each other, an observation that we
do not believe has been observed in other objects. Because we were
able to separate out the point source and extended emission in the
spectral slit, the conclusions we draw from the PAH ratios are
detailed where those presented in \citet{Sandstrom2012} are more
general: we find that the PAHs are more neutral in the densest parts
of the PDR (including the entire point source catalogue), and more
ionized in the diffuse ISM while \citeauthor{Sandstrom2012} determined
that the PAHs were on average more neutral than in higher-metallicity
galaxies. On average, \citeauthor{Sandstrom2012} found that PAHs in
the SMC are smaller compared to those in more metal rich galaxies.  We
show that on small scales, the size distribution of the PAHs in N66
(based on the 8.6~\micron\ PAH feature flux) does not appear to be
correlated with their ionization state and is therefore independent of
the ratio of UV field strength to electron density. We stress that, on
average, our results match the work of \citet{Sandstrom2012} extremely
well, and that the spread in ionization state and lack of clear trend
with minimum PAH size on small scales is made possible by our spectral
extraction techniques. Lastly, we have determined using spectral maps
of the region that only $\sim$~20-35 \% of the PAH emission in N66 is
coming from the unresolved point sources, and it is therefore
dominated by the diffuse emission. This percentage suggests that
studies of unresolved low-metallicity star-forming regions in nearby
BCDs and starburst galaxies will also have PAH spectra dominated by
the diffuse H{\sc ii} region and PDR, not the unresolved point
sources, and the observed PAH signatures of the point source spectra
are therefore not representative of the entire H{\sc ii} region's
properties.

\acknowledgements The authors would like to thank the anonymous
referee, whose comments helped to improve the manuscript. DGW would
like to thank K. Sandstrom for discussions concerning the PAH spectra
of N66, E. Muller for discussions about the molecular gas content, and
P. Martini and A. Leroy for helpful suggestions. KEJ gratefully
acknowledges support for this paper provided by NSF through CAREER
award 0548103 and the David And Lucile Packard Foundation through a
Packard Fellowship. JSB wishes to acknowledge the support from a Marie
Curie Intra-European Fellowship within the 7th European Community
Framework Program under project number 272820.

\clearpage
\appendix
\section{The Atlas of Reduced Spectra}

The spectra used in the analysis are presented here. 
The reduction method for producing the point source and extended emission
is discussed in \S~\ref{srcext}.

\begin{figure}
\includegraphics[scale=0.6]{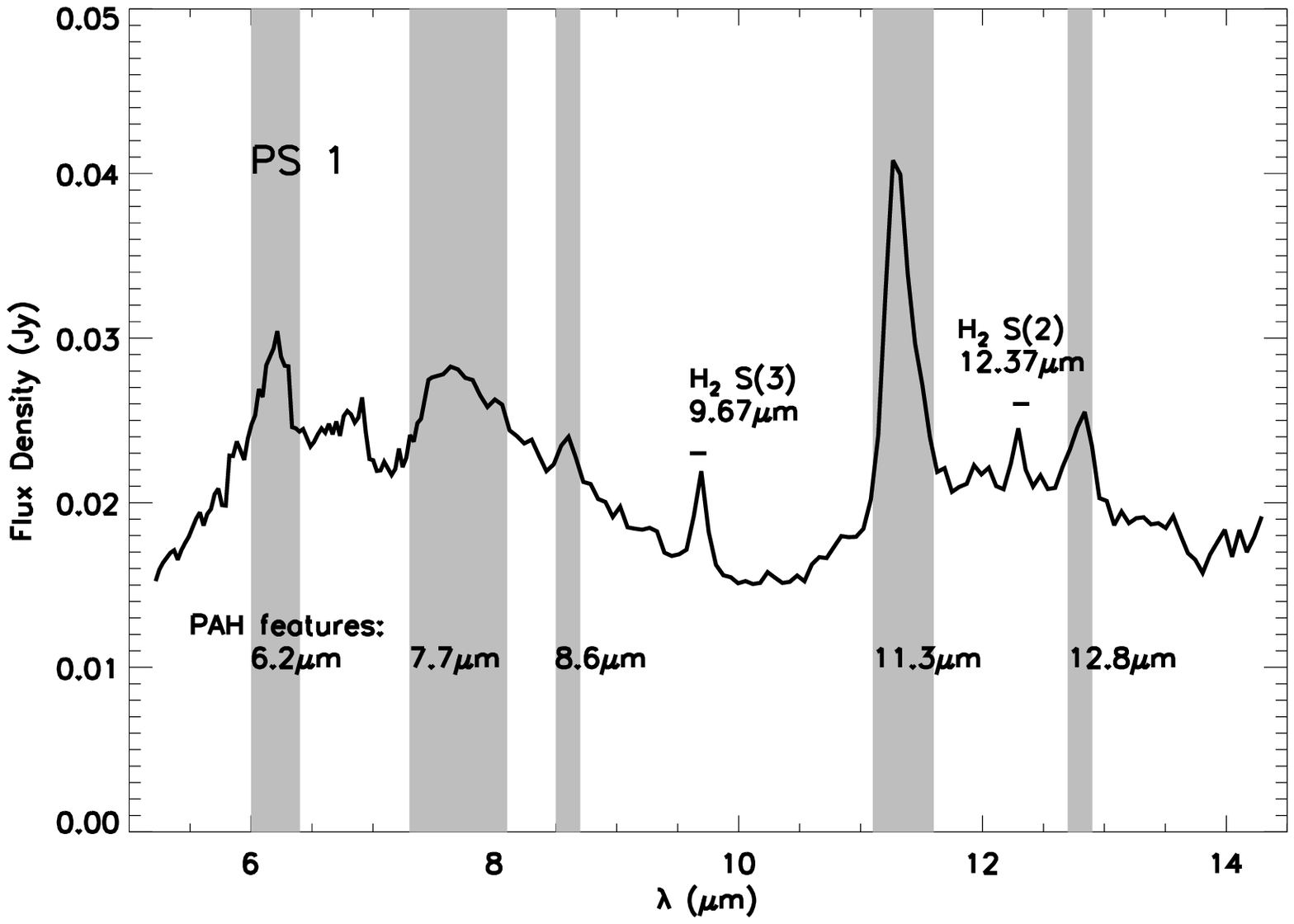}
\includegraphics[scale=0.3]{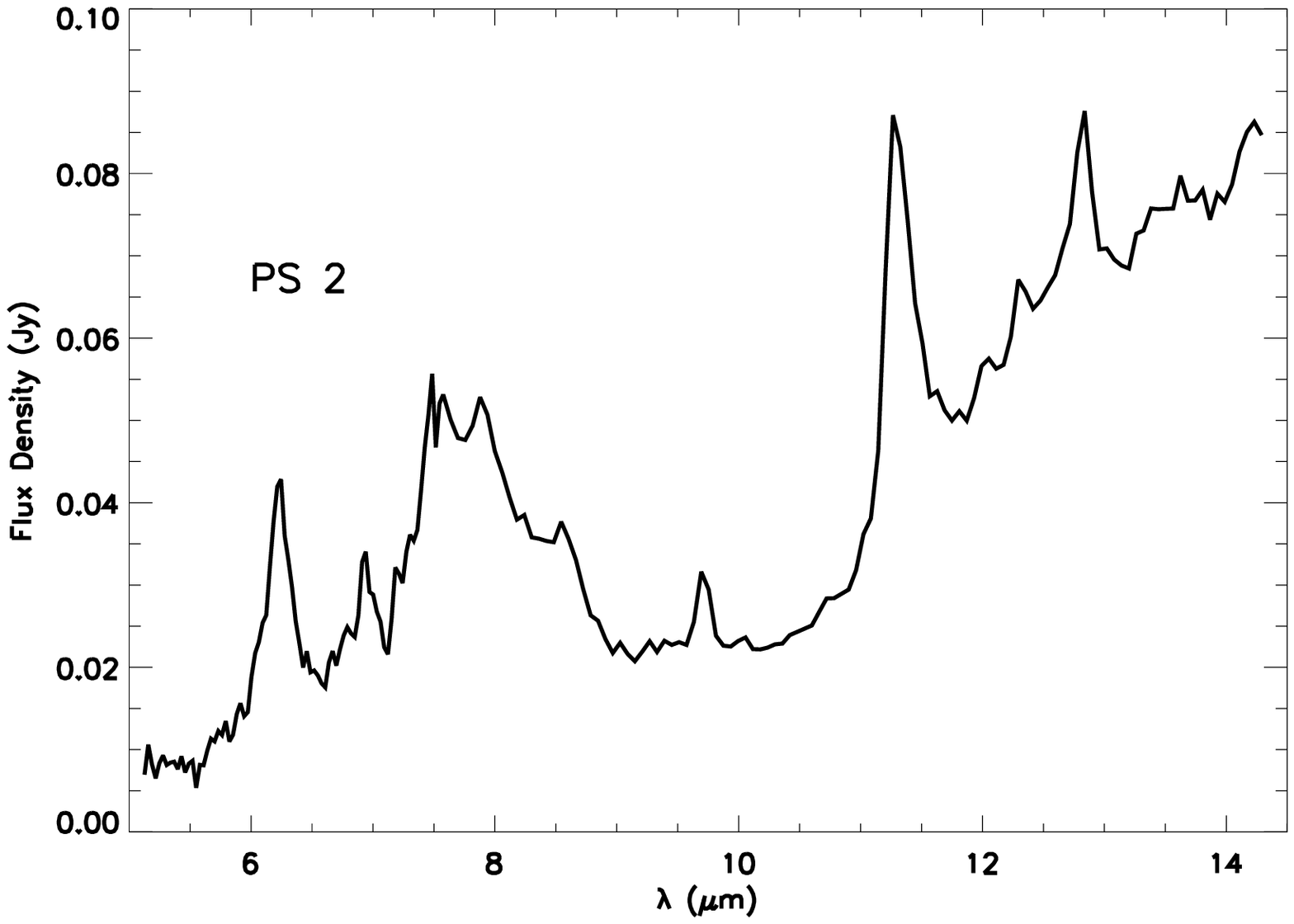}
\includegraphics[scale=0.3]{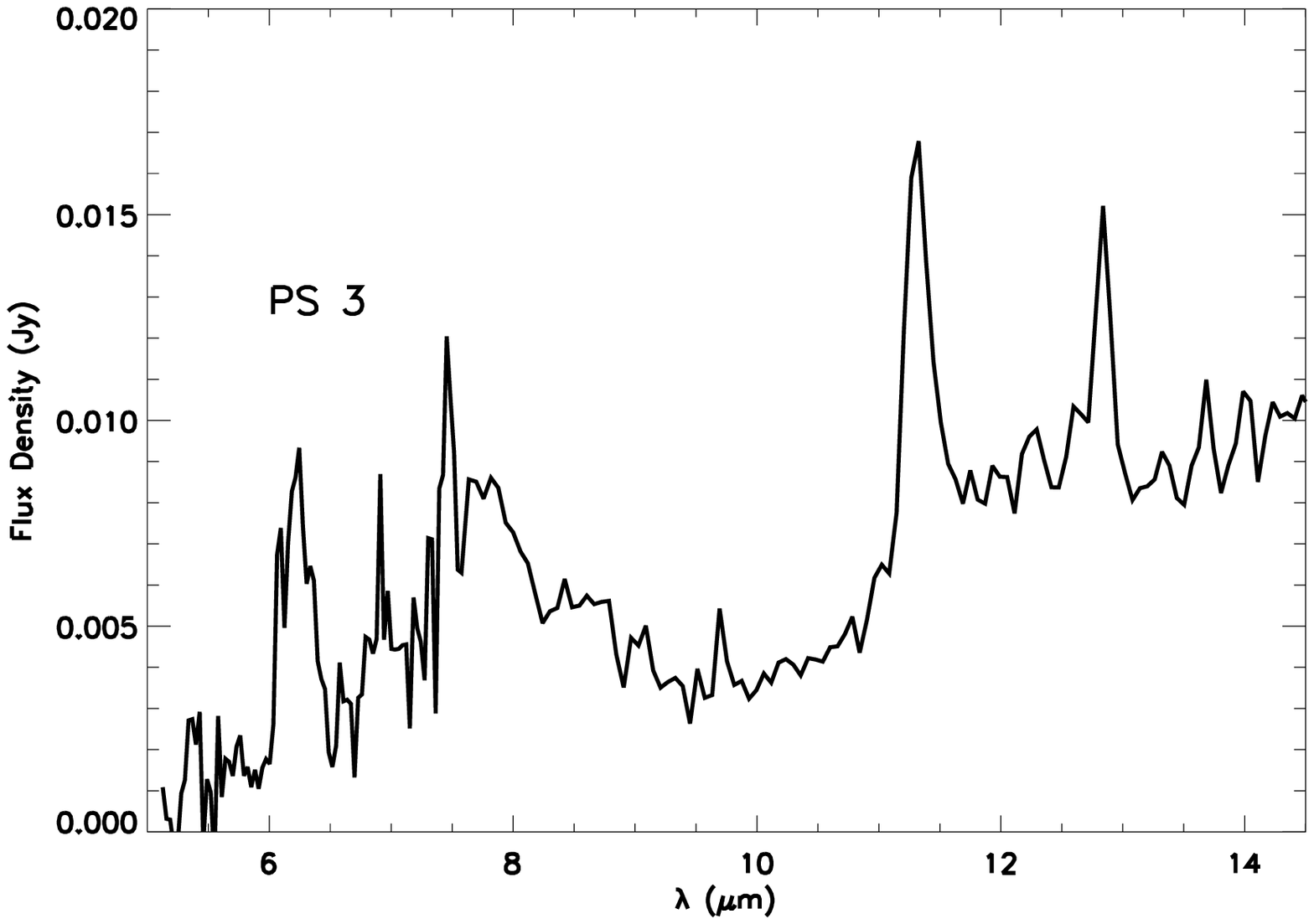}
\includegraphics[scale=0.3]{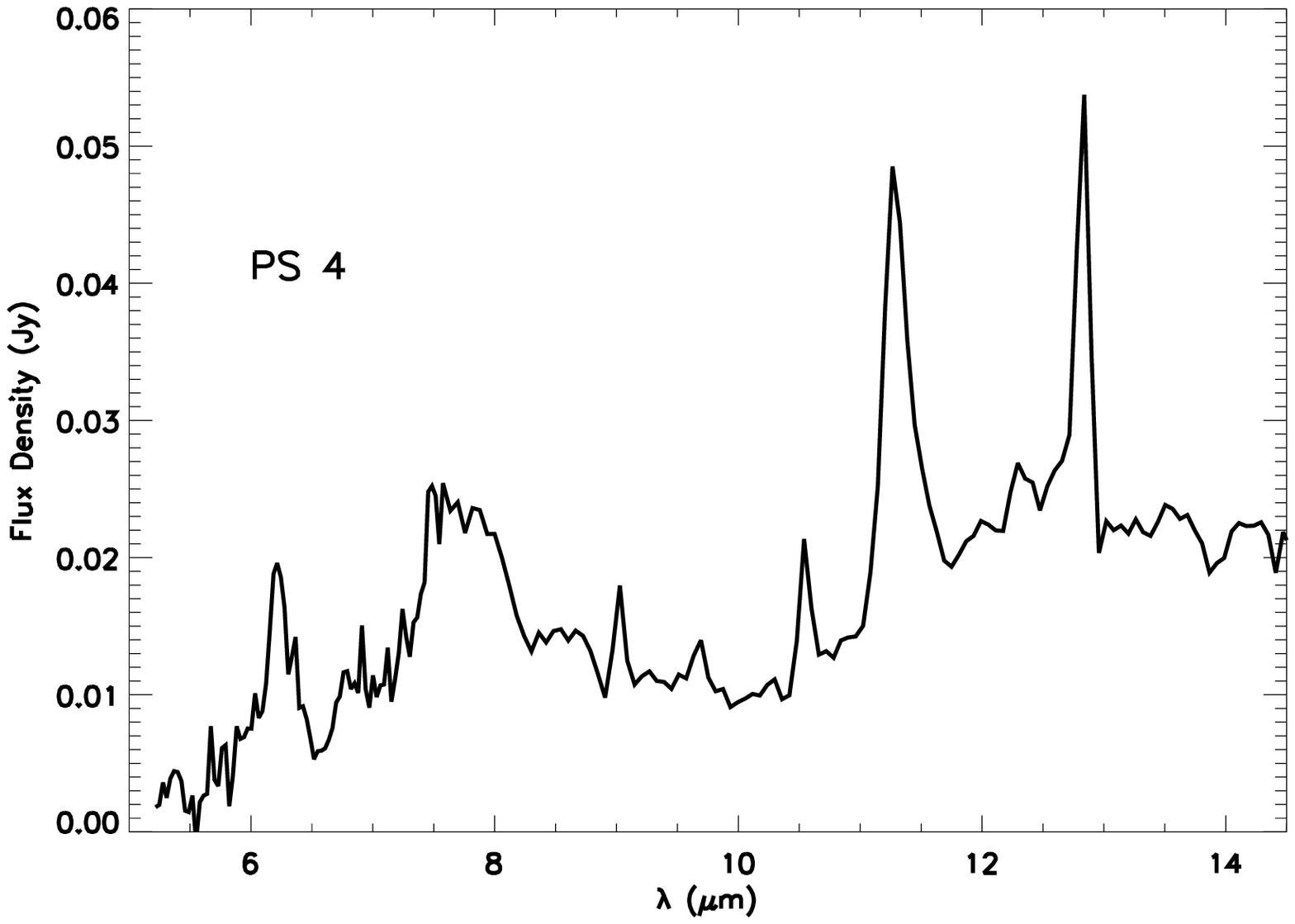}
\includegraphics[scale=0.3]{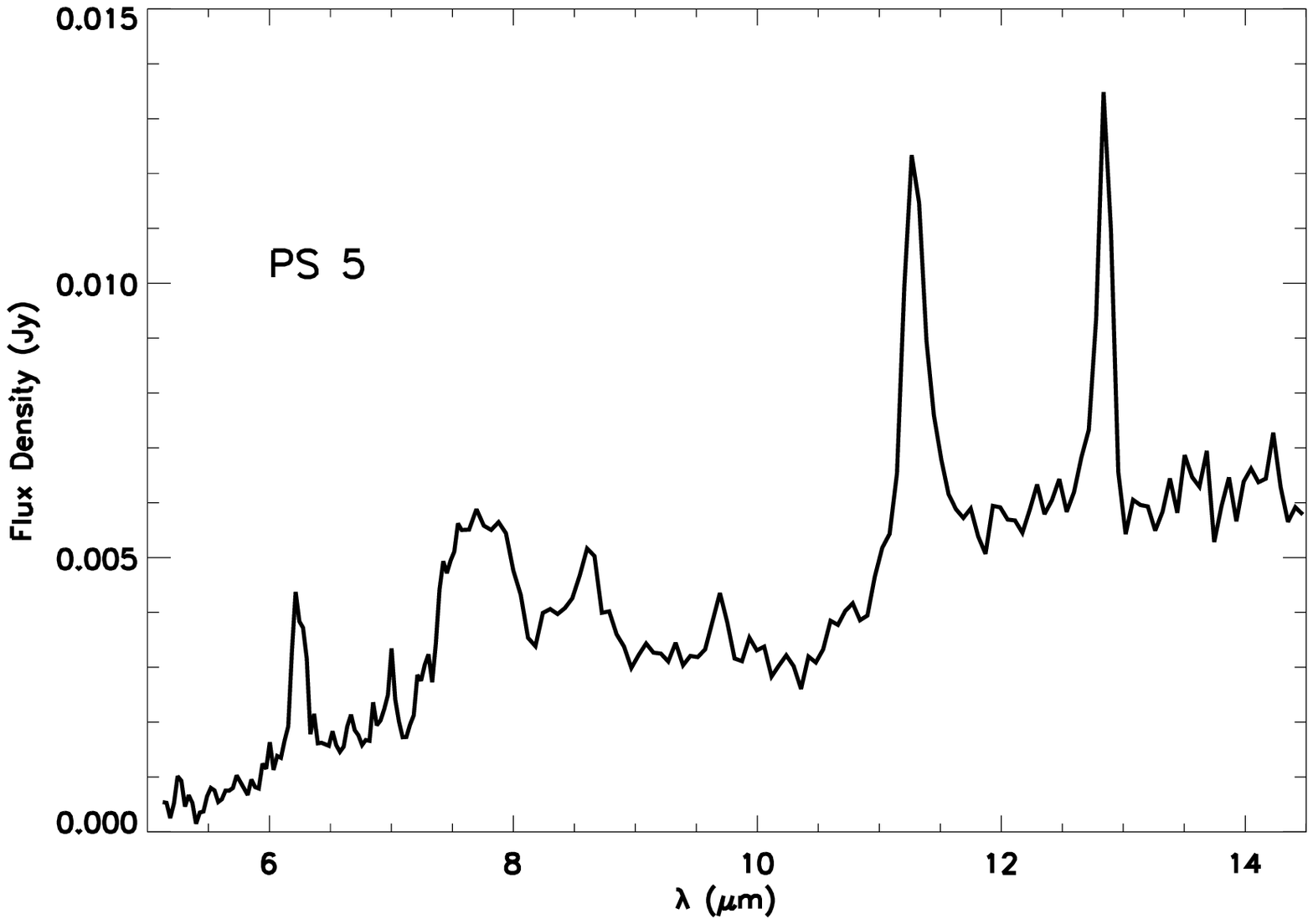}
\includegraphics[scale=0.3]{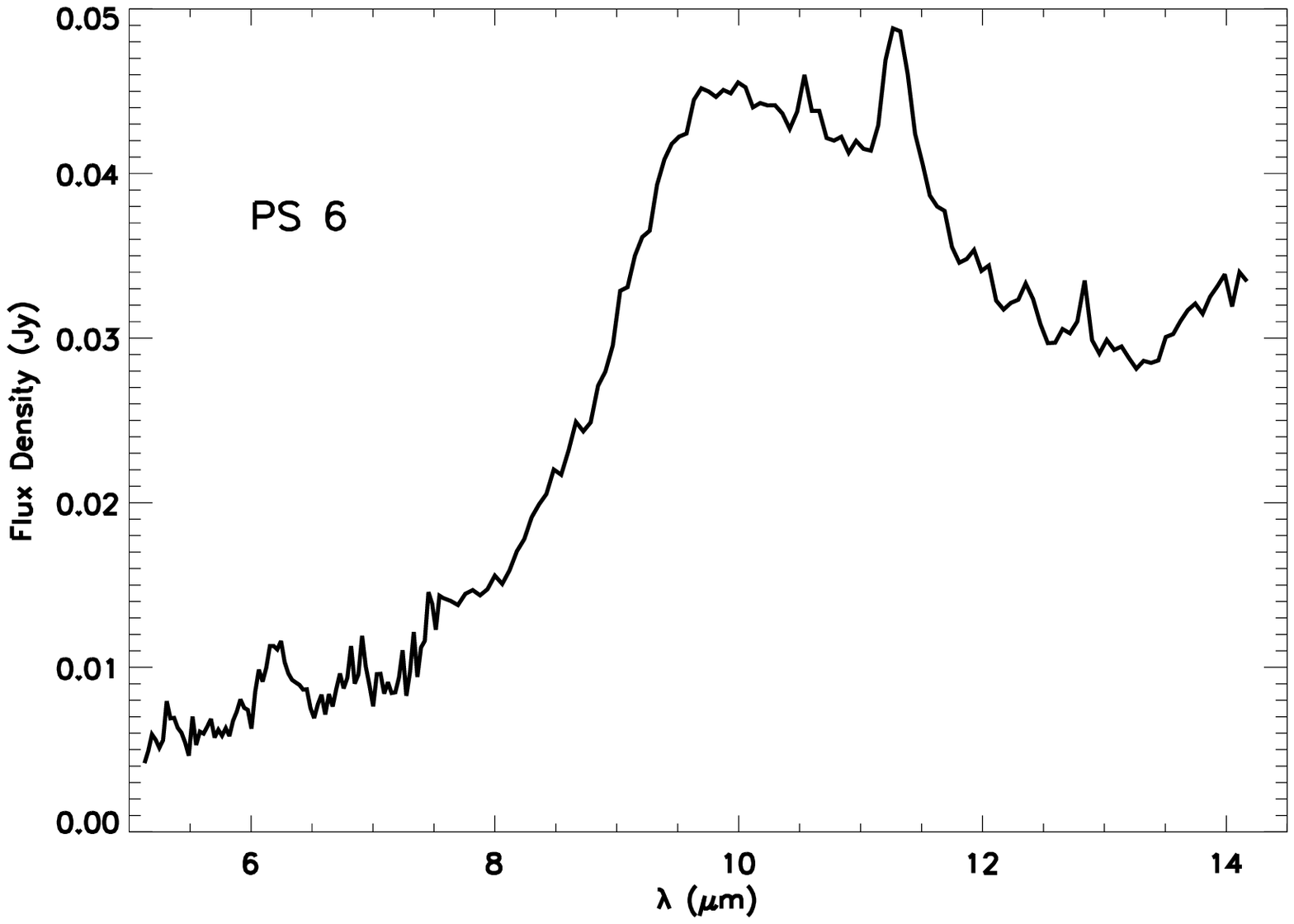}
\includegraphics[scale=0.3]{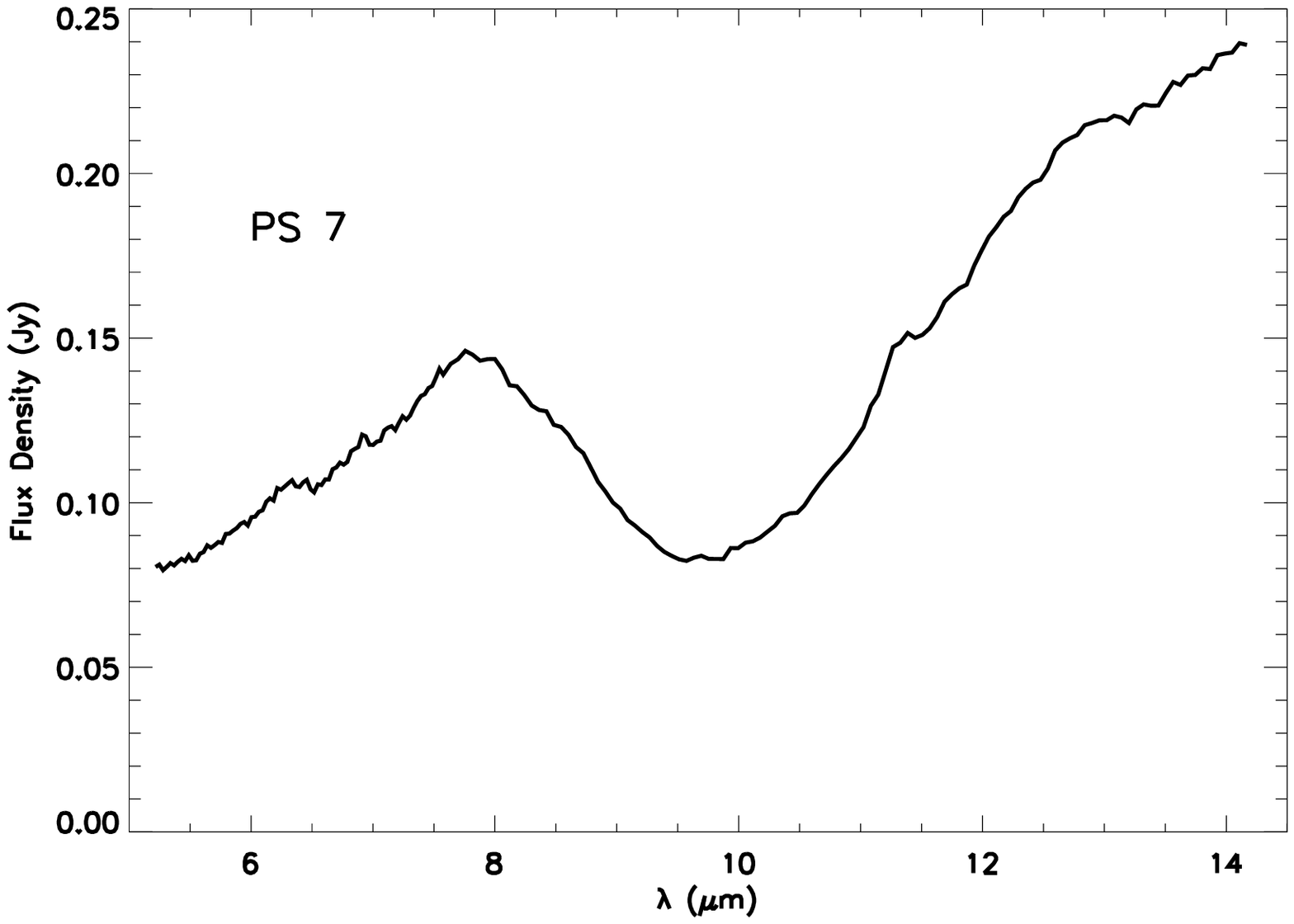}
\caption{{\it Spitzer} spectra of point sources PS1-PS7. The spectrum
  of PS1 is enlarged so that the PAH and H$_2$ feature labels can be
  seen clearly.\label{Spec1}}
\end{figure}

\begin{figure}
\includegraphics[scale=0.3]{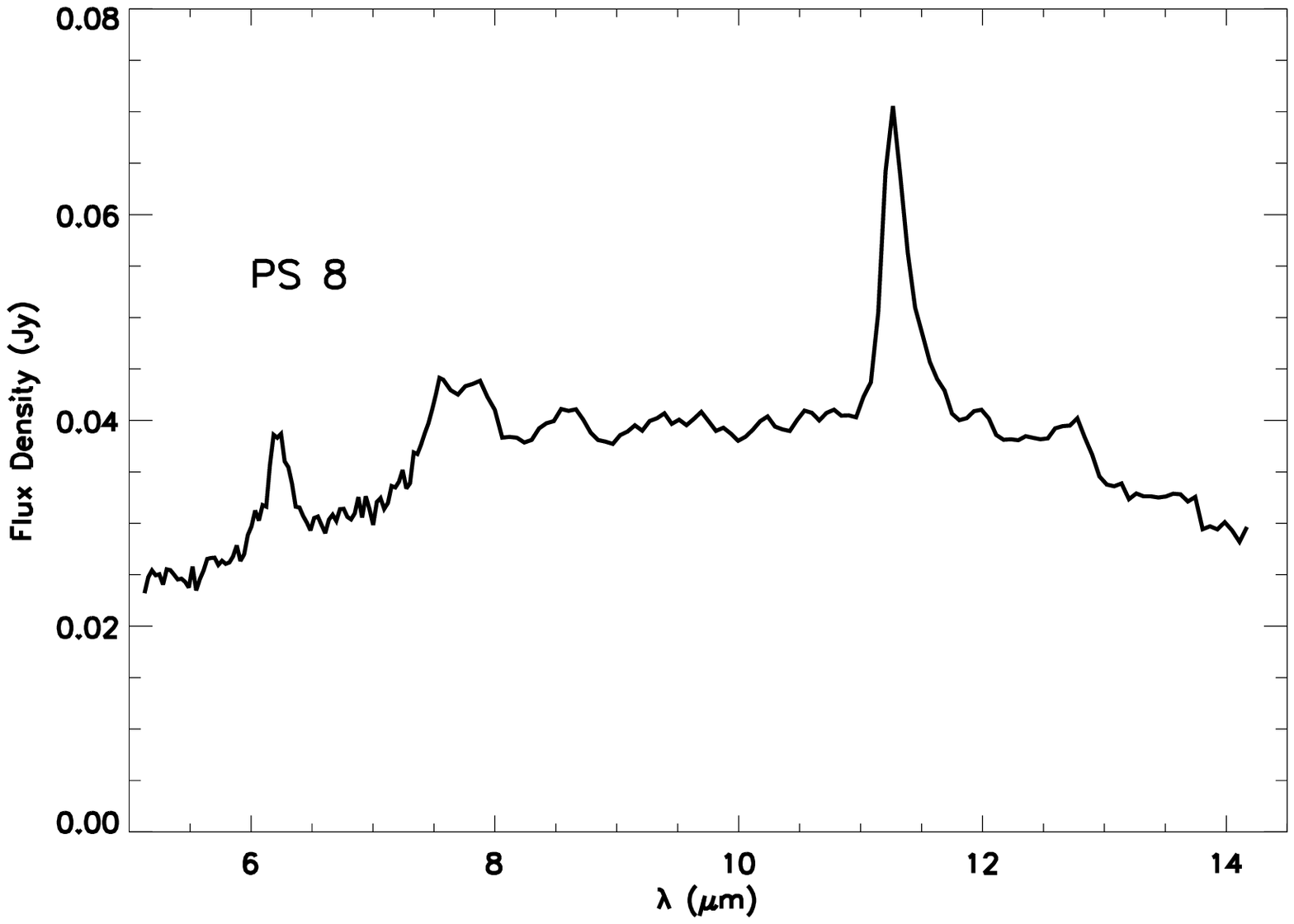}
\includegraphics[scale=0.3]{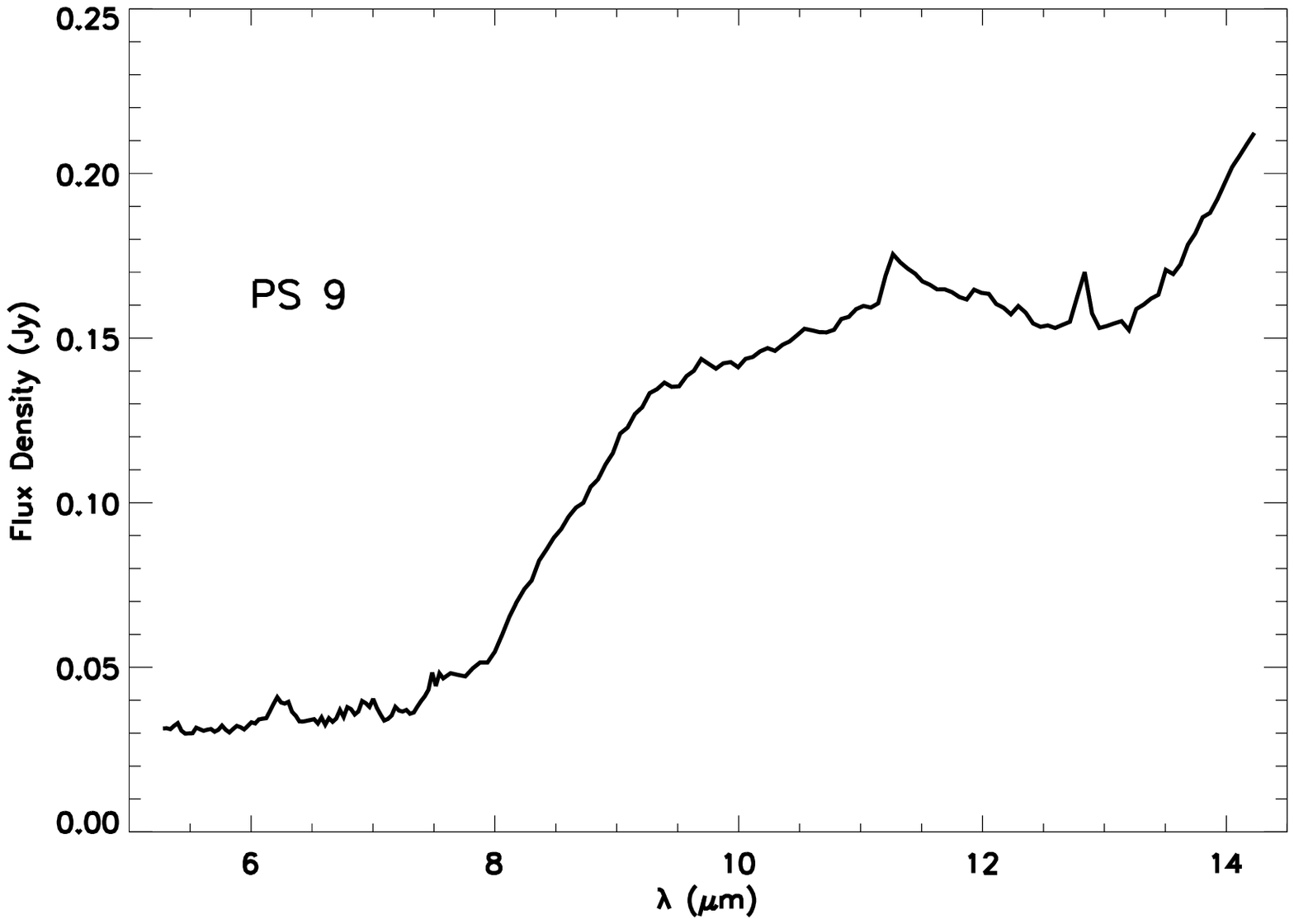}
\includegraphics[scale=0.3]{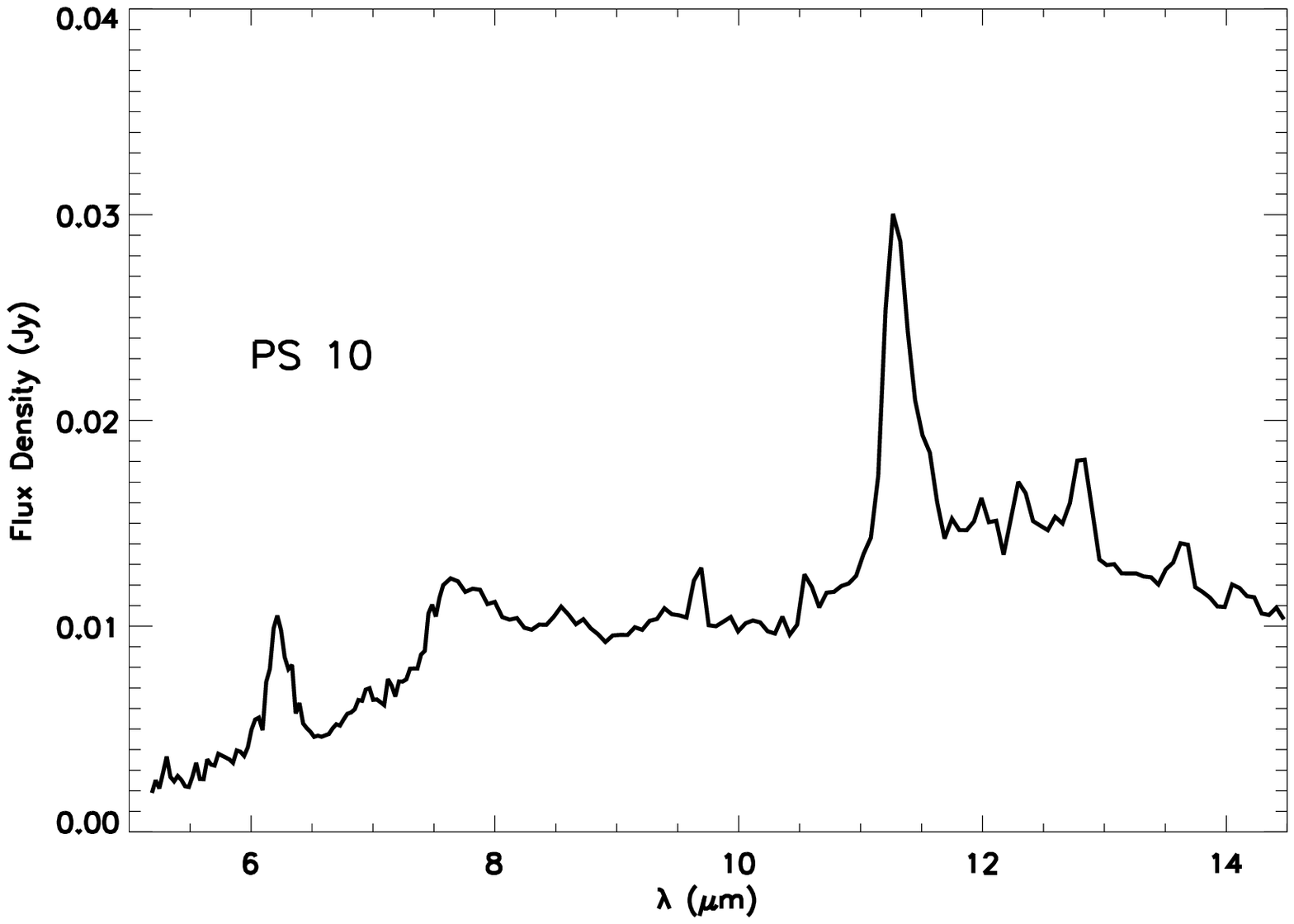}
\includegraphics[scale=0.3]{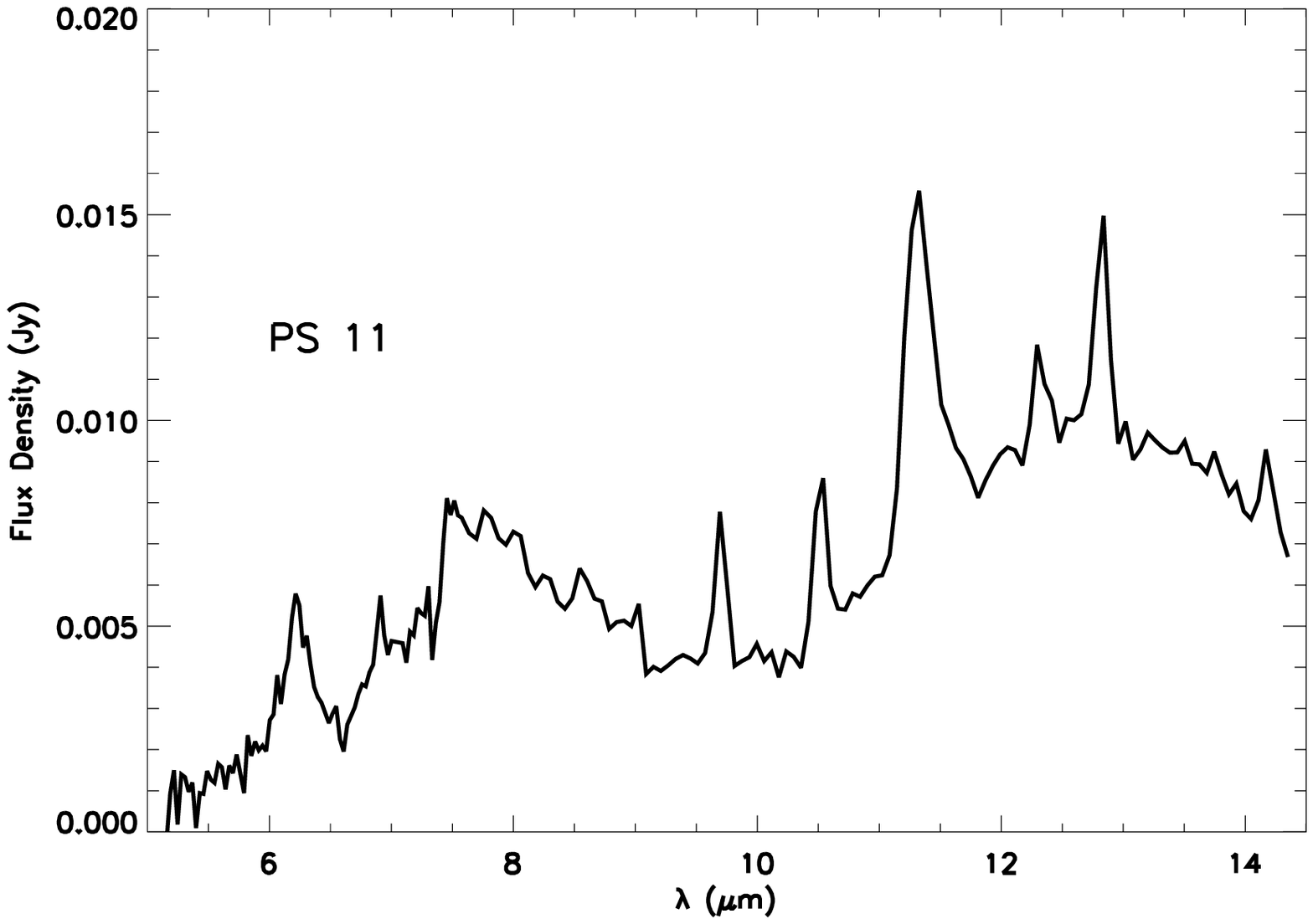}
\includegraphics[scale=0.3]{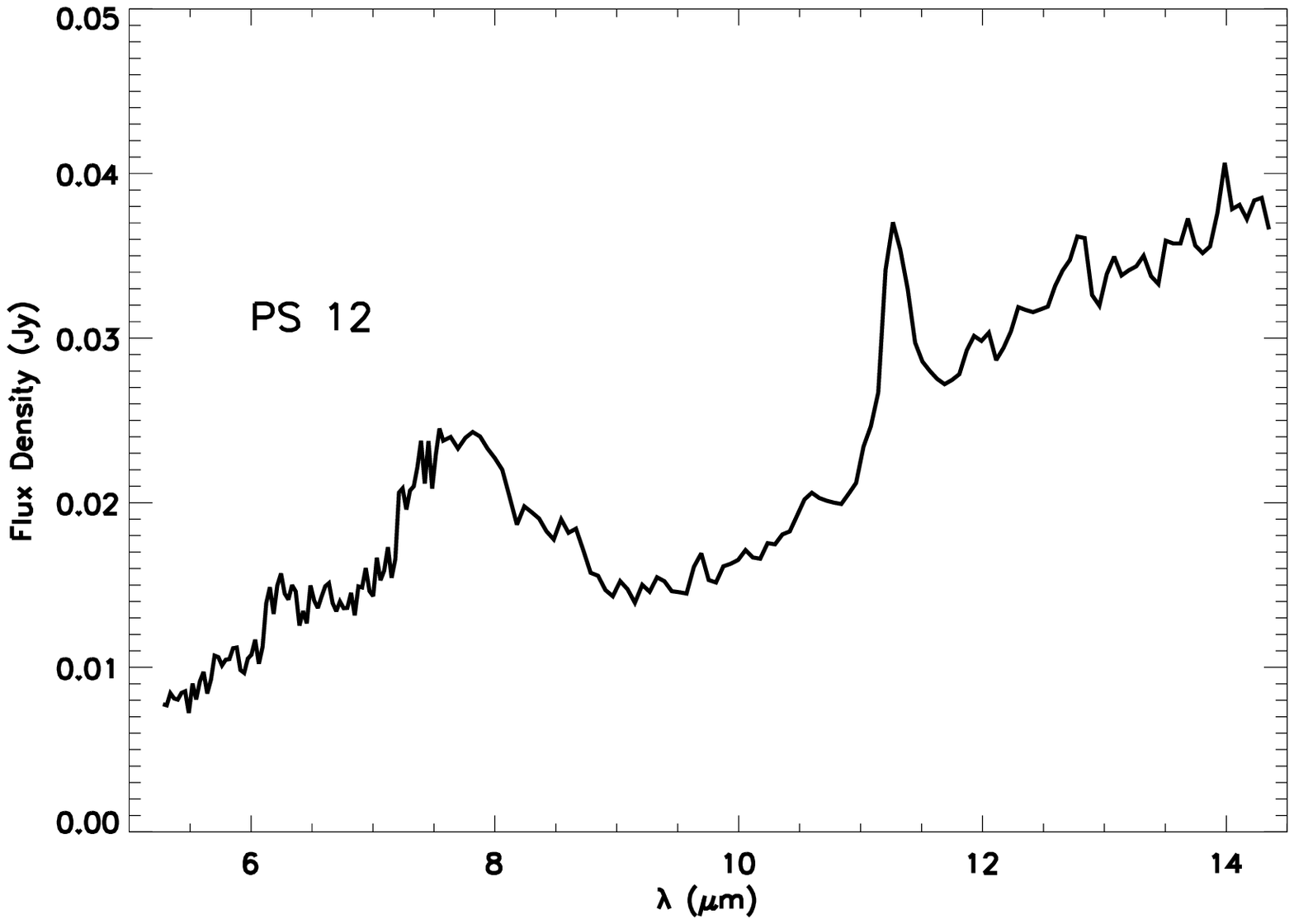}
\includegraphics[scale=0.3]{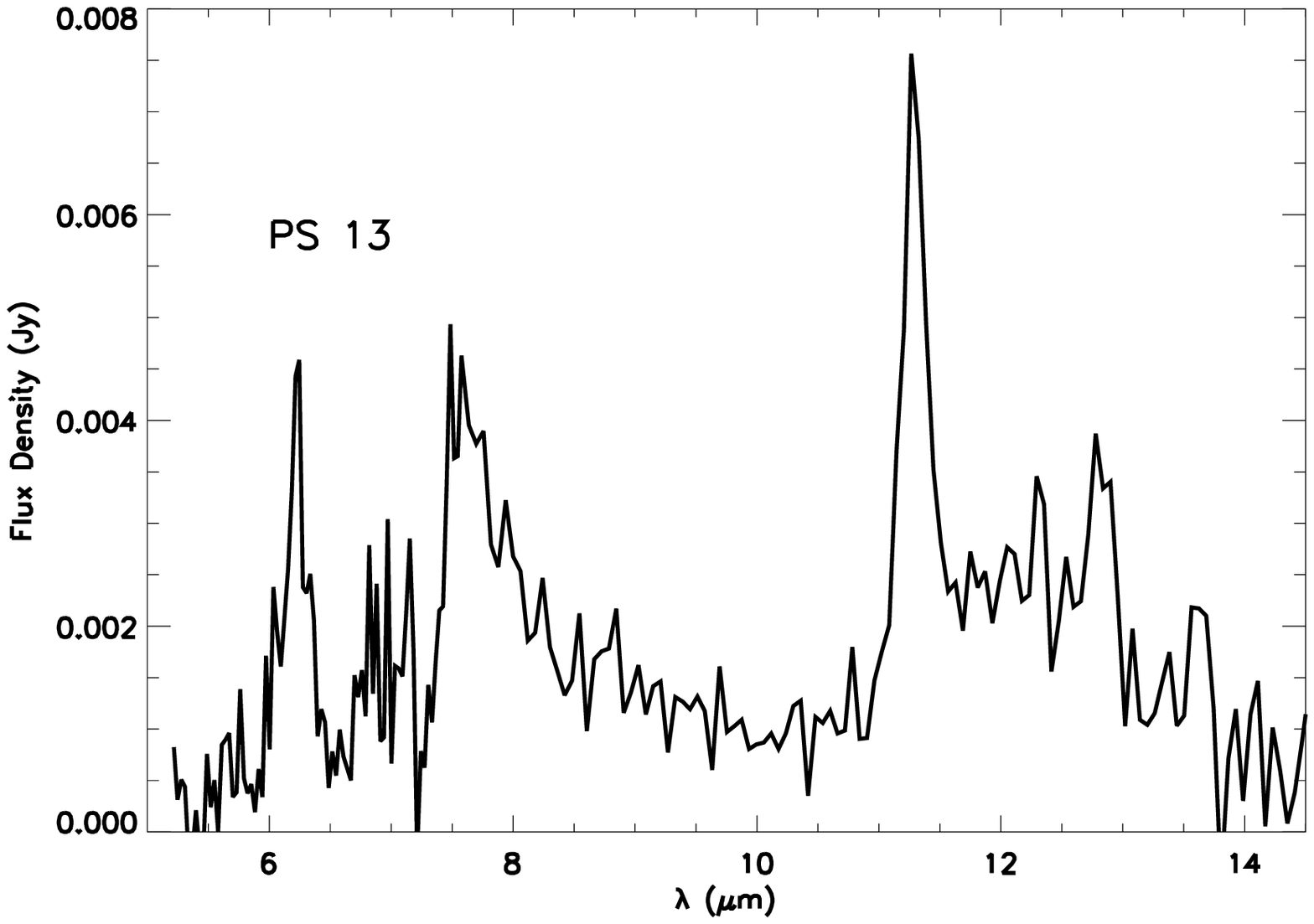}
\includegraphics[scale=0.3]{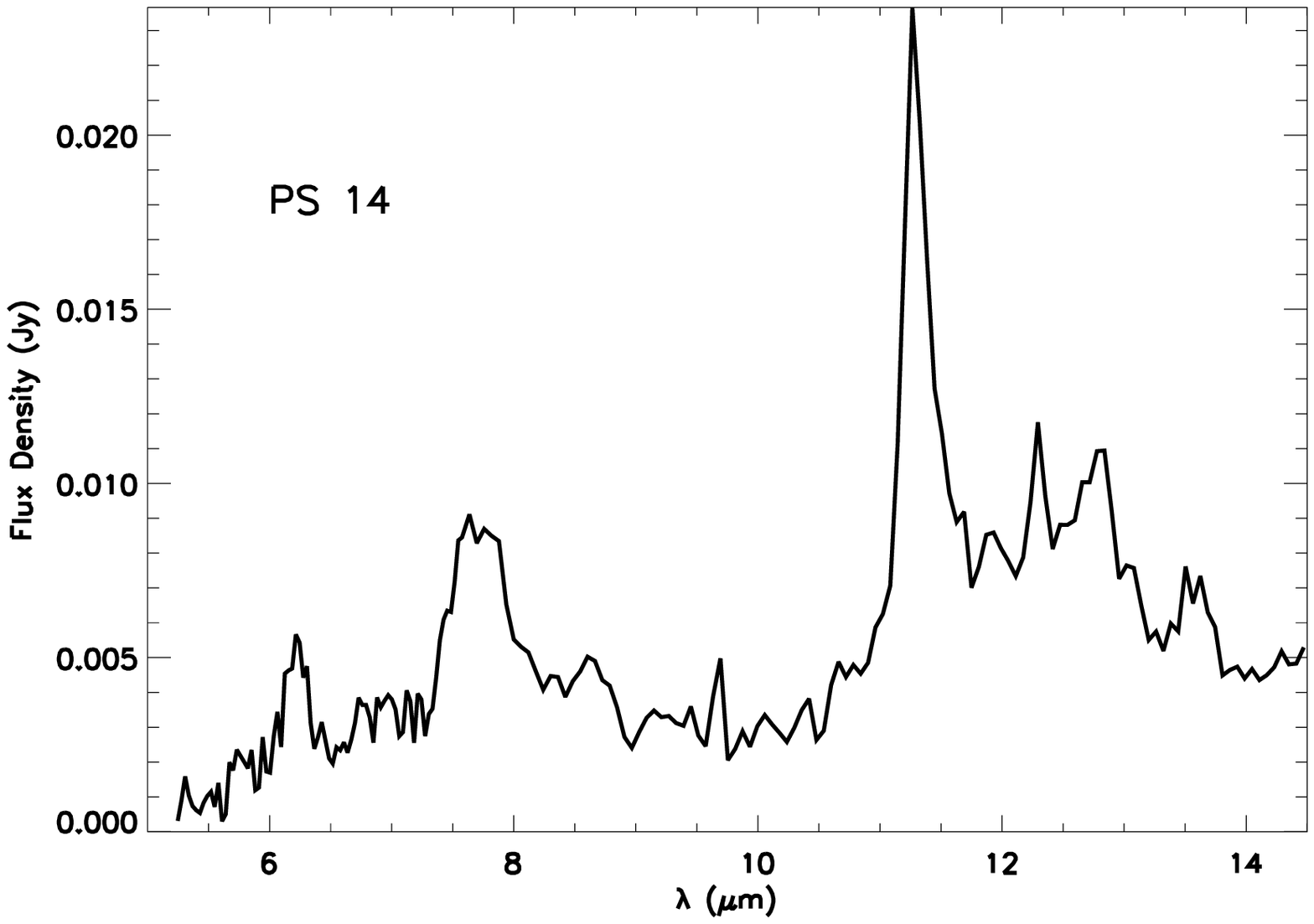}
\caption{{\it Spitzer} spectra of point Sources PS8-PS14.\label{Spec2}}
\end{figure}

\begin{figure}
\includegraphics[scale=0.6]{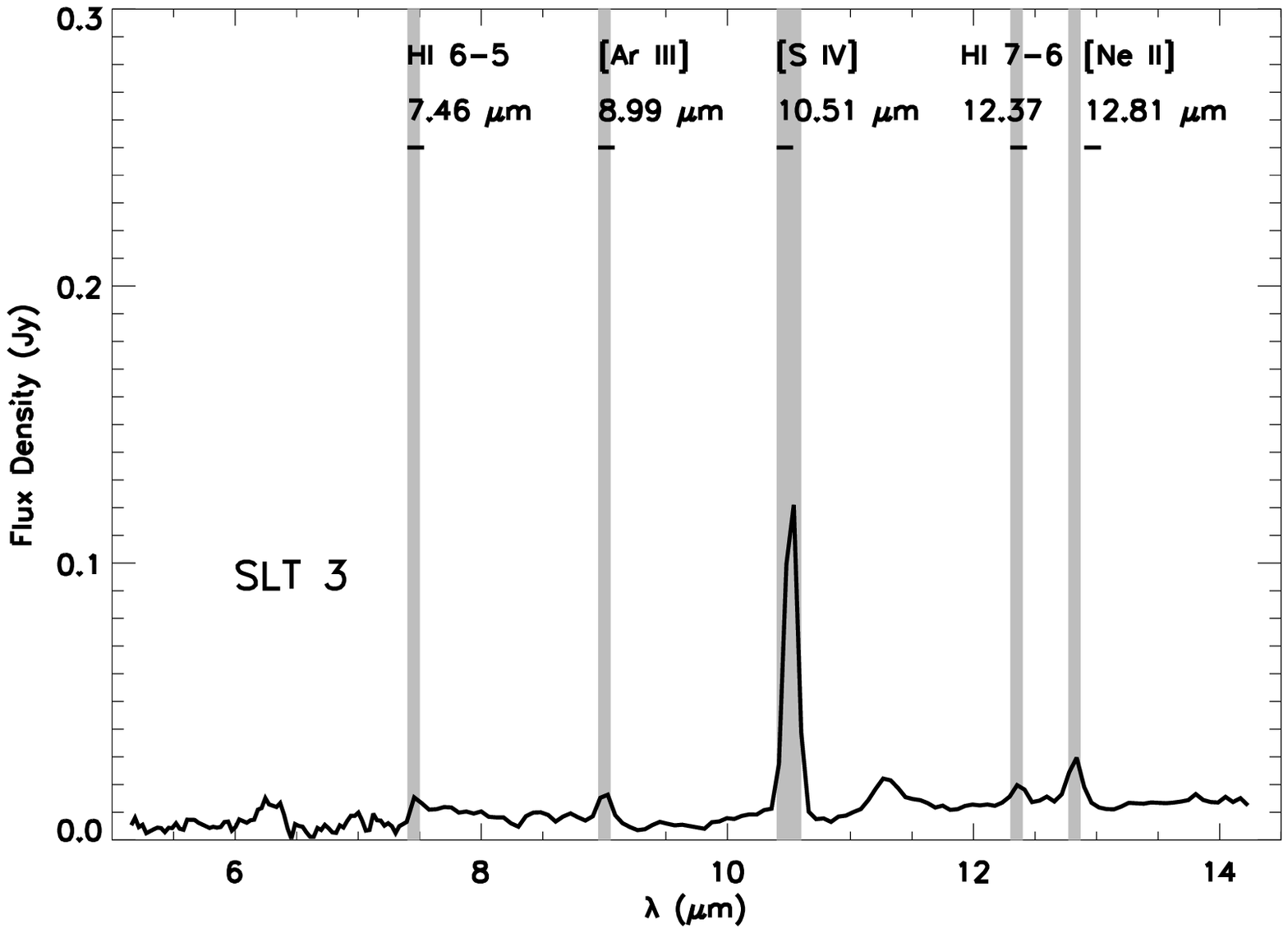}
\includegraphics[scale=0.3]{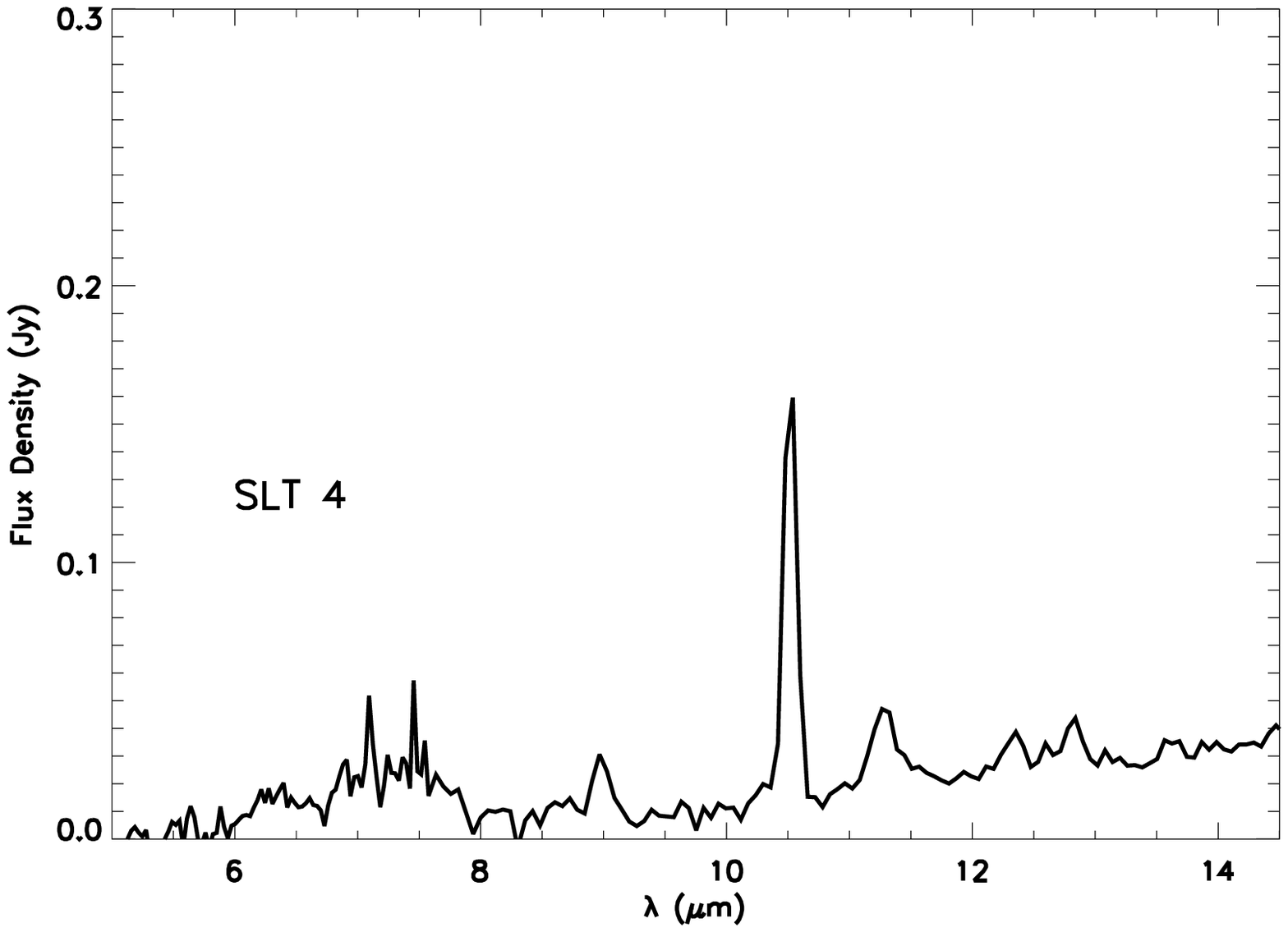}
\includegraphics[scale=0.3]{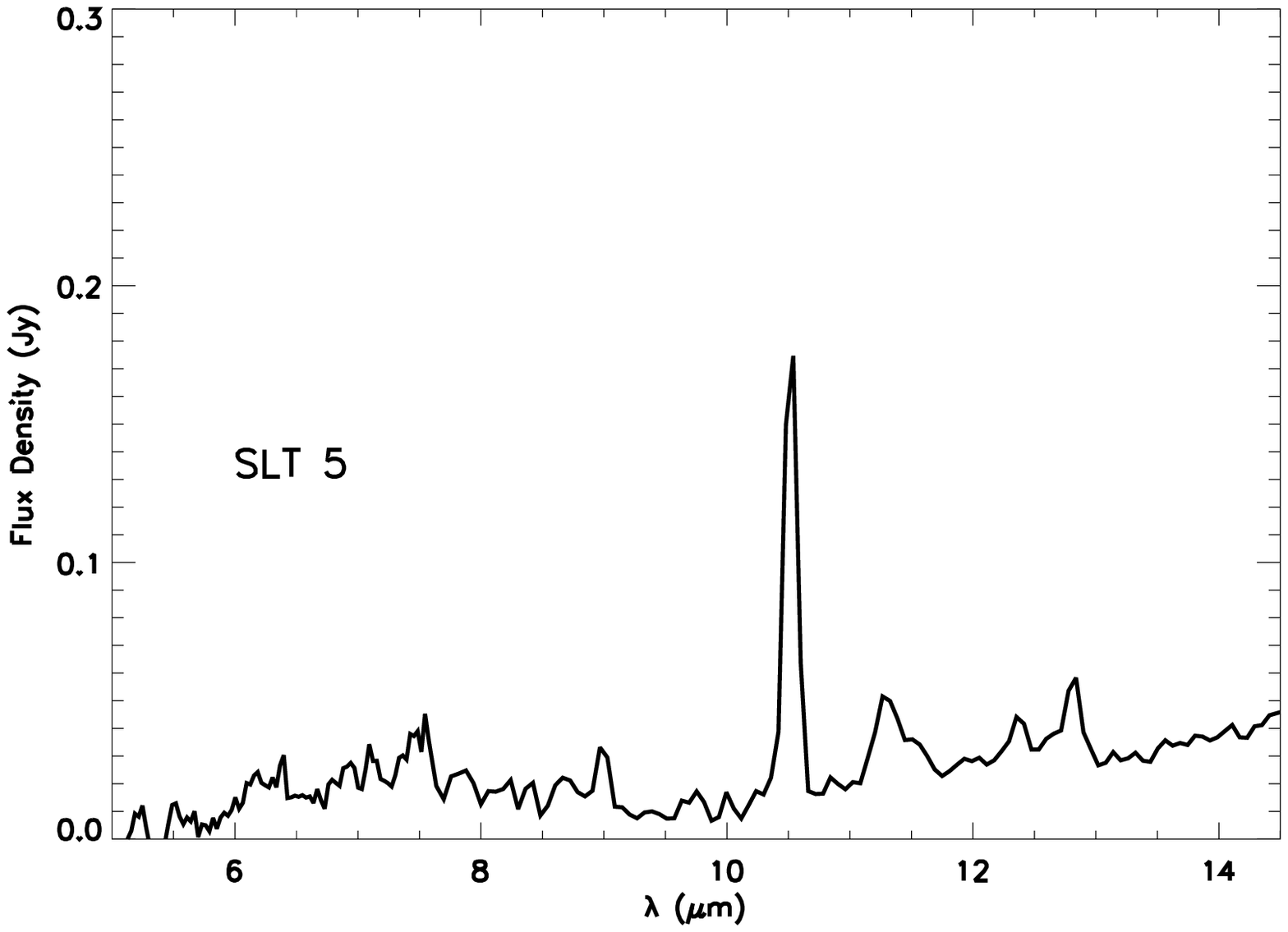}
\includegraphics[scale=0.3]{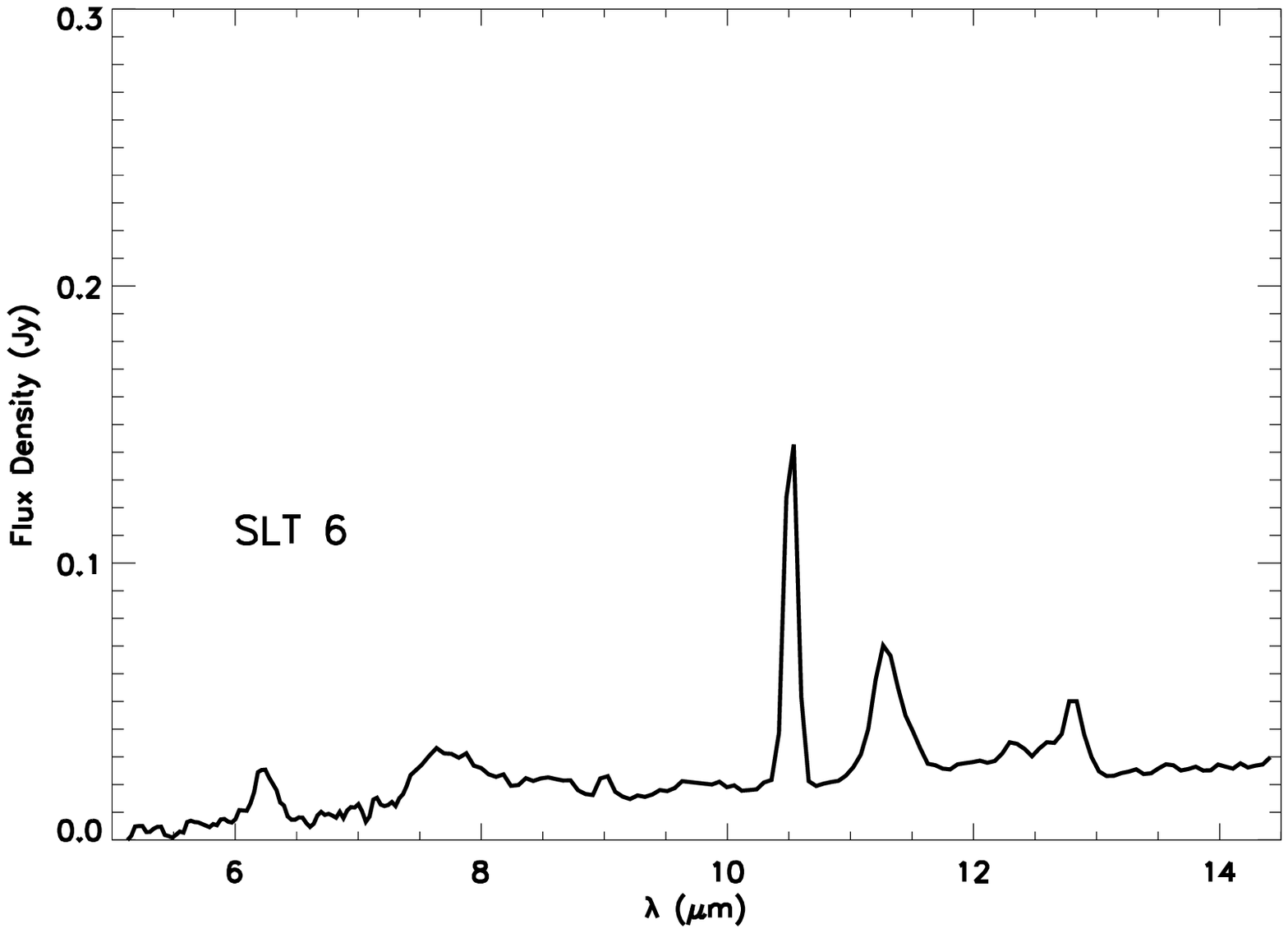}
\includegraphics[scale=0.3]{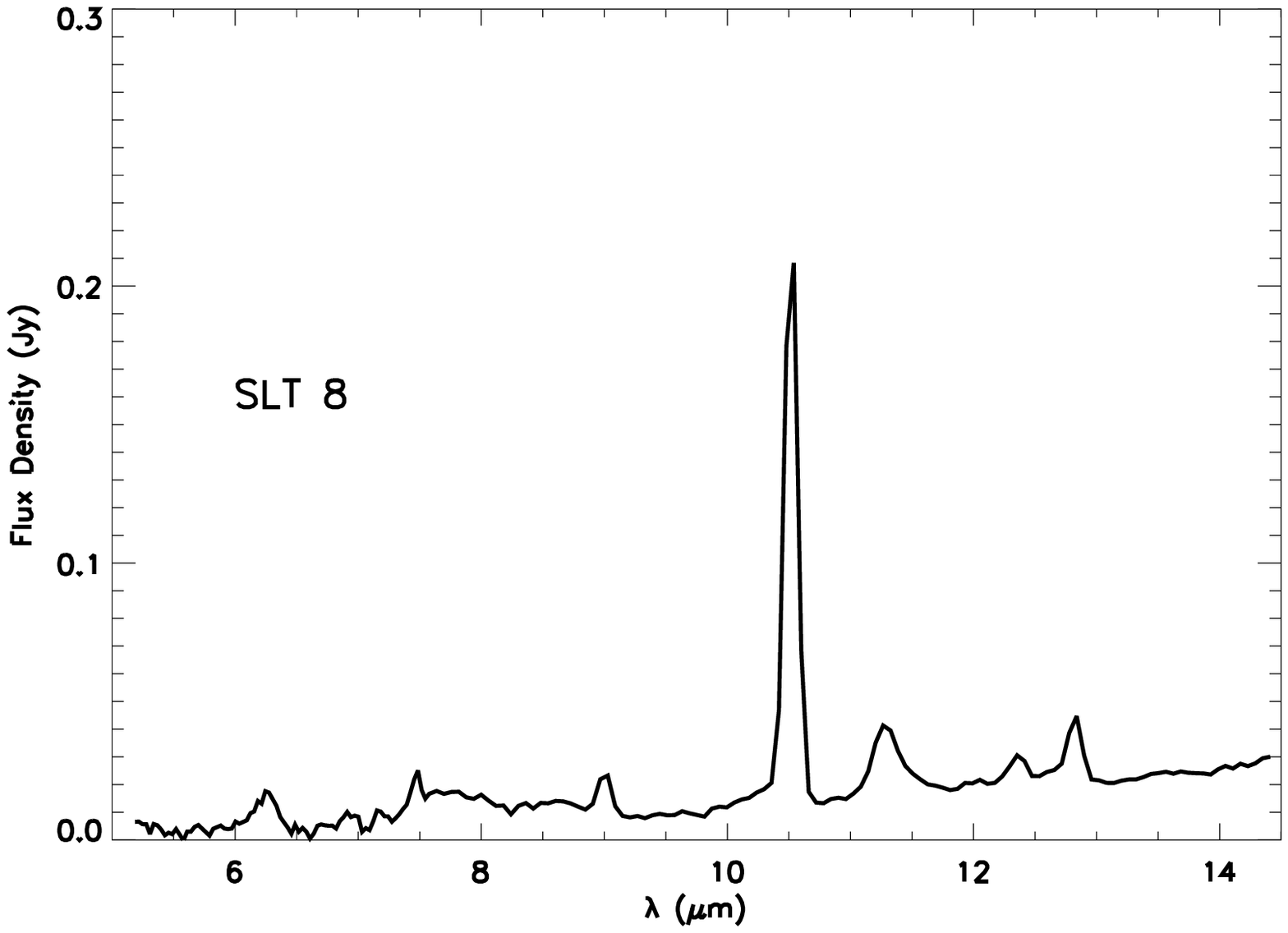}
\includegraphics[scale=0.3]{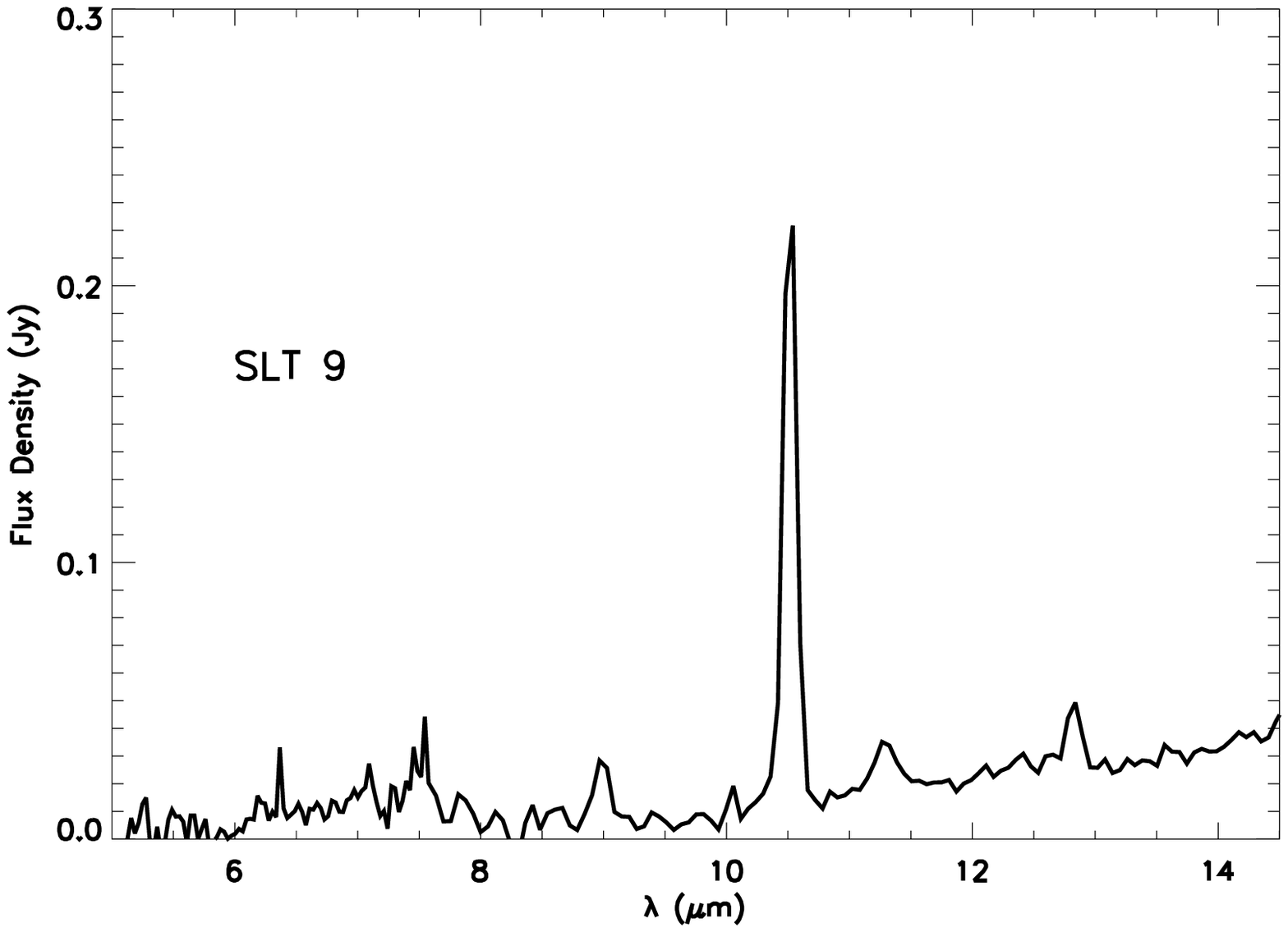}
\includegraphics[scale=0.3]{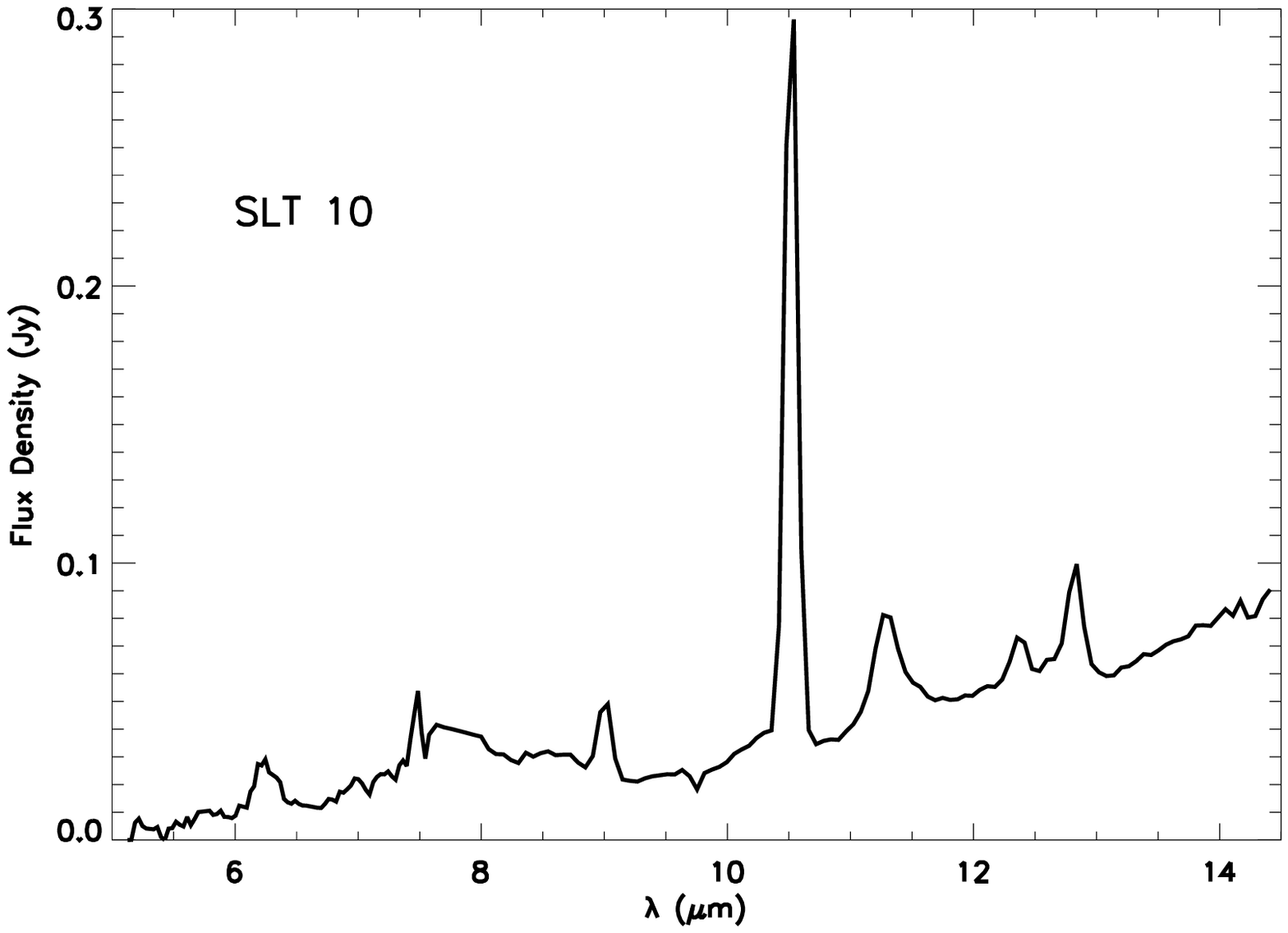}
\includegraphics[scale=0.3]{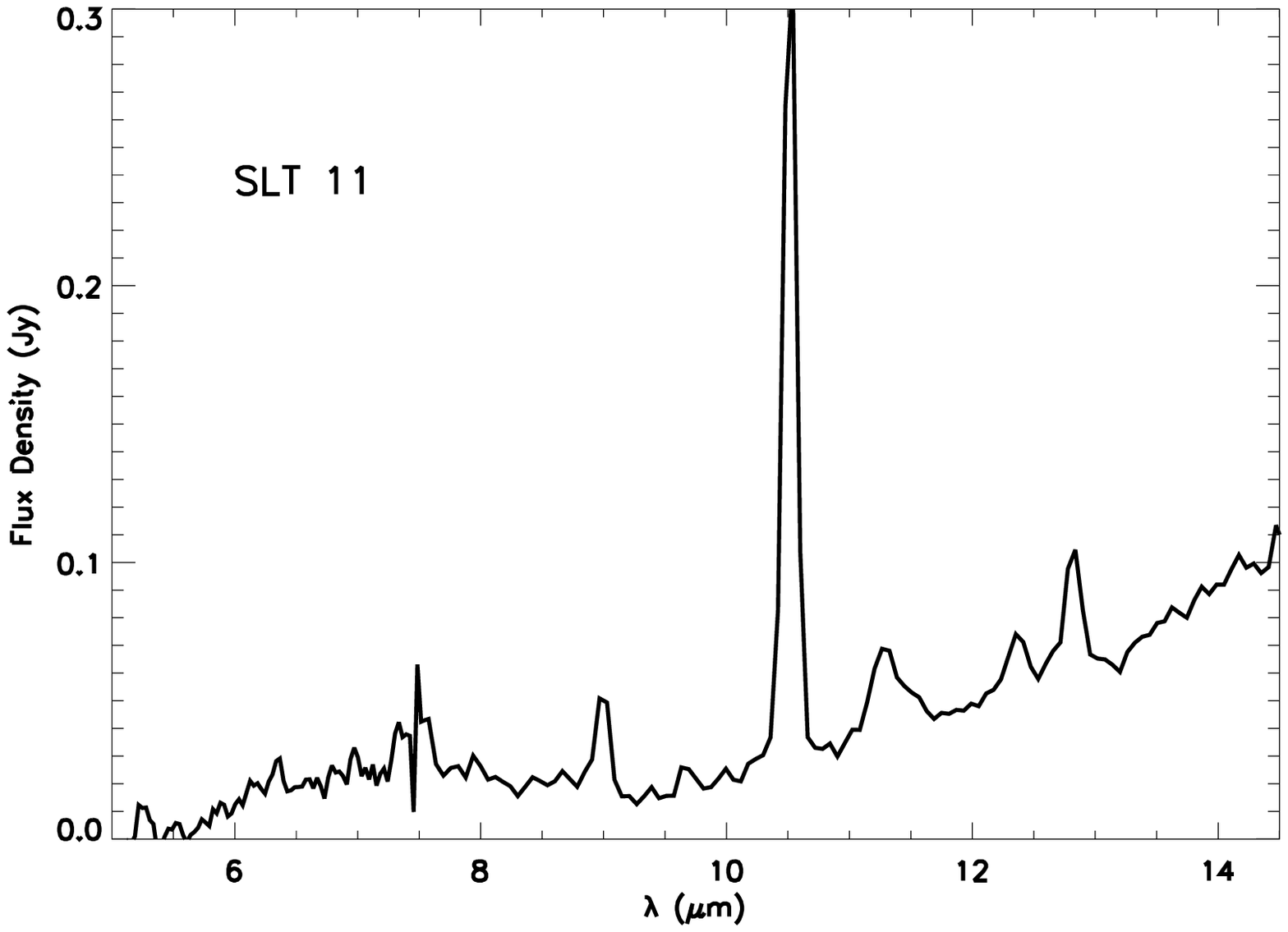}
\caption{{\it Spitzer} extended emission spectra SLT3-SLT11. The
  spectra are all plotted on the same scale, so that the relative flux
  in each slit can be seen. The spectrum for SLT3 has been enlarged so
  that the atomic and ionic line labels can be see
  clearly.\label{Spec3}}
\end{figure}

\begin{figure}
\includegraphics[scale=0.3]{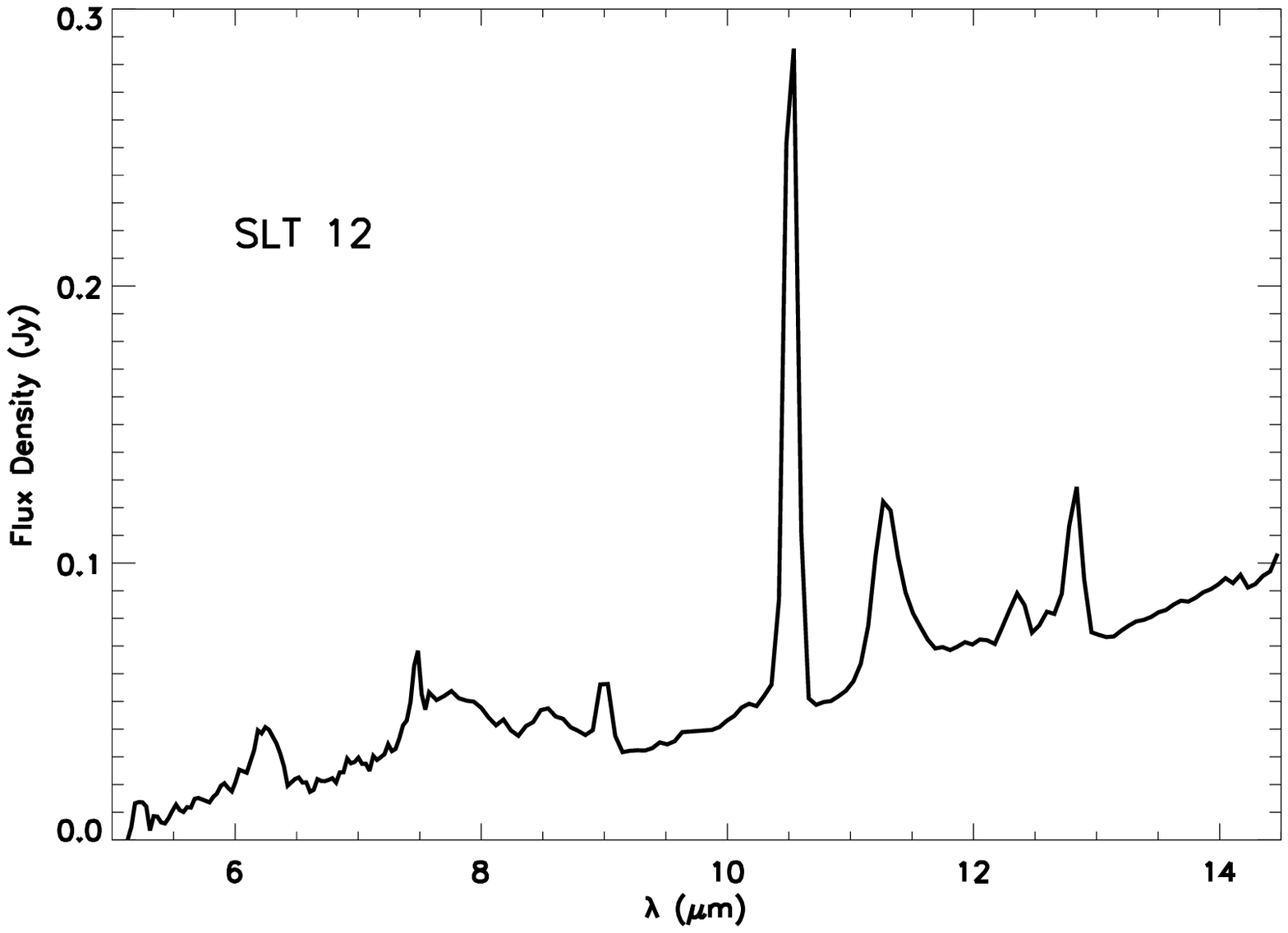}
\includegraphics[scale=0.3]{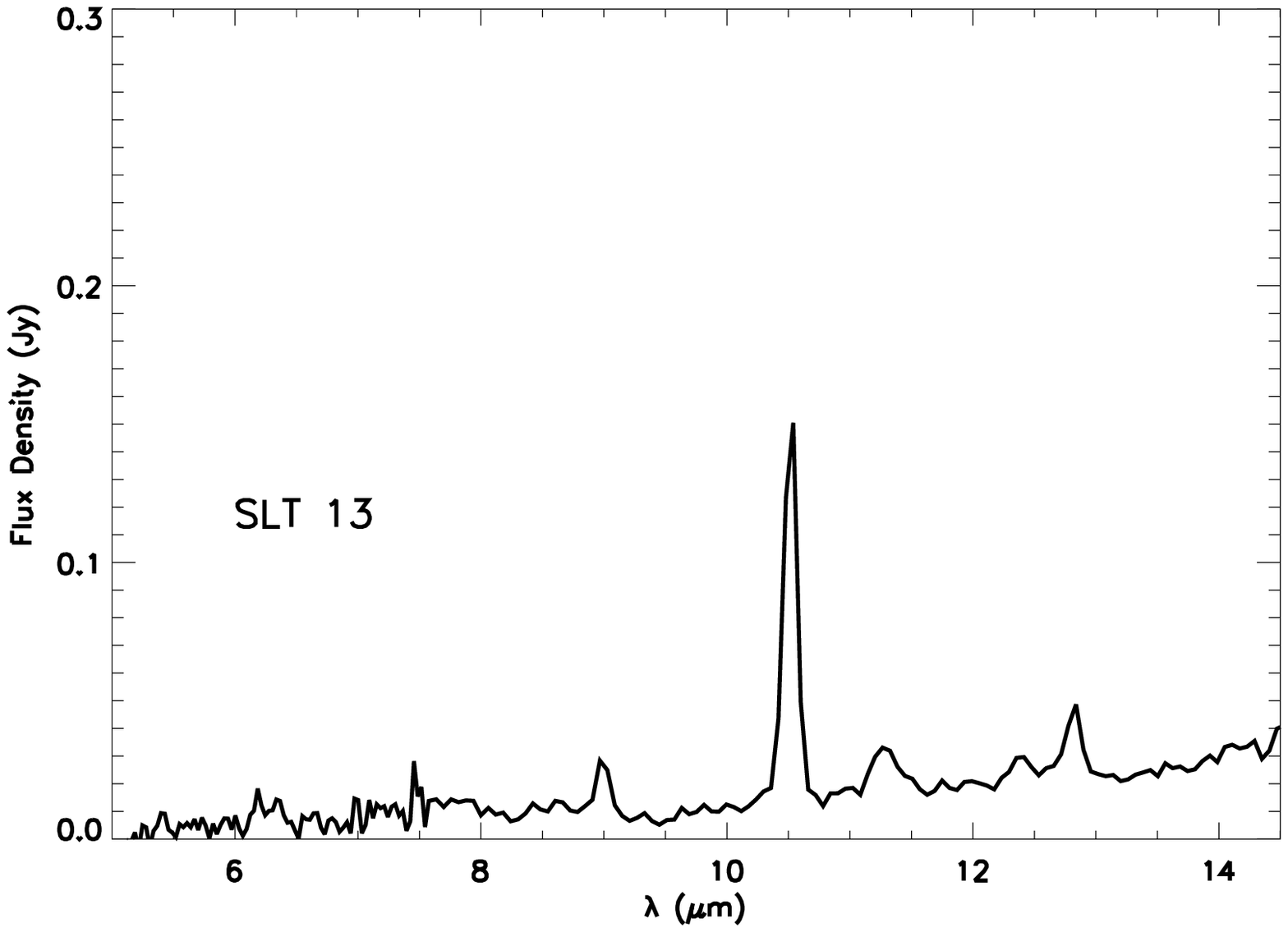}
\includegraphics[scale=0.3]{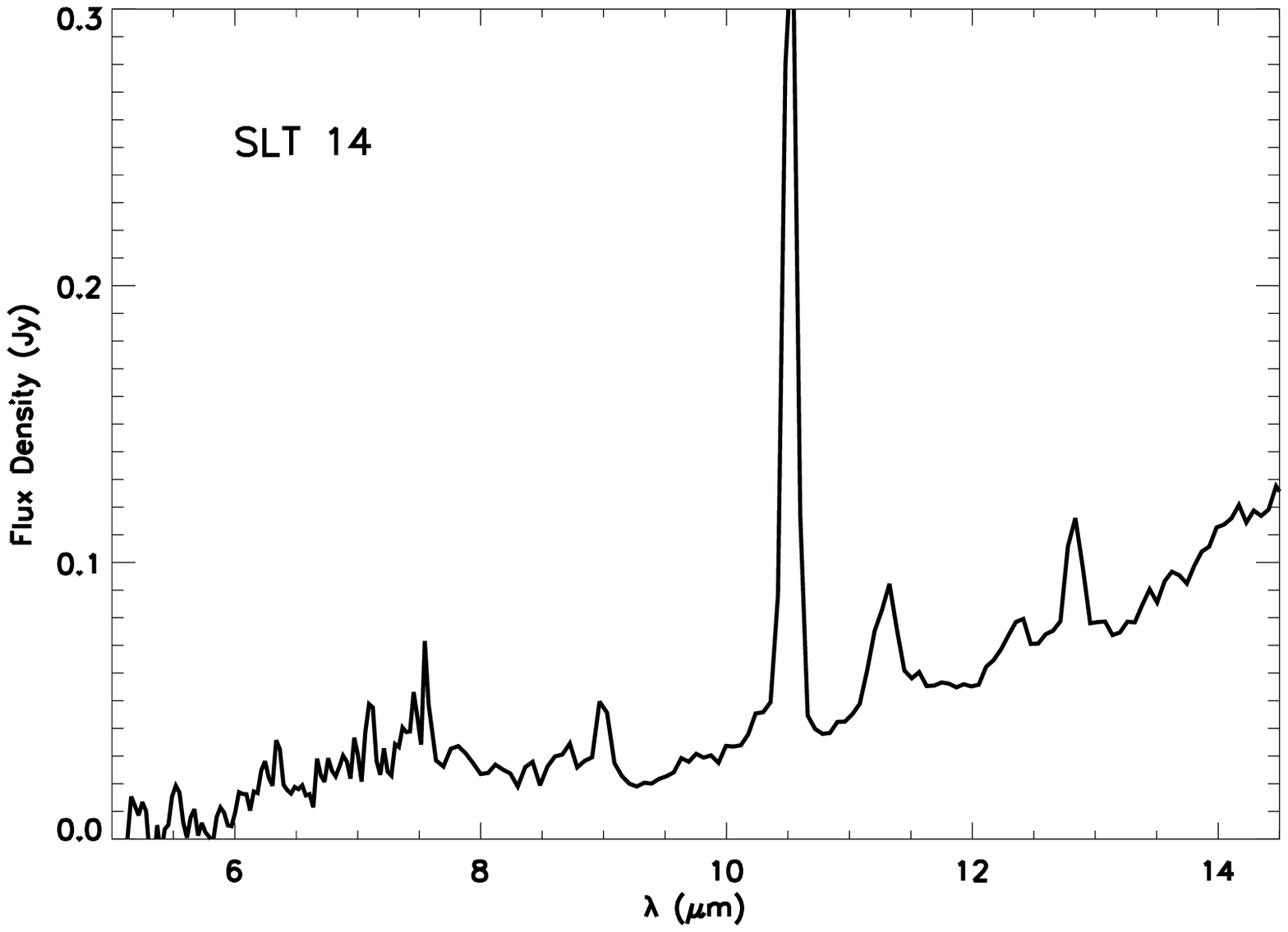}
\includegraphics[scale=0.3]{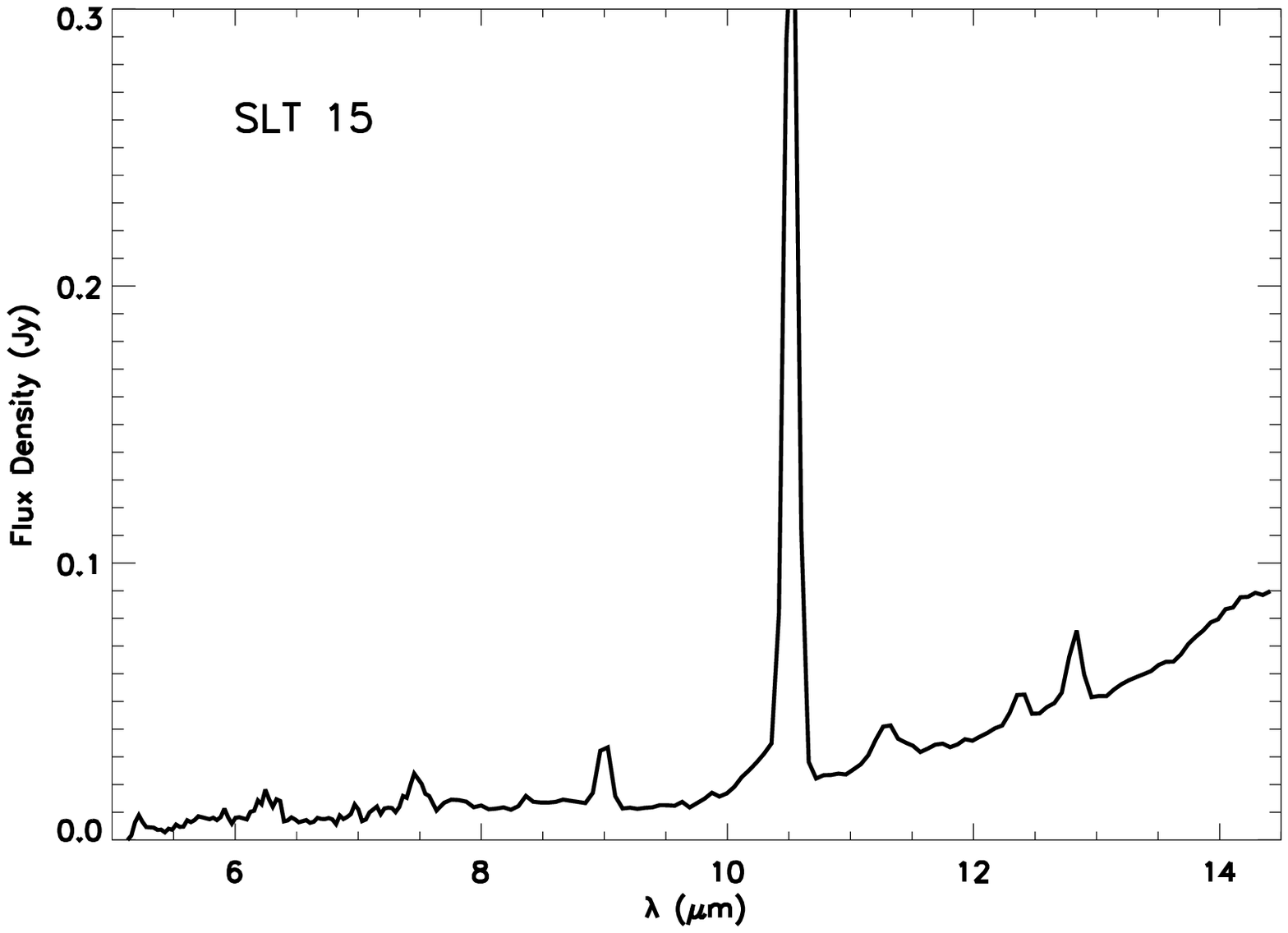}
\includegraphics[scale=0.3]{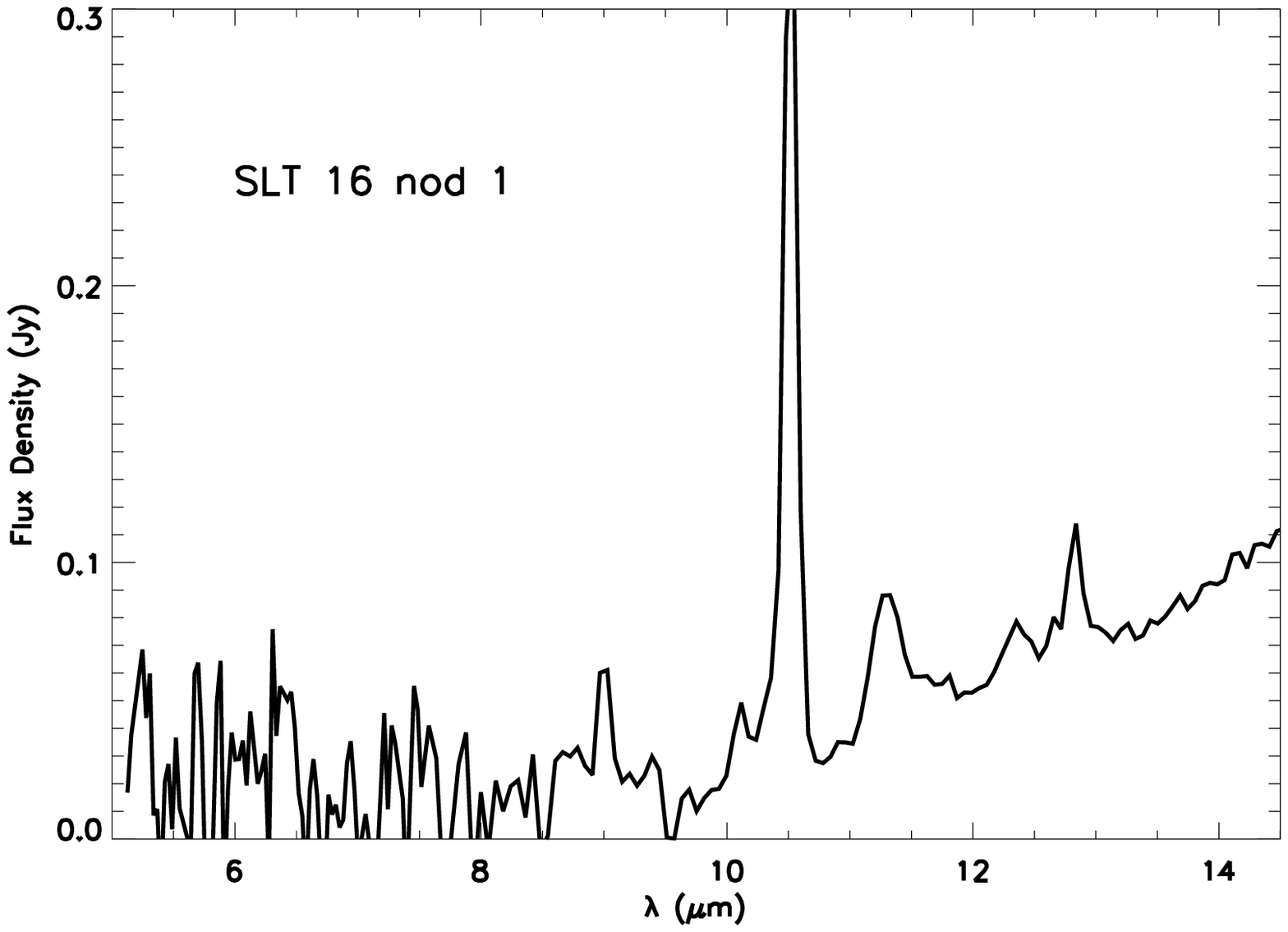}
\includegraphics[scale=0.3]{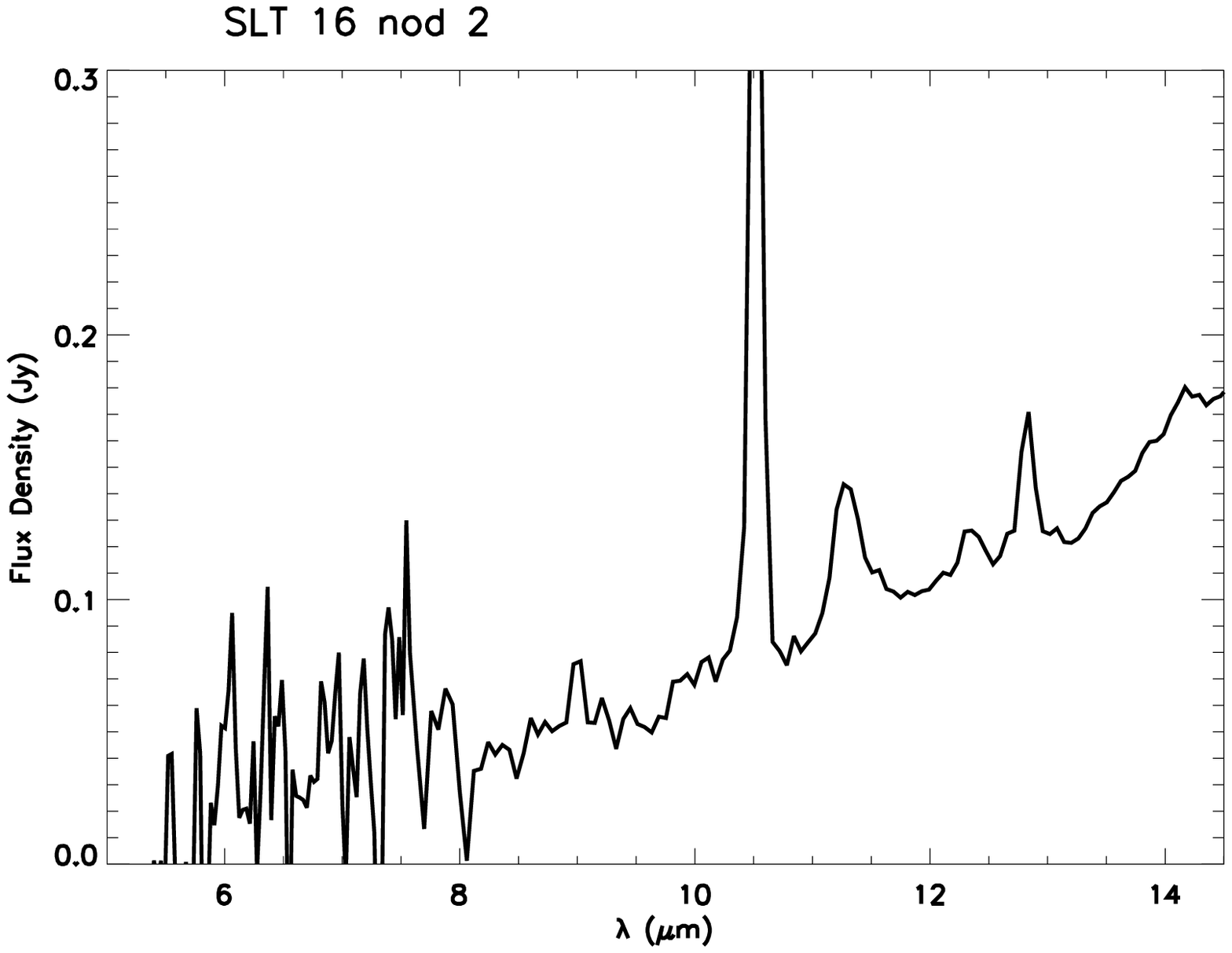}
\includegraphics[scale=0.3]{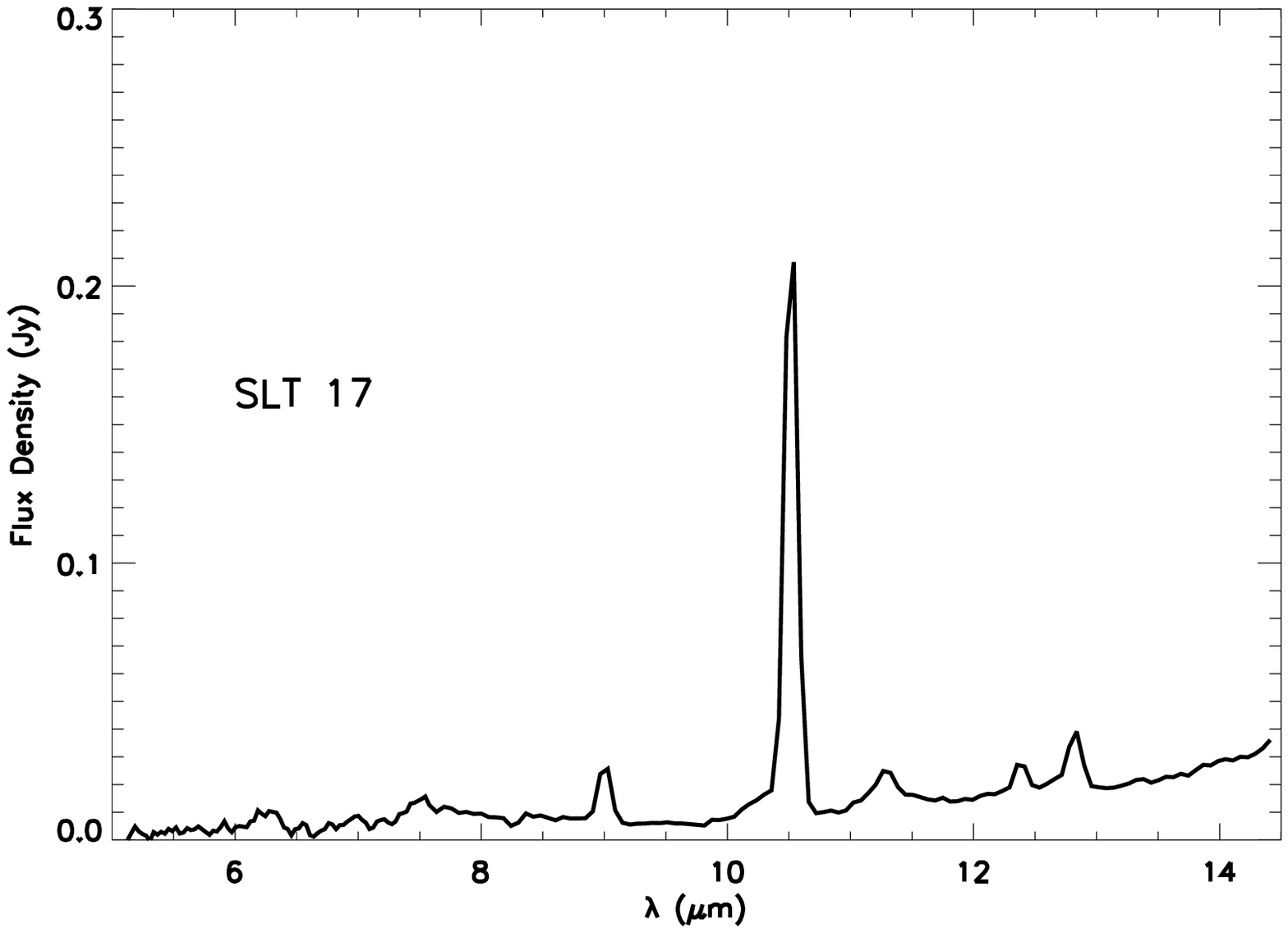}
\includegraphics[scale=0.3]{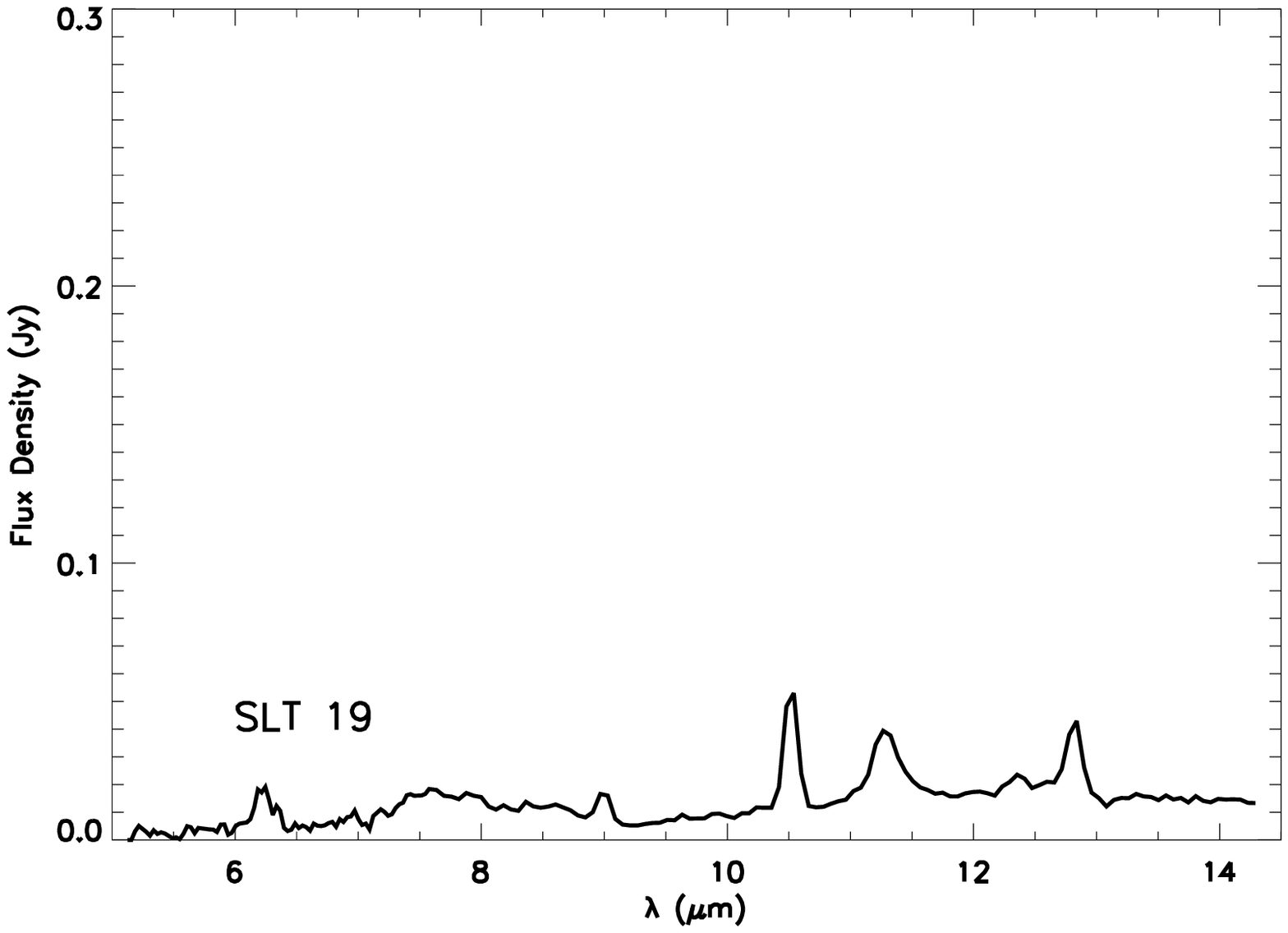}
\caption{{\it Spitzer} extended emission spectra SL12-SLT19. The
  spectra are all plotted on the same scale, so that the relative flux
  in each slit can be seen.\label{Spec4}}
\end{figure}

\clearpage

\begin{deluxetable}{cccccc}
\tabletypesize{\footnotesize}
\tablecaption{Point Source Positions\label{targets}}

\tablehead{ \colhead{RA} & \colhead{Dec} & \multicolumn{3}{c}{Designations} & Source \\ (J2000) & (J2000) & This work & \citet{Lebouteiller2008} & \citet{Contursi2000} & Identification}

\startdata
00h59m21.8s & -72d11m13.3s & PS1  & -- & -- & \\
00h59m20.0s & -72d11m21.0s & PS2  &  9 & I  & \\
00h59m17.1s & -72d11m24.3s & PS3  & 13 & -- & \\
00h59m14.8s & -72d11m03.8s & PS4  &  8 & H  & 1 class I YSO\tablenotemark{b} \\
00h59m14.0s & -72d09m27.1s & PS5  & -- & G  & N66C\tablenotemark{a}; 1 class II YSO\tablenotemark{b} \\
00h59m12.3s & -72d09m58.7s & PS6  & 11 & F  & N66B\tablenotemark{a}; 1 class I YSO\tablenotemark{b} \\
00h59m09.3s & -72d10m57.8s & PS7  &  5 & E  & N66A\tablenotemark{a}; 2 class I YSOs\tablenotemark{b} \\
00h59m05.8s & -72d11m27.5s & PS8  &  7 & D  & KWBBE 200, a Be star \\
00h59m05.4s & -72d10m36.1s & PS9  &  6 & C  & N66\tablenotemark{a}; NGC 346 \\
00h58m59.8s & -72d10m16.7s & PS10 & -- & -- & \\
00h58m57.6s & -72d09m56.7s & PS11 & -- & -- & \\
00h58m57.0s & -72d09m54.5s & PS12 & 10 & A  & \\
00h58m54.7s & -72d09m51.5s & PS13 & -- & -- & \\
00h58m51.7s & -72d09m29.2s & PS14 & -- & -- & \\
\enddata

\tablenotetext{a}{H{\sc ii} regions from \citet{Henize1956} ordered by
  brightness.}

\tablenotetext{b}{YSO classifications are from \citet{Simon2007}.
  Where multiple YSOs contribute to the point source flux, only class
  I YSOs are listed because they contribute the most to the point
  source flux.}

\end{deluxetable}

\clearpage

\begin{deluxetable}{cccc}
\tabletypesize{\footnotesize}
\tablecaption{Slit Positions\label{slitcentroids}}

\tablehead{ \colhead{RA} & \colhead{Dec} &
  Designation\tablenotemark{a} & Notes }

\startdata
00h59m49.99s & -72d13m00.1s & SLT1  & sky position \\
00h59m42.00s & -72d11m09.9s & SLT2  & sky position \\
00h59m21.62s & -72d11m17.2s & SLT3  & \\
00h59m20.42s & -72d11m22.2s & SLT4  & \\
00h59m17.30s & -72d11m25.1s & SLT5  & \\
00h59m15.26s & -72d09m19.1s & SLT6  & \\
00h59m14.69s & -72d11m03.1s & SLT7  & \\
00h59m13.49s & -72d09m54.7s & SLT8  & \\
00h59m12.29s & -72d09m58.3s & SLT9  & \\
00h59m10.13s & -72d10m51.2s & SLT10 & \\
00h59m09.24s & -72d10m57.0s & SLT11 & \\
00h59m06.74s & -72d10m25.3s & SLT12 & \\
00h59m05.98s & -72d11m26.9s & SLT13 & \\
00h59m05.52s & -72d10m35.8s & SLT14 & \\
00h58m59.11s & -72d10m23.2s & SLT15 & \\
00h58m59.02s & -72d10m28.6s & SLT16 & \\
00h58m58.32s & -72d09m50.0s & SLT17 & \\
00h58m56.95s & -72d09m54.0s & SLT18 & \\
00h58m52.42s & -72d09m24.1s & SLT19 & \\
\enddata

\tablenotetext{a}{The slits are labeled in order of decreasing slit centroid R.A. N.b. The slit labels do not necessarily correspond to the point source labels of the same number.}

\end{deluxetable}

\clearpage

\begin{deluxetable}{ccrrrrrrr}
\tabletypesize{\scriptsize}

\tablecaption{Atomic, Ionic, and Molecular Hydrogen Line
  Measurements \label{lines}}

\tablehead{ \colhead{Slit position} & \colhead{Type\tablenotemark{a}}
  & \colhead{HI 6-5} & \colhead{[Ar{\sc iii}]} & \colhead{H$_2$S(3)} &
  \colhead{[S{\sc iv}]} & \colhead{H$_2$S(2)} & \colhead{HI 7-6} &
  \colhead{[Ne{\sc ii}]} \\ & & \colhead{7.46\micron} &
  \colhead{8.99\micron} & \colhead{9.67\micron} &
  \colhead{10.51\micron} & \colhead{12.28\micron} &
  \colhead{12.37\micron} & \colhead{12.81\micron} \\ & &
  \multicolumn{7}{c}{$\times 10^{-21}$ W cm$^{-2}$} }

\startdata
PS1  & ps  &$<$0.47        &$<$0.52        &   2.2$\pm$0.3 &$<$0.11        &  0.77$\pm$0.12&$<$0.22        &   1.5$\pm$0.3 \\
PS2  & ps  &   3.3$\pm$0.7 &$<$0.47        &   3.8$\pm$0.5 &$<$0.41        &   1.6$\pm$0.2 &$<$0.12        &   4.7$\pm$0.8 \\
PS3  & ps  &   0.8$\pm$0.2 &$<$0.32        &  0.64$\pm$0.12&$<$0.18        &  0.41$\pm$0.08&$<$0.12        &   1.3$\pm$0.3 \\
PS4  & ps  &$<$0.80        &   1.6$\pm$0.4 &   1.3$\pm$0.2 &   2.6$\pm$0.5 &$<$0.91        &$<$0.30        &   6.2$\pm$1.0 \\
PS5  & ps  &$<$0.05        &$<$0.09        &   5.4$\pm$1.1 &$<$0.09        &$<$0.056       &$<$0.15        &   1.9$\pm$0.4 \\
PS6  & ps  &$<$0.29        &$<$0.09        &$<$0.38        &  0.39$\pm$0.10&$<$0.097       &$<$0.30        &   0.55$\pm$0.13 \\
PS7  & ps  &$<$1.0         &$<$0.80        &  0.54$\pm$0.11&$<$0.68        &$<$0.52        &$<$0.38        &$<$0.70 \\
PS8  & ps  &$<$0.40        &$<$0.42        &$<$0.59        &$<$0.45        &$<$0.11        &$<$0.10        &$<$0.38 \\
PS9  & ps  &$<$0.67        &$<$0.85        &$<$2.1         &$<$0.81        &$<$0.68        &$<$0.68        &   3.5$\pm$0.6 \\
PS10 & ps  &$<$0.62        &$<$0.17        &   1.3$\pm$0.2 &$<$0.42        &   0.54$\pm$0.9&$<$0.052       &   1.3$\pm$0.3 \\
PS11 & ps  &$<$0.45        &$<$0.05        &   1.4$\pm$0.2 &   1.5$\pm$0.3 &  0.46$\pm$0.08&  0.31$\pm$0.09&   1.3$\pm$0.3 \\
PS12 & ps  &$<$0.45        &$<$0.38        &  0.67$\pm$0.13&$<$0.39        &  0.29$\pm$0.06&$<$0.032       &   0.45$\pm$0.11 \\
PS13 & ps  &$<$0.37        &$<$0.33        &$<$0.37        &$<$0.14        &$<$0.35        &$<$0.35        &   0.44$\pm$0.10 \\
PS14 & ps  &$<$0.15        &$<$0.21        &   1.2$\pm$0.2 &$<$0.40        &  0.89$\pm$0.14&$<$0.037       &   0.98$\pm$0.21 \\
\hline
SLT3        & ee  &$<$4.0         &   4.3$\pm$0.9 &$<$0.52        &   41$\pm$5    &$<$0.39        &   1.7$\pm$0.4 &   4.2$\pm$0.8 \\
SLT4        & ee  &$<$2.9         &   1.2$\pm$2   &$<$1.2         &   55$\pm$6    &$<$0.79        &   3.7$\pm$0.7 &   4.2$\pm$0.7 \\
SLT5        & ee  &$<$3.9         &   8.3$\pm$1.7 &$<$4.7         &   60$\pm$0.7  &$<$1.3         &   3.5$\pm$0.7 &   5.3$\pm$0.9 \\
SLT6        & ee  &   2.0$\pm$0.5 &   3.5$\pm$0.8 &$<$0.26        &   54$\pm$6    &$<$1.6         &  0.60$\pm$0.16&   9.1$\pm$1.4 \\
SLT8        & ee  &   4.5$\pm$0.9 &   7.0$\pm$1.4 &$<$0.57        &   72$\pm$8    &$<$0.31        &   2.6$\pm$0.5 &   5.7$\pm$1.0 \\
SLT9        & ee  &$<$0.82        &   9.6$\pm$1.9 &$<$0.77        &   79$\pm$9    &$<$0.82        &$<$1.0         &   5.5$\pm$0.9 \\
SLT10       & ee  &   7.0$\pm$1.4 &   13$\pm$2    &$<$0.40        &   94$\pm$10   &$<$0.46        &   4.2$\pm$0.8 &   8.7$\pm$1.4 \\
SLT11       & ee  &$<$2.3         &   19$\pm$ 3   &$<$1.4         &   104$\pm$11  &$<$0.38        &   6.3$\pm$1.2 &   11$\pm$2 \\
SLT12       & ee  &   8.9$\pm$1.6 &   11$\pm$2    &$<$0.38        &   87$\pm$9    &$<$0.60        &   4.0$\pm$0.8 &   13$\pm$2 \\
SLT13       & ee  &$<$8.3         &   10$\pm$2    &$<$0.67        &   50$\pm$6    &$<$0.19        &   2.7$\pm$0.6 &   5.8$\pm$1.0 \\
SLT14       & ee  &$<$3.9         &   11$\pm$2    &$<$0.32        &   106$\pm$11  &   1.3$\pm$0.2 &   2.6$\pm$0.5 &   9.8$\pm$1.5 \\
SLT15       & ee  &   8.4$\pm$1.6 &   11$\pm$2    &$<$0.28        &   153$\pm$15  &$<$0.27        &   2.7$\pm$0.6 &   5.9$\pm$1.0 \\
SLT16 nod 1 & ee  &$<$11          &   18$\pm$3    &$<$1.8         &   116$\pm$12  &$<$1.3         &   6.4$\pm$1.2 &   8.0$\pm$1.3 \\
SLT16 nod 2 & ee  &$<$19          &   11$\pm$2    &$<$3.5         &   123$\pm$13  &   2.7$\pm$0.3 &   2.6$\pm$0.6 &   12$\pm$2 \\
SLT17       & ee  &$<$0.40        &   9.6$\pm$1.9 &$<$0.19        &   73$\pm$8    &$<$0.27        &   2.5$\pm$0.5 &   5.2$\pm$0.9 \\
SLT19       & ee  &$<$0.24        &   5.1$\pm$1.1 &$<$0.10        &   16$\pm$2    &$<$0.23        &   1.7$\pm$0.4 &   6.3$\pm$1.1 \\
\enddata

\tablenotetext{a}{``Type'' refers to the extraction method used to
  produce the spectrum: {\bf ps} means an optimally extracted point
  source, and {\bf ee} means extended emission extracted as a
  polynomial fit to the full-slit background.}

\end{deluxetable}

\begin{deluxetable}{ccrrrrc}
\tabletypesize{\scriptsize}
\tablecaption{PAH and continuum measurements\label{PAHs}}
\tablehead{

\colhead{Slit position} & \colhead{Type\tablenotemark{a}} &
\colhead{PAH$_{6.2\micron}$} & \colhead{PAH$_{7.7\micron}$} &
\colhead{PAH$_{8.6\micron}$} & \colhead{PAH$_{11.3\micron}$} &
\colhead{14\micron\ continuum} \\ & & &
& & & \\ & & \multicolumn{4}{c}{$\times 10^{-20}$ W cm$^{-2}$} &
\colhead{$\times 10^{-20}$ W cm$^{-2}$ \micron$^{-1}$} }

\startdata
PS1  & ps & 2.9$\pm$0.1   & 2.4$\pm$0.27  & 0.67$\pm$0.15 & 2.38$\pm$0.07 &  1.6 \\
PS2  & ps & 7.4$\pm$0.3   & 18$\pm$2      & 3.5$\pm$0.5   & 6.2$\pm$0.4   &  5.9 \\
PS3  & ps & 2.1$\pm$0.2   & 4.9$\pm$0.3   & 0.89$\pm$0.22 & 1.2$\pm$0.2   &  0.98 \\
PS4  & ps & 2.9$\pm$0.2   & 7.4$\pm$1.3   & 1.4$\pm$0.4   & 3.3$\pm$0.3   &  2.1 \\
PS5  & ps & 0.71$\pm$0.03 & 2.10$\pm$0.9  & 0.71$\pm$0.03 & 0.64$\pm$0.02 &  0.68 \\
PS6  & ps & 0.28$\pm$0.02 & 0.96$\pm$0.04 & 0.21$\pm$0.05 & 0.52$\pm$0.01 &  0.65 \\
PS7  & ps & 1.6$\pm$0.3   & 11$\pm$1      & 0.99$\pm$0.43 & 1.7$\pm$0.1   &  18 \\
PS8  & ps & 2.5$\pm$0.4   & 6.3$\pm$1.1   & 1.2$\pm$0.3   & 2.26$\pm$0.06 &  3.0 \\
PS9  & ps & 2.8$\pm$0.5   & 0.05$\pm$0.06 & 2.8$\pm$0.5   & 1.7$\pm$0.1   &  16 \\
PS10 & ps & 1.61$\pm$0.05 & 3.2$\pm$0.3   & 0.59$\pm$0.08 & 1.50$\pm$0.03 &  1.2 \\
PS11 & ps & 0.99$\pm$0.05 & 2.6$\pm$0.2   & 0.54$\pm$0.06 & 1.01$\pm$0.03 &  0.85 \\
PS12 & ps & 0.81$\pm$0.08 & 8.1$\pm$0.7   & 1.0$\pm$0.2   & 1.45$\pm$0.07 &  3.5 \\
PS13 & ps & 0.62$\pm$0.02 & 1.67$\pm$0.07 & 0.25$\pm$0.04 & 0.50$\pm$0.01 &  0.11 \\
PS14 & ps & 1.20$\pm$0.07 & 3.5$\pm$0.2   & 0.48$\pm$0.08 & 2.1$\pm$0.1   &  0.60 \\
\hline
SLT3        & ee  & 2.6$\pm$0.3   & 3.4$\pm$0.34  & 1.1$\pm$0.2   & 1.3$\pm$0.3   & 2.2 \\
SLT4        & ee  & 3.1$\pm$0.31  & 18.6$\pm$0.9  & 0.81$\pm$0.30 & 2.2$\pm$0.2   & 4.8 \\
SLT5        & ee  & 4.3$\pm$0.4   & 17$\pm$2      & 1.9$\pm$0.6   & 2.9$\pm$0.4   & 5.3 \\
SLT6        & ee  & 5.2$\pm$0.3   & 11.4$\pm$0.8  & 3.3$\pm$0.3   & 4.2$\pm$0.3   & 4.3 \\
SLT8        & ee  & 3.7$\pm$0.3   & 7.3$\pm$0.56  & 1.9$\pm$0.4   & 2.5$\pm$0.5   & 3.8 \\
SLT9        & ee  & 2.4$\pm$0.4   & 10.1$\pm$0.9  & 0.51$\pm$0.35 & 1.4$\pm$0.2   & 4.9 \\
SLT10       & ee  & 6.7$\pm$0.4   & 19$\pm$1      & 2.6$\pm$0.8   & 5.5$\pm$0.6   & 8.6 \\
SLT11       & ee  & 3.5$\pm$0.3   & 19$\pm$1      & 1.7$\pm$0.5   & 2.5$\pm$0.2   & 9.7 \\
SLT12       & ee  & 8.0$\pm$0.5   & 17$\pm$2      & 3.0$\pm$0.8   & 7.4$\pm$0.7   & 9.1 \\
SLT13       & ee  & 2.3$\pm$0.3   & 5.2$\pm$0.7   & 1.1$\pm$0.3   & 1.3$\pm$0.1   & 4.4 \\
SLT14       & ee  & 4.0$\pm$0.5   & 21$\pm$2      & 1.9$\pm$0.6   & 3.3$\pm$0.3   & 7.9 \\
SLT15       & ee  & 3.0$\pm$0.2   & 6.9$\pm$0.69  & 1.2$\pm$0.4   & 1.0$\pm$0.2   & 8.6 \\
SLT16 nod 1 & ee  & 6.8$\pm$2.2   & 1.9$\pm$2.4   & 2.2$\pm$1.8   & 4.0$\pm$1.0   & 9.6 \\
SLT16 nod 2 & ee  & 4.9$\pm$2.0   & 9.6$\pm$4.0   & 2.3$\pm$1.9   & 4.4$\pm$1.0   & 12 \\
SLT17       & ee  & 1.9$\pm$0.1   & 4.6$\pm$0.3   & 0.60$\pm$0.14 & 1.20$\pm$0.07 & 1.7 \\
SLT19       & ee  & 3.8$\pm$1.0   & 9.5$\pm$0.3   & 2.6$\pm$0.2   & 2.02$\pm$0.06 & 2.3 \\
\enddata

\tablecomments{The PAH measurements here are those determined using
  PAHFIT. 15\% errors are assumed on continuum fluxes.}

\tablenotetext{a}{``Type'' refers to the extraction method used to
  produce the spectrum: {\bf ps} means an optimally extracted point
  source, and {\bf ee} means extended emission extracted as a
  polynomial fit to the full-slit background.}

\end{deluxetable}

\end{document}